\documentclass[prb,twocolumn]{revtex4}

\usepackage{graphicx}
\usepackage{amssymb}
\usepackage{amsmath}
\usepackage{subfigure}

\usepackage{color}

\begin{document}

\title{Quantum kinetic approach to the calculation of the Nernst effect}
\author{Karen~Michaeli$^1$ and Alexander~M.~Finkel'stein$^{1,2}$}
\affiliation{$^1$ Department of Condensed Matter Physics, The Weizmann Institute of Science,
Rehovot 76100, Israel \\ $^2$ Department of Physics, Texas A\&M University, College Station, TX $77843-4242$, USA}

\begin{abstract}
We show that the strong Nernst effect observed recently in amorphous superconducting films far above the critical temperature is caused by the fluctuations of the superconducting order parameter.  We employ the quantum kinetic approach~\cite{QKE} for the derivation of the Nernst coefficient.  We present here the main steps of the calculation and discuss some subtle issues  that we encountered while calculating the Nernst coefficient. In particular, we demonstrate that in the limit $T\rightarrow0$ the contribution of the magnetization ensures the vanishing of the Nernst signal in accordance with the third law of thermodynamics. We obtained a striking agreement between our theoretical calculations and the experimental data in a broad region of temperatures and magnetic fields.
\end{abstract}
\maketitle

\section{Introduction}
\label{intro}

After many years in the shade, the Nernst effect  (the transverse thermoelectric signal) entered the spotlight in condensed matter physics as well as other fields of research, such as the theory of gravitation.~\cite{Zaanen2007,Sachdev2007} The "rediscovery" of the Nernst effect by the condensed matter community occurred after the measurement of the effect in high-$T_c$ materials above the superconducting transition temperatures.~\cite{Ong2000,Ong2005} Since then, the Nernst effect was also observed in conventional amorphous superconducting films far above $T_c$.~\cite{Aubin2006,Aubin2007} The Nernst effect in high-$T_c$ superconductors~\cite{Ong2000,Ong2005} has been attributed to the motion of vortices~\cite{Anderson2007,Podolsky2007,Huse2004} existing even above $T_c$ (the vortex-liquid regime). In conventional amorphous superconducting films the strong Nernst signal observed deep in the normal state~\cite{Aubin2006,Aubin2007} cannot be explained by the vortex-like fluctuations. Rather, the authors of Refs.~\onlinecite{Aubin2006,Aubin2007} suggested that the effect is caused by fluctuations of the superconducting order parameter. Here we present a comprehensive analysis of this mechanism using the quantum kinetic technique and demonstrate a quantitative agreement between the theoretical expressions and the experiment.~\cite{Aubin2007} No fitting parameters have been used; the values of $T_c$ and the diffusion coefficient were taken from independent measurements (see Refs.~\onlinecite{Aubin2006,Aubin2007}). In particular, we succeeded in reproducing the non-trivial dependence of the signal on the magnetic field. Our results imply that in the quest for understanding the thermoelectric phenomena in high-$T_c$ materials the fluctuations of the order parameter should not be ignored.

The Nernst effect and its counterpart, the Ettingshausen effect, are effective tools for studding the superconducting fluctuations because in metallic conductors the contribution of the quasi-particle excitations is negligible. Under the approximation of a constant density of states at the Fermi energy, which is a standard approximation for the Fermi liquid theory, this contribution vanishes completely.~\cite{Sondheimer1948} On the other hand, the collective modes describing all kinds of fluctuations can in general generate significant contributions to the Nernst effect. Since the neutral modes are not deflected by the Lorentz force, they do not contribute to the transverse thermoelectric current. The charged modes, such as fluctuations of superconducting order parameter, are a possible source for the giant Nernst effect even far from the superconducting transition. The fact that the main contribution to the Nernst signal originates from the superconducting fluctuations is in contrast to other transport phenomena such as the electric conductivity. The contributions to the electric conductivity caused by the superconducting fluctuations (paraconductivity~\cite{Aslamazov1968,Maki1968,Varlamov}) can be observed close enough to the superconducting transition where the paraconductivity increases rapidly and may even overcome the Drude conductivity. Far from the transition the superconducting fluctuations produce only one among many corrections to the conductivity and, therefore, can hardly be identified. Owing to the fact that in the absence of fluctuations the Nernst effect is negligible, measurements of the Nernst signal provide a unique opportunity to study the superconducting fluctuations deep inside the normal state.

\begin{figure}[pt]
\begin{flushright}\begin{minipage}{0.5\textwidth}  \centering
        \includegraphics[bb={34 237 578 554},width=0.6\textwidth]{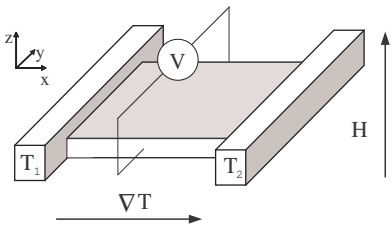} \hspace{0.05in}
                 \caption[0.4\textwidth]{\small The setup of the Nernst effect measurement. The sample is placed between two thermal baths of different temperatures. The temperature gradient is in the $x$-direction, the magnetic field is along the $z$-direction and the electric field is induced in the $y$-direction.} \label{fig:NernstSetup}
\end{minipage}\end{flushright}
\end{figure}

The transport coefficients for the electric and thermal currents are defined via the standard conductivity tensor:
\begin{equation}
\left(\begin{array}{c}
\mathbf{j}_{e} \\
\mathbf{j}_{h}\end{array}\right)=
\left(
\begin{array}{cc}
\hat{\sigma} & \hat{\alpha} \\
\hat{\tilde{\alpha}} & \hat{\kappa} \\
\end{array}\right)\left(\begin{array}{c}
\mathbf{E} \\
-\boldsymbol{\nabla}{T}
\end{array}\right).
\label{eq:Intro-ConductivityTensor}
\end{equation}%
When the thermo-magnetic phenomena are studied in films (or layered conductors) the magnetic field is conventionally directed perpendicularly to the conducting plane, see Fig.~\ref{fig:NernstSetup}. Then,  each element of the conductivity tensor corresponds to a $2\times2$ matrix describing the conductivity components in the $x-y$ plane (see Fig.~\ref{fig:NernstSetup}). The different components of the conductivity tensor are connected through the Onsager relations. In particular, $\sigma_{ij}(\mathbf{H})=\sigma_{ji}(-\mathbf{H})$ and $\tilde{\alpha}_{ij}(\mathbf{H})=T{\alpha}_{ji}(-\mathbf{H})$. In an open circuit setup, from the condition $\mathbf{j}_e=0$ one gets that the Nernst coefficient is:
\begin{equation}
e_{N}=\frac{E_{y}}{-\boldsymbol{\nabla}_{x}{T}}=\frac{\sigma_{xx}\alpha_{xy}-\sigma_{xy}\alpha_{xx}}{\sigma_{xx}^{2}+\sigma_{xy}^2}.
\label{eq:Intro-NernstCoefficient}
\end{equation}%
We checked that the second term in the numerator is negligible in comparison to the first one (see the comment below Eq.~\ref{eq:QKE-L}). This observation has been experimentally verified as it follows from Fig.2(a) in  Ref.~\onlinecite{Aubin2006}. Therefore, the leading order term in the expression for the Nernst coefficient is $e_{N}\approx\alpha_{xy}/\sigma_{xx}$ and our goal is to find the transverse Peltier coefficient, $\alpha_{xy}$.

The electric current generated as a response to an external force, such as the electric field, can be found in the linear regime by the Kubo formula~\cite{Kubo1957} which expresses the response in terms of a corresponding correlation function. Extending the Kubo formalism to the calculation of the response to the temperature gradient is not trivial because this gradient is not directly connected to any mechanical force. Following the scheme used in the derivation of the Einstein relation, Luttinger~\cite{Luttinger1964} made a connection between the responses to the temperature gradient and to an auxiliary gravitational field. As a result, Luttinger succeeded in relating all transport coefficients with various current-current correlation functions. A main ingredient in the Kubo formula is the quantum mechanical expression for the current operators that enter the correlation function; e.g., the electric and heat currents in case of the thermoelectric transport. When the electron-electron interactions are neglected, the expression for the heat current operator is
\begin{align}\label{eq:CurrentNonInt}
\mathbf{j}_h(\mathbf{q}=0,\tau)&=\sum_{\mathbf{p},\sigma}\frac{\partial\varepsilon_{\mathbf{p}}}{\partial\mathbf{p}}(\varepsilon_{\mathbf{p}}-\mu)c_{\mathbf{p},\sigma}^{\dag}(\tau)c_{\mathbf{p},\sigma}(\tau)\\\nonumber
&+\sum_{\mathbf{p},\mathbf{p}',\sigma}\frac{\partial\varepsilon_{\mathbf{p}}}{\partial\mathbf{p}}V_{imp}(\mathbf{p},\mathbf{p}')c_{\mathbf{p},\sigma}^{\dag}(\tau)c_{\mathbf{p}',\sigma}(\tau),
\end{align}
where $c_{\mathbf{p},\sigma}^{\dag}(\tau)$ ($c_{\mathbf{p},\sigma}(\tau)$) is the creation (annihilation) operator of an electron in a state with energy $\varepsilon_{\mathbf{p}}$. Here $\mu$ is the chemical potential, $V_{imp}(\mathbf{p},\mathbf{p}')$ is the potential created by the disorder and $\tau$ is the imaginary time. With the help of the equations of motion and after transforming to the Matsubara frequencies, the current operator can be written as:~\cite{commentH}
\begin{align}\label{eq:CurrentNonIntFrequency}
\mathbf{j}_h(\mathbf{q}=0,\omega_n)=\sum_{\mathbf{p},\epsilon_n,\sigma}\frac{\partial\varepsilon_{\mathbf{p}}}{\partial\mathbf{p}}\frac{2i\epsilon_n-i\omega_n}{2}c_{\mathbf{p},\sigma}^{\dag}(\epsilon_n)c_{\mathbf{p},\sigma}(\epsilon_n-\omega_n).
\end{align}
When electron-electron interactions are included, $\mathbf{j}_{h}$ is more complicated function (it contains terms with four fermion operators).  In general, the resulting expression for the heat current is not just the frequency multiplied by the velocity as it is for free electrons. Unfortunately, very often the expression for the heat current of free electrons presented in Eq.~\ref{eq:CurrentNonIntFrequency} is used in the presence of electron-electrons interactions, when there is no real justification for it. In Appendix~B of Ref.~\onlinecite{QKE} we showed that this simplified form of the Kubo formula fails to reproduce  the thermal conductivity of Fermi liquids. The incorrect result that emerges from Eq.~\ref{eq:CurrentNonIntFrequency} does not imply that the use of the Kubo formula for the thermal transport coefficients is necessarily wrong. The weak point is in replacing the full expression for the heat current by the one in Eq.~\ref{eq:CurrentNonIntFrequency}. The problem with the full expression for the heat current is in its complexity.

In addition, the Kubo formalism meets with some difficulties when the thermoelectric currents are considered in the presence of a magnetic field. Obraztsov~\cite{Obraztsov1965} pointed out that when a magnetic field is applied, the heat current describing the change in the entropy must include a contribution from the magnetization. This is because the thermodynamic expression for the heat contains the magnetization term. Thus, additional problem of the Kubo formula is that the current cannot be expressed entirely by a correlation function. In order to determine the transverse  thermoelectric currents one needs to combine the quantum mechanical response to the external field with the magnetization, which is a thermodynamic quantity.~\cite{Obraztsov1965,Streda1977,Halperin1997}

In the derivation of the thermoelectric currents we decided, instead of applying the Kubo formula, to employ a different approach and to use the quantum kinetic equation.~\cite{Keldysh1964,Rammer1986,Haug} One main advantage of the quantum kinetic approach is that the problem of the magnetization current is solved straightforwardly. We directly obtained the expression for the thermoelectric current which includes the magnetization current. In this way, the electric current generated by the temperature gradient can be related to the flow of entropy. Therefore, according to the third law of thermodynamics the Nernst signal must vanish at $T\rightarrow0$.~\cite{Hu1976}
As we will see, this argument imposes a strict constraint on the magnitude of the Peltier coefficient in a broad range of temperatures. Note that the calculation of the thermoelectric transport using the kinetic equation allows a direct verification of the Onsager relations between the off-diagonal components of the conductivity tensor (see Appendix~\ref{sec:OR}).

The paper is organized as follows: in Secs.~\ref{sec:QKE} and~\ref{sec:Current} we present the main steps in the derivation of the electric current as a response to a temperature gradient in the presence of fluctuations of the superconducting order parameter using the quantum kinetic equation. Then, in Secs.~\ref{sec:Peltier} and~\ref{sec:FinalExpression} we give details of the calculation that are specific to the transverse current. We devote Sec.~\ref{sec:Magnetization} and Appendix~\ref{App:Magnetization} to the contribution of the magnetization current to the transverse thermoelectric current. We demonstrate that the magnetization current ensures the vanishing of the  Peltier coefficient in the limit $T\rightarrow0$. This makes the Nernst signal compatible with the third law of thermodynamics. The result of the calculation of the Nernst effect and comparison with the Nernst signal measured in amorphous superconducting films~\cite{Aubin2007} are presented in Sec.~\ref{sec:phaseDiagram}. The content of Sec.~\ref{sec:phaseDiagram} has already been published~\cite{KM2008}  as a separate letter; we include it here for completeness. In Appendix~\ref{sec:OR} we demonstrate that  the two off-diagonal coefficients of the conductivity tensor that are found independently using the quantum kinetic approach satisfy the Onsager relations, $\alpha_{ij}(\mathbf{B})=T\tilde{\alpha}_{ji}(-\mathbf{B})$. In view of the frequently used argument that the particle-hole symmetry limits the magnitude of the Nernst effect (see e.g. Ref.~\onlinecite{Reizer2008}) we discuss this issue in Appendix~\ref{sec:p-hSymmtery}. We demonstrate that the value of the Nernst coefficient is not constrained by the particle-hole symmetry. Rather, the contribution from the quasi-particle excitations is zero when their density of states is taken to be constant, which is often confused with the particle-hole symmetry.

\section{The quantum kinetic equation above $T_c$ in the presence of a temperature gradient}\label{sec:QKE}

In this paper we extend the scheme developed in Ref.~\onlinecite{QKE} to the case of electrons interacting with superconducting fluctuations in the presence of a magnetic field. Here we describe the system using two fields; one is the quasi-particle field  $\psi$, while the other represents the fluctuations of the superconducting order parameter $\Delta$. The matrix functions $\hat{G}(\mathbf{r},t;\mathbf{r}',t')$ and $\hat{L}(\mathbf{r},t;\mathbf{r}',t')$ written in the Keldysh form~\cite{Keldysh1964,Rammer1986,Haug} describe the propagation of these two fields, respectively. Throughout the paper, we  work in the basis of the retarded, advanced and Keldysh propagators:
\begin{align}
\hat{G}(\mathbf{r},t;\mathbf{r}',t')=\left(
                                       \begin{array}{cc}
                                         G^{R}(\mathbf{r},t;\mathbf{r}',t') & G^{K}(\mathbf{r},t;\mathbf{r}',t') \\
                                         0 & G^{A}(\mathbf{r},t;\mathbf{r}',t') \\
                                       \end{array}
                                     \right),
\end{align}
where a similar expression can be written for $\hat{L}$. [Notice that we use the term propagators when referring to both these functions, while separately we name $\hat{G}(\mathbf{r},t;\mathbf{r}',t')$ the quasi-particle Green function and
$\hat{L}(\mathbf{r},t;\mathbf{r}',t')$ the propagator of the superconducting fluctuations.] The derivation of the transport coefficients in the quantum kinetic equation is separated into two steps. First,  the propagators are found using the quantum kinetic equations. Then, the expression for the current in terms of the propagators is derived.

We now derive the quantum kinetic equations for the propagators $\hat{G}$ and $\hat{L}$ in the presence of a temperature gradient.  Inspired by Luttinger~\cite{Luttinger1964}, we introduce an auxiliary gravitational field  of the form $\gamma(\mathbf{r})=T_0/T(\mathbf{r})$ (where $T_0$ is the constant part of the temperature). The purpose of the  gravitational field is to compensates for the non-uniform temperature at the initial state. In other words, the temperature gradient and the gravitational field are applied in such a way that at $t=-\infty$ the system is in equilibrium. Then, starting at $t=-\infty$, the gravitational field  is adiabatically switched off. From the response to switching off the gravitational field, we can learn about the effect of the temperature gradient on the system (for more details see Ref.~\onlinecite{QKE}). The general expression for the action in the presence of a  gravitational field $\gamma(\mathbf{r})$ and a vector potential $\mathbf{A}(\mathbf{r})$ is
\begin{align}\label{eq:QKE-S}
&\mathcal{S}=\int{d\mathbf{r}dt}\gamma(\mathbf{r})\left\{\sum_{\sigma}\left[\frac{i}{\gamma(\mathbf{r})}\psi_{\sigma}^{\dag}(\mathbf{r},t)\frac{\partial}{\partial{t}}\psi_{\sigma}(\mathbf{r},t)\right.\right.\\\nonumber
&\left.\left.\hspace{25mm}-\frac{1}{2m}\hspace{-0.5mm}\Big{|}\hspace{-0.5mm}\left(\hspace{-0.5mm}\boldsymbol{\nabla}-\frac{ie}{c}\mathbf{A}(\mathbf{r})\hspace{-0.5mm}\right)\hspace{-0.5mm}\psi_{\sigma}(\mathbf{r},t)\Big{|}^2\hspace{-1mm}\right.\right.\\\nonumber
&\left.\left.\hspace{25mm}-V_{imp}(\mathbf{r})\sum_{\sigma}\psi_{\sigma}^{\dag}(\mathbf{r},t)\psi_{\sigma}(\mathbf{r},t)\right.\right.\\\nonumber
&\left.\left.-\frac{\sigma}{2}\left(\Delta(\mathbf{r},t)\psi_{\sigma}^{\dag}(\mathbf{r},t)\psi_{-\sigma}^{\dag}(\mathbf{r},t)+h.c.\right)\phantom{\frac{i\partial_t}{\gamma(\mathbf{r})}}\hspace{-7mm}
\right]-\frac{|\Delta(\mathbf{r},t)|^2}{\lambda}\right\}.
\end{align}
Here $\lambda$ is the coupling constant of the interaction (we are interested in the case of an $s$-wave coupling). The choice of signs is such that $\lambda>0$ corresponds to an attractive interaction. The spin index $\sigma=1(-1)$, or equivalently $\uparrow(\downarrow)$, indicates the spin direction up (down). In the above equation and throughout the paper we set $\hbar=1$.

The Dyson equation for the Green function in the presence of the gravitational field is:
\begin{align}\label{eq:QKET-DEG}
&\left[i\frac{\partial}{\partial{t}}+\frac{1}{2m}\left(\boldsymbol{\nabla}-\frac{ie}{c}\mathbf{A}(\mathbf{r})\right)\gamma(\mathbf{r})\left(\boldsymbol{\nabla}-\frac{ie}{c}\mathbf{A}(\mathbf{r})\right)\right.\\\nonumber
&\left.-\gamma(\mathbf{r})\left(V_{imp}(\mathbf{r})-\mu\right)
\phantom{\frac{E}{E}}\hspace{-4mm}\right]\hat{G}(\mathbf{r},t;\mathbf{r}',t')=\delta(\mathbf{r}-\mathbf{r}')\delta(t-t')\\\nonumber
&+\gamma(\mathbf{r})\int{dt_{\scriptscriptstyle1}d\mathbf{r}_{\scriptscriptstyle1}}\hat{\Sigma}(\mathbf{r},t;\mathbf{r}_{\scriptscriptstyle1},t_{\scriptscriptstyle1})\gamma(\mathbf{r}_{\scriptscriptstyle1})\hat{G}(\mathbf{r}_{\scriptscriptstyle1},t_{\scriptscriptstyle1};\mathbf{r}',t').
\end{align}
In general, the Green function $\hat{G}$ and self-energy $\hat{\Sigma}$ contain spin indices. Since we do not consider scattering mechanisms that flip the spins and ignore the Zeeman splitting, here and in the following we do not indicate the spin indices whenever it is possible. Next, we introduce the following transformation:
\begin{align}\label{eq:QKET-Transformation}
\hat{Y}(\mathbf{r},t;\mathbf{r}',t')=\gamma^{-1/2}(\mathbf{r})\hat{\underline{\underline{Y}}}(\mathbf{r},t;\mathbf{r}',t')\gamma^{-1/2}(\mathbf{r}'),
\end{align}
where $\hat{Y}$ can be either $\hat{G}$ or $\hat{\Sigma}$. For the calculation of the response to switching off the gravitational field  in the linear regime, we set $\gamma(\mathbf{r})=1+\mathbf{r}\cdot\boldsymbol{\nabla}T/T_0$. Then, the quantum kinetic equation for the Green function of the quasi-particles becomes:
\begin{align}\label{eq:QKE-DEG}
&\left[i\left(1-\frac{\mathbf{r}\cdot\boldsymbol{\nabla}T}{T_0}\right)\frac{\partial}{\partial{t}}+\frac{1}{2m}\left(\boldsymbol{\nabla}-\frac{ie}{c}\mathbf{A}(\mathbf{r},t)\right)^2\right.\\\nonumber&\left.-V_{imp}(\mathbf{r})+\mu\phantom{\frac{1}{1}}\hspace{-2mm}\right]\underline{\underline{\hat{G}}}(\mathbf{r},t;\mathbf{r}',t')=\delta(\mathbf{r}-\mathbf{r}')\delta(t-t')\\\nonumber
&+\int{dt_{\scriptscriptstyle1}d\mathbf{r}_{\scriptscriptstyle1}}\underline{\underline{\hat{\Sigma}}}(\mathbf{r},t;\mathbf{r}_{\scriptscriptstyle1},t_{\scriptscriptstyle1})\underline{\underline{\hat{G}}}(\mathbf{r}_{\scriptscriptstyle1},t_{\scriptscriptstyle1};\mathbf{r}',t').
\end{align}
The dependence of this equation on the temperature gradient is much simplified by the transformation to $\hat{\underline{\underline{G}}}$ and $\hat{\underline{\underline{\Sigma}}}$ because $\boldsymbol{\nabla}T/T_0$ was eliminated from all the terms in the equation except the derivative with respect to time. In field theories which include a non-trivial space-time metric $\hat{g}$ the transformation of the kind $\underline{\underline{Y}}(\mathbf{x},\mathbf{x}')=\sqrt{-\det{\hat{g}(\mathbf{x})}}Y(\mathbf{x},\mathbf{x}')\sqrt{-\det{\hat{g}(\mathbf{x}')}}$ (where $\mathbf{x}$ is a $4$-vector) is standard. The success of this transformation in simplifying the quantum kinetic equation is due to the relatively simple structure of the metric.

We write Eq.~\ref{eq:QKE-DEG} in coordinate space because all the propagators and self energies that enter the kinetic equation are not translationally invariant. There are three sources for the lack of
translation invariance. The first one is because the propagators and
self energies  depend on the magnetic field through
the vector potential, which is a function of the coordinate. The second reason is due to the fact that we did not yet
perform the averaging over the disorder. Finally, and most important, in the presence of a temperature gradient (even in the
absence of a magnetic field) the propagators become functions of
the center of mass coordinate.

\begin{figure}[pt]
\begin{flushright}\begin{minipage}{0.5\textwidth}  \centering
        \includegraphics[width=0.85\textwidth]{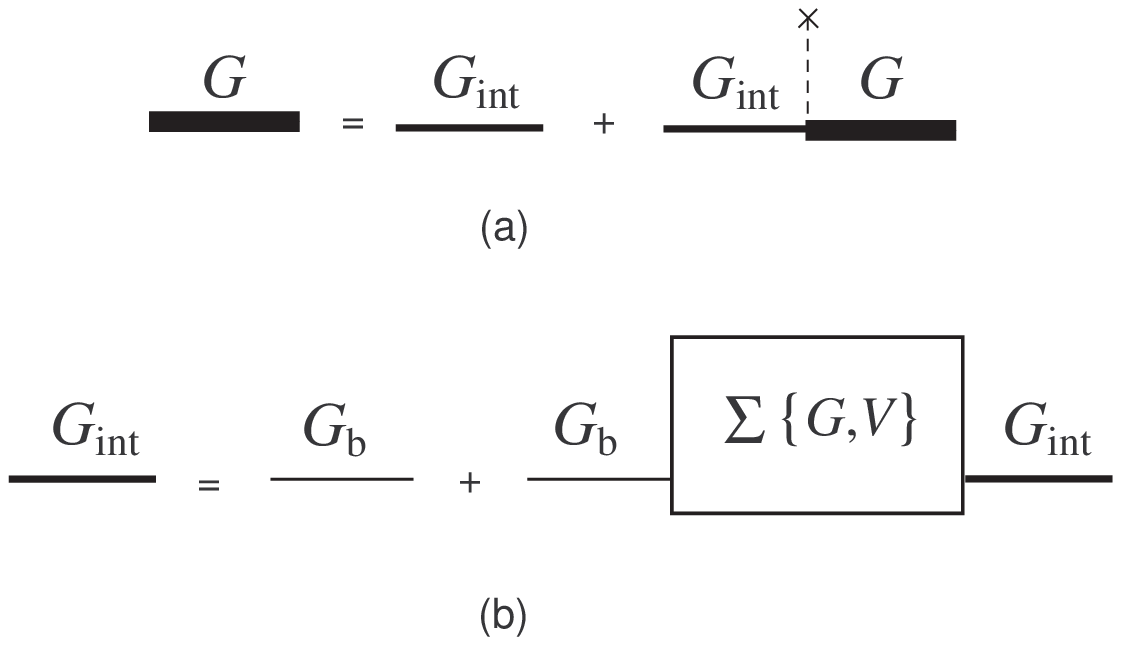}
                 \caption[0.4\textwidth]{\small (a) Illustration of Eq.~\ref{eq:EC-OpenDisorderG1} for the full Green function $\hat{G}$. (b) The Dyson equation for $\hat{G}_{int}$ (see Eq.~\ref{eq:EC-OpenDisorderG2}). Note that $\hat{G}_{int}$ includes scattering by impurities only through $\hat{\Sigma}(G)$ which is a function of the full Green function $\hat{G}$. The bare Green function, i.e., free from the interactions and the scattering by impurities, is denoted by $\hat{G}_{b}$. } \label{fig:GreenFunction}
\end{minipage}\end{flushright}
\end{figure}

We choose to postpone the averaging over impurities until the last stage of the derivation of the current. Therefore,  the Green function of
the quasi-particles contains open impurity lines as illustrated in the two coupled equations presented in Fig.~\ref{fig:GreenFunction}:
\begin{subequations}\label{eq:EC-OpenDisorderG}
\begin{align}\label{eq:EC-OpenDisorderG1}
&\hat{G}(\mathbf{r},t;\mathbf{r}',t')=\hat{G}_{int}(\mathbf{r},t;\mathbf{r}',t')\\\nonumber
&\hspace{5mm}+\int{d\mathbf{r}_{\scriptscriptstyle1}dt_{\scriptscriptstyle1}}\hat{G}_{int}(\mathbf{r},t;\mathbf{r}_{\scriptscriptstyle1},t_{\scriptscriptstyle1})V_{imp}(\mathbf{r}_{\scriptscriptstyle1})\hat{G}(\mathbf{r}_{\scriptscriptstyle1},t_{\scriptscriptstyle1};\mathbf{r}',t');
\end{align}
\begin{align}\label{eq:EC-OpenDisorderG2}\nonumber
&\hat{G}_{int}(\mathbf{r},t;\mathbf{r}',t')=\hat{G}_b(\mathbf{r},t;\mathbf{r}',t')+\hspace{-1.5mm}\int\hspace{-1.5mm}{d\mathbf{r}_{\scriptscriptstyle1}dt_{\scriptscriptstyle1}}{d\mathbf{r}_{\scriptscriptstyle2}dt_{\scriptscriptstyle2}}\hat{G}_b(\mathbf{r},t;\mathbf{r}_{\scriptscriptstyle1},t_{\scriptscriptstyle1})\\
&\times\hat{\Sigma}(\mathbf{r}_{\scriptscriptstyle1},t_{\scriptscriptstyle1};\mathbf{r}_{\scriptscriptstyle2},t_{\scriptscriptstyle2})\hat{G}_{int}(\mathbf{r}_{\scriptscriptstyle2},t_{\scriptscriptstyle2};\mathbf{r}',t').
\end{align}
\end{subequations}
Here, $\hat{G}_{int}(\mathbf{r},t;\mathbf{r}',t')$ is the Green function of interacting electrons, while $\hat{G}_{b}(\mathbf{r},t;\mathbf{r}',t')$ is free from both the interactions and the scattering by impurities. Note, that $\hat{G}_{int}(\mathbf{r},t;\mathbf{r}',t')$ includes partially the scattering by impurities.

Next we write the quantum kinetic equation using the center of mass coordinates for space and time, $\mathbf{R}=(\mathbf{r+r}')/2$, $\mathcal{T}=(t+t')/2$ and the
relative space and time coordinates, $\boldsymbol{\rho}=\mathbf{r-r}'$, $\tau=t-t'$.
Since the gravitational field is independent of time and we are interested in the steady state solution, the Green function will be taken to be independent of $\mathcal{T}$. On the other hand, the dependence of the Green function on  $\mathbf{R}$ remains, because the temperature gradient enters the equation as the product $\mathbf{r}\cdot\boldsymbol{\nabla}T=(\mathbf{R}+\boldsymbol{\rho}/2)\cdot\boldsymbol{\nabla}T$. This dependence on $\mathbf{R}$ is the main difference between the response to a temperature gradient and the response to an electric field. The point is that in the presence of an electric field the quantum kinetic equation can be formulated in such a way that the electric field enters only as a product with the  relative coordinate, $(\mathbf{r-r}')\cdot\mathbf{E}$. Therefore, after averaging over the disorder the electric field dependent Green function becomes translationally invariant.

In order to find the expression for the $\boldsymbol{\nabla}T$-dependent Green function using the quantum kinetic equation, we separate the Green function into three parts:
\begin{align}\label{eq:QKE-G_T}
\hat{\underline{\underline{G}}}=\hat{g}_{eq}+\hat{G}_{loc-eq}+\hat{G}_{\boldsymbol{\nabla}T}.
\end{align}
The first part describes the propagation at equilibrium. The retarded and advanced components of $\hat{g}_{eq}$  are described by Eq.~\ref{eq:QKE-DEG} with $\boldsymbol{\nabla}T=0$:
\begin{align}\label{eq:QKE-G_eq}
&\left[\epsilon+\frac{1}{2m}\left(\boldsymbol{\nabla}-\frac{ie}{c}\mathbf{A}(\mathbf{R}+\boldsymbol{\rho}/2,t)\right)^2
\right.\\\nonumber&\left.-V_{imp}(\mathbf{R}+\boldsymbol{\rho}/2)+\mu\phantom{\frac{1}{1}}\hspace{-2mm}\right]g_{eq}^{R,A}(\boldsymbol{\rho},\epsilon;\mathbf{A},imp)\\\nonumber
&-\int{d\mathbf{r}_{\scriptscriptstyle1}}\sigma_{eq}^{R,A}(\boldsymbol{\rho}-\mathbf{r}_{\scriptscriptstyle1},\epsilon;\mathbf{A},imp)g_{eq}^{R,A}(\mathbf{r}_{\scriptscriptstyle1},\epsilon;\mathbf{A},imp)
=\delta(\boldsymbol{\rho}).
\end{align}
This is the usual Dyson equation for the Green function at equilibrium, in which we performed the Fourier transform of the relative time $\tau$. In the above equation we introduced the equilibrium self-energy, $\hat{\sigma}_{eq}$. The Green function $\hat{g}_{eq}$ depends on the center of mass coordinate through the vector potential and the potential $V_{imp}$ created by the impurities at a specific realization. Correspondingly, we use the notation $g_{eq}(\boldsymbol{\rho},\epsilon;\mathbf{A},imp)$ in which these dependencies on $\mathbf{R}$ are incorporated into $\mathbf{A}$ and $imp$. The gradient, $\boldsymbol{\nabla}=\frac{1}{2}\boldsymbol{\nabla}_R+\boldsymbol{\nabla}_{\rho}$, in the equation for $g_{eq}^{R,A}$ contains the derivatives with respect to both $\mathbf{R}$ and $\boldsymbol{\rho}$.

According to the standard rule, the Keldysh component of the Green function at equilibrium can be written in terms of the Fermi distribution function $n_F(\epsilon)$ and the retarded and advanced Green functions:
\begin{align}\label{eq:QKE-G_eqK}
&g_{eq}^{K}(\boldsymbol{\rho},\epsilon;\mathbf{A},imp)=(1-2n_F(\epsilon))\\\nonumber
&\hspace{10mm}\times\left[g_{eq}^{R}(\boldsymbol{\rho},\epsilon;\mathbf{A},imp)-g_{eq}^{A}(\boldsymbol{\rho},\epsilon;\mathbf{A},imp)\right].
\end{align}

In the presence of a uniform and constant in time magnetic field, the expressions given in Eq.~\ref{eq:QKE-G_eq} can be rewritten as a product of the phase
\begin{align}\label{eq:QKE-Phase}
&\exp\left\{i\frac{e}{c}\int_{\mathbf{r}'}^{\mathbf{r}}\mathbf{A}\cdot(\mathbf{r}_{\scriptscriptstyle1})d\mathbf{r}_{\scriptscriptstyle1}\right\}\\\nonumber
&\hspace{15mm}=\exp\left\{-i\frac{e\mathbf{B}}{4c}\cdot\left[(\mathbf{r-r}')\times(\mathbf{r+r}')\right]\right\},
\end{align}
and the gauge invariant Green functions, $\hat{\tilde{g}}$. The retarded and advanced components of $\hat{\tilde{g}}$ satisfy the equation:
\begin{align}\label{eq:QKE-G_eqGaugeInv}\nonumber
&\left[\epsilon+\frac{1}{2m}\left(\boldsymbol{\nabla}-i\frac{e\mathbf{B}\times\boldsymbol{\rho}}{2c}\right)^2-V_{imp}-\sigma_{eq}^{R,A}\right]\\
&\times\tilde{g}_{eq}^{R,A}(\boldsymbol{\rho},\epsilon;imp)=\delta(\boldsymbol{\rho}),
\end{align}
where the product of the Green function and the self-energy should be understood as a convolution in real space (see Eq.~\ref{eq:QKE-G_eq}). In the following the permeability is taken to be $1$, and correspondingly we will not distinguish between $B$ (the magnetic flux density) and the magnetic field $H$. After averaging over the disorder, the gauge invariant Green functions at equilibrium become translationally invariant, i.e., functions of the relative coordinate $\boldsymbol{\rho}$ alone (see Ref.~\onlinecite{Khodas2003} and references therein):
\begin{align}\label{eq:QKE-G_eqGaugeInv}\nonumber
&\left[\epsilon+\frac{1}{2m}\left(\frac{\partial^2}{\partial\boldsymbol{\rho}^2}-\frac{e^2H^2\rho^2}{4c^2}\right)+\mu\pm\frac{i}{2\tau}-\sigma_{eq}^{R,A}\right]\\
&\times\tilde{g}_{eq}^{R,A}(\boldsymbol{\rho},\epsilon)=\delta(\boldsymbol{\rho}),
\end{align}
where $\tau$ is the elastic mean free time of the electrons.

As we have already discussed, when we turn from the equilibrium Green function to the $\boldsymbol{\nabla}T$-dependent Green function, an additional dependence on the center of mass coordinate appears. We wish to isolate this dependence on $\mathbf{R}$ from the others. Similar to $\hat{g}_{eq}(\boldsymbol{\rho},\epsilon;\mathbf{A},imp)$, we denote the dependencies of $\hat{\underline{\underline{G}}}$ on the center of mass coordinate caused by the impurity potential and the vector potential by $imp$ and $\mathbf{A}$, respectively. Then, the remaining explicit dependence on $\mathbf{R}$ in  $\hat{\underline{\underline{G}}}(\mathbf{R};\boldsymbol{\rho},\epsilon;\mathbf{A},imp)$ arises due to the temperature gradient.  Therefore, in the process of linearizing the equation in $\boldsymbol{\nabla}T/T_0$, we expand $\hat{\underline{\underline{G}}}$ and $\hat{\underline{\underline{\Sigma}}}$ in the collision integral with respect to this explicit dependence on $\mathbf{R}$. In other words, we may write
\begin{widetext}
\begin{align}\label{eq:QKE-Expand}\nonumber
&\int{dr_{\scriptscriptstyle1}}\hat{\underline{\underline{\Sigma}}}\left(\mathbf{R}+\frac{\mathbf{r}_{\scriptscriptstyle1}}{2};\boldsymbol{\rho}-\mathbf{r}_{\scriptscriptstyle1},\epsilon;\mathbf{A},imp\right)
\hat{\underline{\underline{G}}}\left(\mathbf{R}-\frac{\boldsymbol{\rho}-\mathbf{r}_{\scriptscriptstyle1}}{2};\mathbf{r}_{\scriptscriptstyle1},\epsilon;\mathbf{A},imp\right)
\approx \int{dr_{\scriptscriptstyle1}}\hat{\underline{\underline{\Sigma}}}(\mathbf{R};\boldsymbol{\rho}-\mathbf{r}_{\scriptscriptstyle1},\epsilon;\mathbf{A},imp)\hat{\underline{\underline{G}}}(\mathbf{R};\mathbf{r}_{\scriptscriptstyle1},\epsilon;\mathbf{A},imp)\\
&+\int{dr_{\scriptscriptstyle1}}\frac{\mathbf{r}_{\scriptscriptstyle1}}{2}\cdot\frac{\partial\hat{\underline{\underline{\Sigma}}}(\mathbf{R};\boldsymbol{\rho}-\mathbf{r}_{\scriptscriptstyle1},\epsilon;\mathbf{A},imp)}{\partial\mathbf{R}}\hat{\underline{\underline{G}}}(\mathbf{R};\mathbf{r}_{\scriptscriptstyle1},\epsilon;\mathbf{A},imp)
-\int{dr_{\scriptscriptstyle1}}\hat{\underline{\underline{\Sigma}}}(\mathbf{R};\boldsymbol{\rho}-\mathbf{r}_{\scriptscriptstyle1},\epsilon;\mathbf{A},imp)\frac{\boldsymbol{\rho}-\mathbf{r}_{\scriptscriptstyle1}}{2}\cdot\frac{\partial\hat{\underline{\underline{G}}}(\mathbf{R};\mathbf{r}_{\scriptscriptstyle1},\epsilon;\mathbf{A},imp)}{\partial\mathbf{R}}.
\end{align}
\end{widetext}
As we shall see below, the last two terms in the expansion are actually proportional to $\boldsymbol{\nabla}T/T_0$.

The equation for the local equilibrium Green function is:
\begin{align}\label{eq:QKE-G_LocEq}\nonumber
&\int{d\mathbf{r}_{\scriptscriptstyle1}}\hat{g}_{eq}^{-1}\left(\boldsymbol{\rho}-\mathbf{r}_{\scriptscriptstyle1},\epsilon;\mathbf{A},imp\right)
\hat{G}_{loc-eq}\left(\mathbf{R};\mathbf{r}_{\scriptscriptstyle1},\epsilon;\mathbf{A},imp\right)\\\nonumber
&=\hspace{-1mm}\int\hspace{-1mm}{d\mathbf{r}_{\scriptscriptstyle1}}\hat{\Sigma}_{loc-eq}\left(\mathbf{R};\boldsymbol{\rho}-\mathbf{r}_{\scriptscriptstyle1},\epsilon;\mathbf{A},imp\right)\hat{g}_{eq}\left(\mathbf{r}_{\scriptscriptstyle1},\epsilon;\mathbf{A},imp\right)\\
&+\frac{\mathbf{R}\cdot\boldsymbol{\nabla}T}{T_0}\epsilon\hat{g}_{eq}(\boldsymbol{\rho},\epsilon;\mathbf{A},imp).
\end{align}
This equation is solved by:
\begin{align}\label{eq:QKE-G_LocEq2}
&\hat{G}_{loc-eq}(\mathbf{R};\boldsymbol{\rho},\epsilon;\mathbf{A},imp)=-\frac{\mathbf{R}\cdot\boldsymbol{\nabla}T}{T_0}\epsilon\frac{\partial\hat{g}_{eq}(\boldsymbol{\rho},\epsilon;\mathbf{A},imp)}{\partial\epsilon},
\end{align}
where the corresponding self-energy should be taken as  $\hat{\Sigma}_{loc-eq}(\mathbf{R};\boldsymbol{\rho},\epsilon;\mathbf{A},imp)=-(\mathbf{R}\cdot\boldsymbol{\nabla}T/T_0)
\epsilon{\partial}\hat{\sigma}_{eq}(\boldsymbol{\rho},\epsilon;\mathbf{A},imp)/{\partial\epsilon}$. We see that the local equilibrium Green function is a straightforward extension of the equilibrium Green function for a non-uniform temperature. Since the same holds for $\hat{\Sigma}_{loc-eq}$, the equation for $\hat{G}_{loc-eq}$ is a closed equation determined by the equilibrium properties of the system.

The Green function  $\hat{G}_{loc-eq}$ describes the readjustment of quasi-particles to the non-uniform temperature when the system is trying to maintain a local equilibrium. This response of the electrons to the temperature gradient tends to induce modulation of the density. Since for charged particles it is impossible to have a large scale charge modulation, the temperature gradient transforms into a gradient of the electro-chemical potential.  Therefore,  $\mathbf{j}_{e}=\hat\sigma(\mathbf{E}-\boldsymbol{\nabla}\mu/e)=\hat{\sigma}\mathbf{E}^{*}$ where the effective field $\mathbf{E}^{*}$ is the one measured in experiments.

The role of the local-equilibrium Green function is most peculiar when the response to the temperature gradient is considered in the presence of a magnetic field. Under these conditions, as we show in Secs.~\ref{sec:Current} and~\ref{sec:Magnetization}, $G_{loc-eq}(\mathbf{R};\boldsymbol{\rho},\epsilon;\mathbf{A},imp)$ is responsible for the non-vanishing contribution to the electric current from the magnetization.

All the remaining terms in the quantum kinetic equation determine the last term of the Green function, $\hat{G}_{\boldsymbol{\nabla}T}$:
\begin{widetext}
\begin{align}\label{eq:QKE-G_TransInv}
&\int{d\mathbf{r}_{\scriptscriptstyle1}}\hat{g}_{eq}^{-1}\left(\boldsymbol{\rho}-\mathbf{r}_{\scriptscriptstyle1},\epsilon;\mathbf{A},imp\right)
\hat{G}_{\boldsymbol{\nabla}T}\left(\mathbf{R};\mathbf{r}_{\scriptscriptstyle1},\epsilon;\mathbf{A},imp\right)-\frac{\boldsymbol{\rho}\cdot\boldsymbol{\nabla}T}{2T_0}\epsilon\hat{g}_{eq}(\boldsymbol{\rho},\epsilon;\mathbf{A},imp)\\\nonumber
&+\frac{1}{2m}\left(\frac{\partial}{\partial\boldsymbol{\rho}}-\frac{ie}{c}\mathbf{A}(\mathbf{R}+\boldsymbol{\rho}/2)\right)\cdot\frac{\partial\hat{G}_{loc-eq}(\mathbf{R};\boldsymbol{\rho},\epsilon;\mathbf{A},imp)}{\partial\mathbf{R}}
=\int{d\mathbf{r}_{\scriptscriptstyle1}}\hat{\Sigma}_{\boldsymbol{\nabla}T}\left(\mathbf{R};\boldsymbol{\rho}-\mathbf{r}_{\scriptscriptstyle1},\epsilon;\mathbf{A},imp\right)
\hat{g}_{eq}\left(\mathbf{r}_{\scriptscriptstyle1},\epsilon;\mathbf{A},imp\right)\\\nonumber
&+\int{d\mathbf{r}_{\scriptscriptstyle1}}\frac{\mathbf{r}_{\scriptscriptstyle1}}{2}\cdot\frac{\partial\hat{\Sigma}_{loc-eq}\left(\mathbf{R};\boldsymbol{\rho}-\mathbf{r}_{\scriptscriptstyle1},\epsilon;\mathbf{A},imp\right)}{\partial\mathbf{R}}
\hat{g}_{eq}\left(\mathbf{r}_{\scriptscriptstyle1},\epsilon;\mathbf{A},imp\right)\\\nonumber&-
\int{d\mathbf{r}_{\scriptscriptstyle1}}\hat{\sigma}_{eq}\left(\boldsymbol{\rho}-\mathbf{r}_{\scriptscriptstyle1},\epsilon;\mathbf{A},imp\right)
\frac{\boldsymbol{\rho}-\mathbf{r}_{\scriptscriptstyle1}}{2}\cdot\frac{\partial\hat{G}_{loc-eq}\left(\mathbf{R};\mathbf{r}_{\scriptscriptstyle1},\epsilon;\mathbf{A},imp\right)}{\partial\mathbf{R}}.
\end{align}
\end{widetext}
In the above equation, the derivatives with respect to the center of mass coordinate act only on the explicit dependence of $\hat{G}_{loc-eq}(\mathbf{R},\boldsymbol{\rho},\epsilon;\mathbf{A},imp)$ and $\hat{\Sigma}_{loc-eq}(\mathbf{R},\boldsymbol{\rho},\epsilon;\mathbf{A},imp)$ on $\mathbf{R}$ (i.e., through the spatially dependent temperature). Recall that the derivatives with respect to the center of mass coordinate which act on $V_{imp}$ and $\mathbf{A}$ in the local equilibrium Green function was already included in $g_{eq}^{-1}$ that appears in Eq.~\ref{eq:QKE-G_LocEq}.

Once the explicit expressions for $\hat{G}_{loc-eq}$ and $\hat{\Sigma}_{loc-eq}$  are inserted, the equation becomes much simpler:
\begin{align}\label{eq:QKE-G_TransInv2}
&\hat{G}_{\boldsymbol{\nabla}T}(\boldsymbol{\rho},\epsilon;\mathbf{A},imp)=\hat{g}_{eq}\left(\epsilon\right)
\hat{\Sigma}_{\boldsymbol{\nabla}T}\left(\epsilon\right)\hat{g}_{eq}\left(\epsilon\right)\\\nonumber
&-i\epsilon\frac{\boldsymbol{\nabla}T}{2T_0}\cdot\left[\frac{\partial\hat{g}_{eq}\left(\epsilon\right)}{\partial\epsilon}\mathbf{\hat{v}}_{eq}(\epsilon)\hat{g}_{eq}\left(\epsilon\right)
-\hat{g}_{eq}\left(\epsilon\right)\mathbf{\hat{v}}_{eq}(\epsilon)\frac{\partial\hat{g}_{eq}\left(\epsilon\right)}{\partial\epsilon}
\right].
\end{align}
The product of matrices should be understood as a convolution in real space. The velocity $\mathbf{\hat{v}}_{eq}$ is the renormalized velocity at equilibrium:
\begin{align}\label{eq:QKE-velocityQP}
\mathbf{\hat{v}}_{eq}&(\mathbf{r},t;\mathbf{r}',t')=-i\lim_{\mathbf{r}'\rightarrow\mathbf{r}}\frac{\boldsymbol{\nabla}-\boldsymbol{\nabla}'}{2m}
-i(\mathbf{r}-\mathbf{r}')\hat{\sigma}_{eq}(\mathbf{r},\mathbf{r}',\epsilon).
\end{align}

Let us point out an important difference between the two parts of the Green function depending on the temperature gradient, $\hat{G}_{loc-eq}$ and $\hat{G}_{\boldsymbol{\nabla}T}$. As was already mentioned, $\hat{G}_{loc-eq}$ and $\hat{\Sigma}_{loc-eq}$ are a straightforward extension of the equilibrium Green function and self-energy for a non-uniform temperature. On the other hand, the equation for $\hat{G}_{\boldsymbol{\nabla}T}$ contains the $\boldsymbol{\nabla}T$-dependent self-energy which by itself is a function of $\hat{G}_{\boldsymbol{\nabla}T}$. Thus, this is a self consistent equation, and in order to find a close expression for $\hat{G}_{\boldsymbol{\nabla}T}$, one has to determine the structure of the self-energy. Once the form of the self-energy is known, one should take into consideration in the coarse of linearization with respect to $\boldsymbol{\nabla}T$ that all the Green functions in $\hat{\Sigma}_{\boldsymbol{\nabla}T}$ may depend on the temperature gradient.

To complete the derivation of the electric current as a response to a temperature gradient, we must also find the dependence of the propagator of the superconducting fluctuations $\hat{L}(\mathbf{r},t;\mathbf{r}',t')$ on $\boldsymbol{\nabla}T$. In the regime of linear response, the explicit dependence on the temperature gradient can be eliminated from the kinetic equation for $\hat{L}$ by transforming to the propagator $\underline{\underline{\hat{L}}}$:
\begin{align}\label{eq:QKE-DEV}
&-\int{d\mathbf{r}_{\scriptscriptstyle1}}\lambda^{-1}(\mathbf{r-r}_{\scriptscriptstyle1})\underline{\underline{\hat{L}}}(\mathbf{r}_{\scriptscriptstyle1},t;\mathbf{r}',t')\\\nonumber
&=\delta(\mathbf{r}-\mathbf{r}')
-\int{d\mathbf{r}_{\scriptscriptstyle1}dt_{\scriptscriptstyle1}}\hat{\underline{\underline{\Pi}}}(\mathbf{r},t;\mathbf{r}_{\scriptscriptstyle1},t_{\scriptscriptstyle1})\hat{\underline{\underline{L}}}(\mathbf{r}_{\scriptscriptstyle1},t_{\scriptscriptstyle1};\mathbf{r}',t').
\end{align}
Thus, the entire dependence of the propagator on the temperature gradient is through the self-energy term $\hat{\underline{\underline{\Pi}}}$, which is a function of the quasi-particle Green functions.

Let us separate the solution of Eq.~\ref{eq:QKE-DEV} into the equilibrium and $\boldsymbol{\nabla}T$-dependent propagators, $\hat{\underline{\underline{L}}}=\hat{L}_{eq}+\hat{L}_{loc-eq}+\hat{L}_{\boldsymbol{\nabla}T}$.  The propagator at equilibrium satisfies the equation:
\begin{align}\label{eq:QKET-V_eq}
\hat{V}_{eq}(\mathbf{R};\boldsymbol{\rho},\omega)=\left[U^{-1}(\boldsymbol{\rho})+\hat{\Pi}_{eq}(\mathbf{R};\boldsymbol{\rho},\omega)\right]^{-1}.
\end{align}
The entire dependence of $\hat{L}_{eq}(\omega)$ on the frequency is due to dressing of the bare propagator by its self-energy $\hat{\Pi}_{eq}(\omega)$. In the above equation the  propagator of the superconducting fluctuations is a function of the temperature $T_0$. Similar to Eq.~\ref{eq:QKE-velocityQP}, we may define the "renormalized velocity" of the collective mode describing the superconducting fluctuations at equilibrium to be:
\begin{align}
\hat{\boldsymbol{\mathcal{V}}}_{eq}(\mathbf{r},t;\mathbf{r}',t')=-i(\mathbf{r-r'})\hat{\Pi}_{eq}(\mathbf{r},t;\mathbf{r}',t').\label{eq:QKE-velocitySC}
\end{align}
Note that in fact $\hat{\boldsymbol{\mathcal{V}}}$ does not have the dimension of a velocity.

The equations for the $\boldsymbol{\nabla}T$-dependent propagators remind the first term in Eq.~\ref{eq:QKE-G_TransInv2} for $\hat{G}_{\boldsymbol{\nabla}T}$:
\begin{align}\label{eq:QKE-V_loceq}
\hat{L}_{loc-eq}(\mathbf{R};\boldsymbol{\rho},\omega)=-\hat{L}_{eq}(\omega)\hat{\Pi}_{loc-eq}(\omega)\hat{L}_{eq}(\omega),
\end{align}
and
\begin{align}\label{eq:QKE-V_GradT}
&\hat{L}_{\boldsymbol{\nabla}T}(\mathbf{R};\boldsymbol{\rho},\omega)=-\hat{L}_{eq}(\omega)\hat{\Pi}_{\boldsymbol{\nabla}T}(\omega)\hat{L}_{eq}(\omega).
\end{align}
Once again, one should understand the product as a convolution of the spatial coordinate.

\section{The electric current as a response to a temperature gradient}\label{sec:Current}

For the calculation of the Nernst effect we have to derive the expression for the electric current as a response to a temperature gradient. In the presence of a magnetic field, the electric current is a sum of two terms:
\begin{align}\label{eq:Current-Jtr+Jmag}
\mathbf{j}_e&=\mathbf{j}_{e}^{con}+\mathbf{j}_{e}^{mag}.
\end{align}
The first one, $\mathbf{j}_{e}^{con}$, is derived using the continuity equation for the electric charge. The second contribution to the electric current originates from the magnetization current. Since the magnetization current is divergenceless it cannot be obtained using the continuity equation and it will be found separately.

As follows from the action in Eq.~\ref{eq:QKE-S}, the fields $\psi$ and $\Delta$ carry electric current. Therefore, the charge continuity equation must include both fields:
\begin{align}\label{eq:Current-ContinuityEq}
&-e\sum_{\sigma}\partial_t|\psi_{\sigma}(\mathbf{r},t)|^2+\boldsymbol{\nabla}\cdot\mathbf{j}_{e}^{con}(\mathbf{r},t)
=-2ie\gamma(\mathbf{r})\\\nonumber&\times\left[\Delta^{\dag}(\mathbf{r},t)\psi_{\downarrow}(\mathbf{r},t)\psi_{\uparrow}(\mathbf{r},t)-\Delta(\mathbf{r},t)\psi_{\uparrow}^{\dag}(\mathbf{r},t)\psi_{\downarrow}^{\dag}(\mathbf{r},t)\right], \end{align}
where $\mathbf{j}_{e}^{con}(\mathbf{r},t)=-ie\sum_{\sigma}\gamma(\mathbf{r})[\psi_{\sigma}^{\dag}(\mathbf{r},t)\boldsymbol{\nabla}\psi_{\sigma}(\mathbf{r},t)
-\boldsymbol{\nabla}\psi_{\sigma}^{\dag}(\mathbf{r},t)\psi_{\sigma}(\mathbf{r},t)-2ie\mathbf{A}|\psi_{\sigma}|^2/c]/2m$.  The terms in the RHS describe absorption and emission of quasi-particles by the superconducting fluctuations; the factor $2$ reflects the fact that the Copper pairs carry charge of $2e$.

To find the expression for $\mathbf{j}_{e}^{con}$ in terms of the Green function, we rewrite the charge density using the lesser component of the Green function:
\begin{align}\label{eq:Current-ChargeDensity}
\rho(\mathbf{r},t)&=-e\hspace{-1mm}\lim_{\begin{array}{c}\scriptstyle{ \mathbf{r}'\rightarrow\mathbf{r}} \\ \scriptstyle{t'\rightarrow{t}^{+}} \\
\end{array}}\sum_{\sigma}\left\langle{\psi_{\sigma}^{\dag}(\mathbf{r}',t')\psi_{\sigma}(\mathbf{r},t)}\right\rangle\\\nonumber
&=ie\hspace{-1mm}\lim_{\begin{array}{c}\scriptstyle{ \mathbf{r}'\rightarrow\mathbf{r}} \\ \scriptstyle{t'\rightarrow{t}} \\
\end{array}}\sum_{\sigma}G^{<}(\mathbf{r},t;\mathbf{r}',t').
\end{align}
We use the notation $t'\rightarrow{t}^{+}$ to indicate that the limit should be taken in such a way that $t$ is on the upper branch of the Keldysh contour, while $t'$ is on the lower branch.~\cite{Rammer1986} The summation over the spin projection results in a factor of $2$.  Here $\langle{A}\rangle$ denotes the quantum mechanical averaging with the action given in Eq.~\ref{eq:QKE-S}. Therefore, the Green function is fully dressed by the interactions and depends on the impurity potential. In addition, $\hat{G}$  is a function of the temperature gradient. Since we find the current by extracting it from the continuity equation, we assume that the temperature gradient has  some spatial modulations that will be set to zero at the end of the procedure.

Next, we insert the above expression into the continuity equation given in Eq.~\ref{eq:Current-ContinuityEq} and rewrite $\boldsymbol{\nabla}\cdot\mathbf{j}_e^{con}$ as a sum of two terms:
\begin{align}\label{eq:Current-Continuity2}
&\boldsymbol{\nabla}\cdot\mathbf{j}_{e}^{con}=\mathcal{I}_{\scriptscriptstyle1}+\mathcal{I}_{\scriptscriptstyle2};\\\nonumber
&\mathcal{I}_{\scriptscriptstyle1}=e\hspace{-1mm}\lim_{\begin{array}{c}\scriptstyle{ \mathbf{r}'\rightarrow\mathbf{r}} \\ \scriptstyle{t'\rightarrow{t}^{+}} \\
\end{array}}\hspace{-2mm}\left(\frac{\partial}{\partial{t}}+\frac{\partial}{\partial{t}'}\right)\sum_{\sigma}\left\langle{\psi_{\sigma}^{\dag}(\mathbf{r}',t')\psi_{\sigma}(\mathbf{r},t)}\right\rangle;\\\nonumber
&\mathcal{I}_{\scriptscriptstyle2}=-2ie\hspace{-1mm}\lim_{\begin{array}{c}\scriptstyle{ \mathbf{r}'\rightarrow\mathbf{r}} \\ \scriptstyle{t'\rightarrow{t}^{+}} \\
\end{array}}\left\langle\phantom{\psi_{\uparrow}^{\dag}(\mathbf{r}',t')\psi_{\downarrow}^{\dag}(\mathbf{r}',t')\gamma(\mathbf{r})\Delta(\mathbf{r},t)}\hspace{-43mm}\gamma(\mathbf{r}')\Delta^{\dag}(\mathbf{r}',t')\psi_{\downarrow}(\mathbf{r},t)\psi_{\uparrow}(\mathbf{r},t)\right.\\\nonumber
&\left.\hspace{20mm}-\psi_{\uparrow}^{\dag}(\mathbf{r}',t')\psi_{\downarrow}^{\dag}(\mathbf{r}',t')\gamma(\mathbf{r})\Delta(\mathbf{r},t)\right\rangle. \end{align}
To resolve the expression for $\mathbf{j}_{e}^{con}$ we need to find the equations of motion for the field $\psi$.  The variational derivative of the action in Eq.~\ref{eq:QKE-S} with respect to the $\psi^{\dag}$ yields the equation of motion for the field $\psi$:
\begin{align}\label{eq:Current-EOMPsi}\nonumber
&i\frac{\partial\psi_{\sigma}(\mathbf{r},t)}{\partial{t}}\hspace{-1mm}=\hspace{-1mm}\frac{-1}{2m}\left(\hspace{-1mm}\boldsymbol{\nabla}-\frac{ie}{c}\mathbf{A}(\mathbf{r})\right)\hspace{-1mm}\gamma(\mathbf{r})\hspace{-1mm}\left(\boldsymbol{\nabla}-\frac{ie}{c}\mathbf{A}(\mathbf{r})\right)\hspace{-1mm}\psi_{\sigma}(\mathbf{r},t)\\
&+\gamma(\mathbf{r})V_{imp}(\mathbf{r})\psi_{\sigma}(\mathbf{r},t)+\sigma\gamma(\mathbf{r})\Delta(\mathbf{r},t)\psi_{-\sigma}^{\dag}(\mathbf{r},t).
\end{align}
Under the average,  the equations of motion allow us to rewrite the expression for $\mathcal{I}_{\scriptscriptstyle1}$ as:
\begin{align}\label{eq:Current-ChargeContinuity3}\nonumber
&\mathcal{I}_{\scriptscriptstyle1}\hspace{-0.5mm}=\hspace{-0.5mm}-ie\hspace{-2mm}\lim_{\begin{array}{c}\scriptstyle{ \mathbf{r}'\rightarrow\mathbf{r}} \\ \scriptstyle{t'\rightarrow{t}^{+}} \\
\end{array}}\hspace{-2mm}\hspace{-1mm}\left\langle\hspace{-0.5mm}\sum_{\sigma}\hspace{-0.5mm}\left[\frac{-1}{2m}\left(\boldsymbol{\nabla}-i\frac{e}{c}\mathbf{A}(\mathbf{r})\right)\gamma(\mathbf{r})
\left(\boldsymbol{\nabla}-i\frac{e}{c}\mathbf{A}(\mathbf{r})\right)\right.\right.\\\nonumber
&\left.\left.+
\frac{1}{2m}\left(\boldsymbol{\nabla}'-i\frac{e}{c}\mathbf{A}(\mathbf{r}')\right)\gamma(\mathbf{r}')\left(\boldsymbol{\nabla}'-i\frac{e}{c}\mathbf{A}(\mathbf{r}')\right)
\right.\right.\\\nonumber
&\left.\left.+\gamma(\mathbf{r})V_{imp}(\mathbf{r})
-\gamma(\mathbf{r}')V_{imp}(\mathbf{r}')
\phantom{\frac{2}{2}}\hspace{-3mm}\right]\psi^{\dag}_{\sigma}(\mathbf{r}',t')\psi_{\sigma}(\mathbf{r},t)
\right.\\\nonumber
&\left.+\sum_{\sigma}\sigma
\psi_{\sigma}^{\dag}(\mathbf{r}',t')\psi_{-\sigma}^{\dag}(\mathbf{r},t)\gamma(\mathbf{r})\Delta(\mathbf{r},t)\right.\\
&\left.
-\sum_{\sigma}\sigma\gamma(\mathbf{r}')\Delta^{\dag}(\mathbf{r}',t')\psi_{-\sigma}(\mathbf{r}',t')\psi_{\sigma}(\mathbf{r},t)\right\rangle.
\end{align}

Now, we wish to express the electric current in the presence of a gravitational field in terms of the propagators. The expression $\langle{\Delta(\mathbf{r},t)\psi_{\sigma}^{\dag}(\mathbf{r}',t')\psi_{-\sigma}^{\dag}(\mathbf{r},t)}\rangle$ and its counterpart are averaged with respect to the Hamiltonian that includes both the interactions and the gravitational field. These expressions can be written in terms of the self-energy of the quasi-particles. For example, $\langle{\sigma\gamma(\mathbf{r})\Delta(\mathbf{r},t)\psi_{\sigma}^{\dag}(\mathbf{r}',t')\psi_{-\sigma}^{\dag}(\mathbf{r},t)}\rangle=-i\gamma(\mathbf{r})\int{d\mathbf{r}_{\scriptscriptstyle1}dt_{\scriptscriptstyle1}}\hat{\Sigma}_{\sigma}(\mathbf{r},t;\mathbf{r}_{\scriptscriptstyle1},t_{\scriptscriptstyle1})\gamma(\mathbf{r}_{\scriptscriptstyle1})\hat{G}_{\sigma}(\mathbf{r}_{\scriptscriptstyle1},t_t;\mathbf{r}',t')$. (Here, the factor $i$ appears because in real time the evolution operator is of the form $e^{-i\mathcal{H}t}$ and because of the conventional definition of the propagators and self-energies.~\cite{Rammer1986}). As a result, we obtain:
\begin{align}\label{eq:Current-ChargeContinuity3}\nonumber
&\mathcal{I}_{\scriptscriptstyle1}=-2e\lim_{\begin{array}{c}\scriptstyle{ \mathbf{r}'\rightarrow\mathbf{r}} \\ \scriptstyle{t'\rightarrow{t}} \\
\end{array}}\left[-\frac{\boldsymbol{\nabla}\gamma(\mathbf{r})(\boldsymbol{\nabla}-ie\mathbf{A}(\mathbf{r})/c)}{2m}\hat{G}(\mathbf{r},t;\mathbf{r}',t')\right.\\\nonumber
&\left.+
\frac{\boldsymbol{\nabla}'\gamma(\mathbf{r}')(\boldsymbol{\nabla}'+ie\mathbf{A}(\mathbf{r}')/c)}{2m}\hat{G}(\mathbf{r},t;\mathbf{r}',t')\right.\\\nonumber
&\left.+\gamma(\mathbf{r})\int{d\mathbf{r}_{\scriptscriptstyle1}dt_{\scriptscriptstyle1}}\hat{\Sigma}(\mathbf{r},t;\mathbf{r}_{\scriptscriptstyle1},t_{\scriptscriptstyle1})\gamma(\mathbf{r}_{\scriptscriptstyle1})\hat{G}(\mathbf{r}_{\scriptscriptstyle1},t_t;\mathbf{r}',t')\right.\\
&\left.-\int{d\mathbf{r}_{\scriptscriptstyle1}dt_{\scriptscriptstyle1}}\hat{G}(\mathbf{r},t;\mathbf{r}_{\scriptscriptstyle1},t_{\scriptscriptstyle1})\gamma(\mathbf{r}_{\scriptscriptstyle1})\hat{\Sigma}(\mathbf{r}_{\scriptscriptstyle1},t_t;\mathbf{r}',t')\gamma(\mathbf{r}')\right]^{<}.
\end{align}
The factor $2$ is a consequence of the sum over  the spin index. Similarly, we can express the averages in the equation for $\mathcal{I}_{\scriptscriptstyle2}$ in terms of the self-energy of the superconducting fluctuations, e.g.,
$\langle\gamma(\mathbf{r}')\Delta^{\dag}(\mathbf{r}',t')\psi_{\downarrow}(\mathbf{r},t)\psi_{\uparrow}(\mathbf{r},t)\rangle
=-i\int{d\mathbf{r}_{\scriptscriptstyle1}dt_{\scriptscriptstyle1}}\hat{\Pi}(\mathbf{r},t;\mathbf{r}_{\scriptscriptstyle1},t_{\scriptscriptstyle1})\gamma(\mathbf{r}_{\scriptscriptstyle1})\hat{L}(\mathbf{r}_{\scriptscriptstyle1},t_t;\mathbf{r}',t')\gamma(\mathbf{r}')$.

In the regime of linear response we may eliminate the explicit dependence of the current on $\gamma(\mathbf{r})$ by expressing the current in terms of $\hat{\underline{\underline{G}}}$, $\hat{\underline{\underline{L}}}$, $\hat{\underline{\underline{\Sigma}}}$ and $\hat{\underline{\underline{\Pi}}}$ (as defined in Eq.~\ref{eq:QKET-Transformation}). Then, the sum of the two contributions to $\boldsymbol{\nabla}\cdot\mathbf{j}_{e}^{con}$ becomes:
\begin{align}\label{eq:Current-ContinuityEq4}\nonumber
\boldsymbol{\nabla}\cdot\mathbf{j}_{e}^{con}&=-2e\lim_{\begin{array}{c}\scriptstyle{ \mathbf{r}'\rightarrow\mathbf{r}} \\ \scriptstyle{t'\rightarrow{t}} \\
\end{array}}\left[-\frac{\boldsymbol{\nabla}(\boldsymbol{\nabla}-ie\mathbf{A}(\mathbf{r})/c)}{2m}\hat{\underline{\underline{G}}}(\mathbf{r},t;\mathbf{r}',t')\right.\\\nonumber
&\left.+\frac{\boldsymbol{\boldsymbol{\nabla}'}(\boldsymbol{\nabla}'+ie\mathbf{A}(\mathbf{r}')/c)}{2m}\hat{G}(\mathbf{r},t;\mathbf{r}',t')\right.\\\nonumber
&\left.+\int{d\mathbf{r}_{\scriptscriptstyle1}dt_{\scriptscriptstyle1}}\hat{\underline{\underline{\Sigma}}}(\mathbf{r},t;\mathbf{r}_{\scriptscriptstyle1},t_{\scriptscriptstyle1})\hat{\underline{\underline{G}}}(\mathbf{r}_{\scriptscriptstyle1},t_{\scriptscriptstyle1};\mathbf{r}',t')\right.\\\nonumber
&\left.-\int{d\mathbf{r}_{\scriptscriptstyle1}dt_{\scriptscriptstyle1}}\hat{\underline{\underline{G}}}(\mathbf{r},t;\mathbf{r}_{\scriptscriptstyle1},t_{\scriptscriptstyle1})\hat{\underline{\underline{\Sigma}}}(\mathbf{r}_{\scriptscriptstyle1},t_{\scriptscriptstyle1};\mathbf{r}',t')\right.\\\nonumber
&\left.+\int{d\mathbf{r}_{\scriptscriptstyle1}dt_{\scriptscriptstyle1}}\hat{\underline{\underline{\Pi}}}(\mathbf{r},t;\mathbf{r}_{\scriptscriptstyle1},t_{\scriptscriptstyle1})\hat{\underline{\underline{L}}}(\mathbf{r}_{\scriptscriptstyle1},t_{\scriptscriptstyle1};\mathbf{r}',t')\right.\\
&\left.-\int{d\mathbf{r}_{\scriptscriptstyle1}dt_{\scriptscriptstyle1}}\hat{\underline{\underline{L}}}(\mathbf{r},t;\mathbf{r}_{\scriptscriptstyle1},t_{\scriptscriptstyle1})\hat{\underline{\underline{\Pi}}}(\mathbf{r}_{\scriptscriptstyle1},t_{\scriptscriptstyle1};\mathbf{r}',t')
\right]^{<}\hspace{-3mm}.
\end{align}
Note that the current still depends on the gravitational field (i.e., on the temperature gradient) through the propagators and self-energies.

In the final step of the derivation one has to resolve the expression for the current out of the gradient. In other words, to reformulate the products of propagators and self-energies as a derivative with respect to the center of mass coordinate. As discussed in the previous section, we can isolate the dependencies on the center of mass coordinate created by the vector potential and by the impurities. After we average the current over the disorder and transform  to the gauge invariant propagators and self-energies, these dependencies vanish.  Pay attention that when the limit $\mathbf{r}'\rightarrow\mathbf{r}$ is taken, one may rewrite Eq.~\ref{eq:Current-ContinuityEq4} in terms of the gauge invariant quantities alone.  Therefore,  we expand the products of the propagators and the self-energies with respect to the deviation from $\mathbf{R}$ exactly in the same way as performed in Eq.~\ref{eq:QKE-Expand}. As a consequence of the symmetric form of the terms in Eq.~\ref{eq:Current-ContinuityEq4}, $[\hat{\Sigma}\hat{G}-\hat{G}\hat{\Sigma}]^{<}$ and $[\hat{\Pi}\hat{L}-\hat{L}\hat{\Pi}]^{<}$, one may check that all even orders in the expansion vanish. In the regime of linear response it is enough to keep only the lowest non-vanishing order in the expansion. Eventually, the expression for the current becomes:
\begin{align}\label{eq:Current-Je}
&\mathbf{j}_{e}^{con}(\mathbf{r},t)=ie\hspace{-1mm}\int{d\mathbf{r}'dt'}\left[\mathbf{\hat{\underline{\underline{v}}}}(\mathbf{r},t;\mathbf{r}',t')\hat{\underline{\underline{G}}}(\mathbf{r}',t';\mathbf{r},t)\right]^{<}\\\nonumber
&+ie\int{d\mathbf{r}'dt'}\hspace{-1mm}\left[\boldsymbol{\hat{\underline{\underline{\mathcal{V}}}}}(\mathbf{r},t;\mathbf{r}',t')\hat{\underline{\underline{L}}}(\mathbf{r}',t';\mathbf{r},t)\right]^{<}+h.c.
\end{align}
We use the notation $\big[...\big]^{<}$ to remind that the expression inside the square brackets is a product of  matrices and to indicate that the current corresponds to the lesser component of the resulting matrix.  The matrices $\mathbf{\hat{\underline{\underline{v}}}}$ and $\hat{\underline{\underline{\boldsymbol{\mathcal{V}}}}}$ are the renormalized velocities defined in Eqs.~\ref{eq:QKE-velocityQP} and~\ref{eq:QKE-velocitySC} with the $\boldsymbol{\nabla}T$-dependent self-energies $\mathbf{\hat{\underline{\underline{\Sigma}}}}$ and $\mathbf{\hat{\underline{\underline{\Pi}}}}$  replacing the equilibrium ones.

The velocity of the quasi-particles $\mathbf{\hat{v}}$ is renormalized by the self-energy,
$\delta\hat{\mathbf{v}}(\mathbf{r},t;\mathbf{r}',t')=-i(\mathbf{r}-\mathbf{r}')\hat{\Sigma}(\mathbf{r},t;\mathbf{r}',t')$. We find it useful to rewrite this expression as follows: $-[i(\mathbf{r}'-\mathbf{r})+2i(\mathbf{r}-\mathbf{r}')]\hat{\Sigma}(\mathbf{r},t;\mathbf{r}',t')$. The idea behind this representation can be explained using, as an example, the first order expansion of the self-energy with respect to the superconducting fluctuations presented in Fig.~\ref{fig:SelfEnergy1st}. In this example, the self-energy contains a quasi-particle Green function propagating from $\mathbf{r}'$ to $\mathbf{r}$ and a propagator of the superconducting fluctuations that goes from $\mathbf{r}$ to $\mathbf{r}'$. Correspondingly, the first difference of the coordinates in the square brackets acts on $\hat{G}$, while the second (which appears with the factor $2$) acts on $\hat{L}$.

As mentioned in the beginning of this section, there is another contribution to the electric current arising from the magnetization:
\begin{align}\label{eq:Current-Jmag}
\mathbf{j}_{e}^{mag}&=-{2ic}\boldsymbol{\nabla}\times\mathbf{M}(\mathbf{r})\lim_{\begin{array}{c}\scriptstyle{ \mathbf{r}'\rightarrow\mathbf{r}} \\ \scriptstyle{t'\rightarrow{t}} \\
\end{array}}\left[\hat{\underline{\underline{G}}}(\mathbf{r}',t';\mathbf{r},t)\right]^{<}
\end{align}
where $\mathbf{M}(\mathbf{r})=e\mathbf{r}\times\mathbf{v}/2mc$  denotes the magnetization and the factor of $2$ is due to the summation over the spin index. We would like to emphasize that since the magnetization is created by itinerant electrons, the magnetization current can equally contribute to the transverse transport electric current.

\section{Derivation of the transverse component of $\mathbf{j}_{e}^{con}$}\label{sec:Peltier}

At this stage of the derivation we shall consider only $\mathbf{j}_e^{con}$ keeping for a while the magnetization current aside.  Inserting the expressions for the $\boldsymbol{\nabla}T$-dependent propagators given in Eqs.~\ref{eq:QKE-G_TransInv2},~\ref{eq:QKE-G_LocEq2} and~\ref{eq:QKE-V_GradT} into Eq.~\ref{eq:Current-Je} and extracting the lesser component, we obtain $\mathbf{j}_e^{con}$ as a response to the temperature gradient. First of all, one may observe that the contributions of the local equilibrium functions $\hat{G}_{loc-eq}$ and $\hat{L}_{loc-eq}$ to $\mathbf{j}_e^{con}$ vanish upon averaging the current over the sample. Since we are not interested in terms that vanish after averaging over the volume, the non-zero part of $\mathbf{j}_e^{con}$ becomes:
\begin{widetext}
\begin{align}\label{eq:Pelt-JETransInv}
j_{e\hspace{1.5mm}i}^{con}&=
-\frac{e\boldsymbol{\nabla}_{j}T}{2T_0}\int\frac{d\epsilon}{2\pi}\epsilon\frac{\partial{n_F(\epsilon)}}{\partial\epsilon}\left[v_{i}^R(\epsilon)g^{R}(\epsilon)v_{j}^A(\epsilon)g^A(\epsilon)
+v_{i}^R(\epsilon)g^{R}(\epsilon)v_{j}^R(\epsilon)g^A(\epsilon)-
v_{i}^R(\epsilon)g^{R}(\epsilon)v_{j}^R(\epsilon)g^R(\epsilon)\right.\\\nonumber&\left.-
g^{R}(\epsilon)v_{j}^R(\epsilon)g^R(\epsilon)v_{i}^A(\epsilon)
\right]
-\frac{e\boldsymbol{\nabla}_jT}{T_0}\int\frac{d\epsilon}{2\pi}\epsilon{n}_F(\epsilon)\left[v_{i}^R(\epsilon)\frac{\partial{g}^R(\epsilon)}{\partial\epsilon}v_{j}^R(\epsilon)g^R(\epsilon)-
v_{i}^R(\epsilon){g}^R(\epsilon)v_{j}^R(\epsilon)\frac{\partial{g}^R(\epsilon)}{\partial\epsilon}\right]\\\nonumber
&-{ie}\int\frac{d\epsilon}{2\pi}v_{i}^R(\epsilon)g^R(\epsilon)\left[\Sigma_{\boldsymbol{\nabla}T}^{<}(\epsilon)(1-n_F(\epsilon))+\Sigma_{\boldsymbol{\nabla}T}^{>}(\epsilon)n_F(\epsilon)\right](g^{R}(\epsilon)-g^A(\epsilon))\\\nonumber
&+ie\hspace{-1mm}\int\hspace{-1mm}\frac{d\omega}{2\pi}\boldsymbol{\mathcal{V}}_{i}^R(\omega)L^R(\omega)\left[\Pi_{\boldsymbol{\nabla}T}^{<}(\omega)(1+n_P(\omega))-\Pi_{\boldsymbol{\nabla}T}^{>}(\omega)n_P(\omega)\right](L^{R}(\omega)-L^A(\omega))+c.c.
\end{align}
\end{widetext}
Here and from now on we omit the notation $eq$ from the equilibrium quantities such as the propagators, self-energies and velocities.

\begin{figure}[pb]
\begin{flushright}\begin{minipage}{0.5\textwidth}  \centering
        \includegraphics[width=0.35\textwidth]{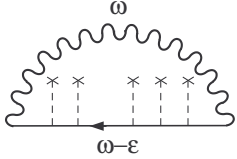}
                 \caption[0.4\textwidth]{\small The self-energy in the first order with respect to the propagator of superconducting fluctuations  before averaging over the disorder.}
                 \label{fig:SelfEnergy1st}
\end{minipage}\end{flushright}
\end{figure}

\begin{figure}[pb]
\begin{flushright}\begin{minipage}{0.5\textwidth}  \centering
        \includegraphics[width=01\textwidth]{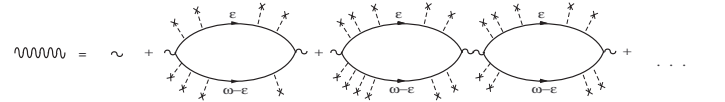}
                 \caption[0.4\textwidth]{\small The geometrical series describing the fluctuations propagator in the Cooper channel.}
                 \label{fig:PolarizationOperator}
\end{minipage}\end{flushright}
\end{figure}

As we are interested in the Gaussian fluctuations, we replace the equilibrium Green function by   $\hat{g}(\mathbf{r},\mathbf{r}',\epsilon)=\hat{g}_0(\mathbf{r},\mathbf{r}',\epsilon)+\int{d\mathbf{r}_{\scriptscriptstyle1}d\mathbf{r}_{\scriptscriptstyle2}}\hat{g}_{0}(\mathbf{r},\mathbf{r}_{\scriptscriptstyle1},\epsilon)\hat{\sigma}(\mathbf{r}_{\scriptscriptstyle1},\mathbf{r}_{\scriptscriptstyle2},\epsilon)$
$\hat{g}_0(\mathbf{r}_{\scriptscriptstyle2},\mathbf{r}',\epsilon)$. Besides, we keep only the contribution to the self-energy with one propagator of the superconducting fluctuations as illustrated in Fig.~\ref{fig:SelfEnergy1st}:
\begin{align}\label{eq:Pelt-SelfEnergy}
\Sigma^{<,>}(\mathbf{r},\mathbf{r}',\epsilon)&=-i\int\frac{d\omega}{2\pi}G^{>,<}(\mathbf{r}',\mathbf{r},\omega-\epsilon)L^{<,>}(\mathbf{r},\mathbf{r}',\omega);\\\nonumber
\Sigma^{R,A}(\mathbf{r},\mathbf{r}',\epsilon)&=-i\int\frac{d\omega}{2\pi}\left[G^{<}(\mathbf{r}',\mathbf{r},\omega-\epsilon)L^{R,A}(\mathbf{r},\mathbf{r}',\omega)\right.\\\nonumber
&\left.
+G^{A,R}(\mathbf{r}',\mathbf{r},\omega-\epsilon)L^{<}(\mathbf{r},\mathbf{r}',\omega)\right].
\end{align}
The propagator of the superconducting fluctuations (see the end of Sec.~\ref{sec:QKE}) is determined by the standard geometrical series $\hat{L}(\omega)=[-\lambda^{-1}+\hat{\Pi}(\omega)]^{-1}$, where $\hat{\Pi}$ is approximated by the particle-particle polarization operator as shown in Fig.~\ref{fig:PolarizationOperator}:
\begin{align}\label{eq:Pelt-Pi}
\Pi^{<,>}(\mathbf{r},\mathbf{r}',\epsilon)&=-i\int\frac{d\epsilon}{2\pi}G^{<,>}(\mathbf{r},\mathbf{r}',\omega-\epsilon)G^{<,>}(\mathbf{r},\mathbf{r}',\epsilon);\\\nonumber
\Pi^{R,A}(\mathbf{r},\mathbf{r}',\epsilon)&=-i\int\frac{d\epsilon}{2\pi}\left[G^{<}(\mathbf{r},\mathbf{r}',\omega-\epsilon)G^{R,A}(\mathbf{r},\mathbf{r}',\epsilon)\right.\\\nonumber
&\left.
+G^{R,A}(\mathbf{r},\mathbf{r}',\omega-\epsilon)G^{>}(\mathbf{r},\mathbf{r}',\epsilon)\right].
\end{align}
One may check that at equilibrium $\Pi_{eq}^{K}=(1+2n_P(\omega))(\Pi_{eq}^{R}-\Pi_{eq}^{A})$, where $n_P(\omega)$ is the Bose distribution function. After averaging over the disorder, $\Pi_{eq}^{R}$ and $\Pi_{eq}^{A}$ are given by the standard expressions.

\begin{figure}[pt]
\begin{flushright}\begin{minipage}{0.5\textwidth}  \centering
        \includegraphics[width=0.85\textwidth]{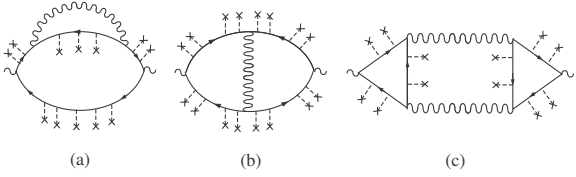} \hspace{0.05in}
                 \caption[0.4\textwidth]{\small  The diagrammatic contributions to the transverse component of $\mathbf{j}_{e}^{con}$ before averaging over the disorder. (The obvious counterpart diagram for (a) is not shown.)} \label{fig:DiagramsOpenImp}
\end{minipage}\end{flushright}
\end{figure}

\begin{figure}[pt]
\begin{flushright}\begin{minipage}{0.5\textwidth}  \centering
        \includegraphics[width=0.85\textwidth]{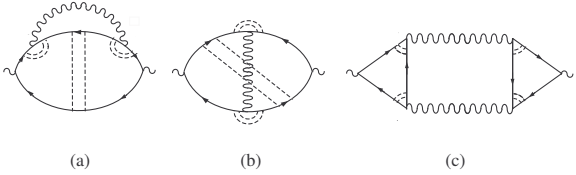} \hspace{0.05in}
                 \caption[0.4\textwidth]{\small  The diagrammatic contributions to the transverse component of the $\mathbf{j}_{e}^{con}$. Diagrams (a) and (b) describe the fluctuation of the superconducting order parameter decorated by three Cooperons and (c) is the Aslamazov-Larkin diagram. (The obvious counterpart diagrams for (a) and (b) are not shown.)} \label{fig:diagrams}
\end{minipage}\end{flushright}
\end{figure}

We may now obtain the leading order corrections in the interaction to the electric current as a response to a temperature gradient in the linear regime.  We should consider all possibilities to linearize the expressions for $\Sigma_{\boldsymbol{\nabla}T}$ and $\Pi_{\boldsymbol{\nabla}T}$ with respect to $\boldsymbol{\nabla}T$ in Eq.~\ref{eq:Pelt-JETransInv}. The diagrammatic interpretation for the different contributions to the transverse electric current obtained in the quantum kinetic equation technique corresponds to the three diagrams shown in Fig.~\ref{fig:DiagramsOpenImp}. After averaging over the disorder the leading contributions to the Nernst signal in the diffusive regime are obtained from the diagrams with three Cooperons~\cite{Varlamov} presented in Figs.~\ref{fig:diagrams}(a) and~\ref{fig:diagrams}(b) and the Aslamazov-Larkin diagram~\cite{Aslamazov1968} shown in Fig.~\ref{fig:diagrams}(c). [The Cooperon is a singular diffusion propagator which describes the rescattering on impurities in the particle-particle channel.] Since we generate these terms using the quantum kinetic equation, the analytic structure of the diagrams is given by the equation.

To get the explicit expression for the current we return to the gauge invariant equilibrium Green functions $\tilde{g}$ given in Eq.~\ref{eq:QKE-G_eqGaugeInv}. Since we restrict our calculation to the limit $\omega_c\tau\ll1$ (where $\omega_c=eH/m^{*}c$ is the cyclotron frequency of the quasi-particles), we may neglect the dependence of $\tilde{g}_{eq}$ on the magnetic field entering through the Landau quantization of the quasi-particles states. Therefore, the entire dependence of the quasi-particle Green functions on the magnetic field is through the phase. Unlike the quasi-particles, the Landau quantization of the collective modes (both the Cooperons and the fluctuations of the superconducting order parameter) cannot be neglected because the quantization condition for these modes is $\Omega_c/T_0>1$, where in the diffusive regime $\Omega_c=4eHD/c$ is the cyclotron frequency in the Cooper channel. Note that $\Omega_c\propto\omega_c(\varepsilon_F\tau)\gg\omega_c$ because the product of the Fermi energy and the mean free time is assumed to be a large parameter. [In $\Omega_c$ the effective charge is equal to $2e$ and the diffusion coefficient $D$ replaces $1/2m$ because in the Cooperons and the fluctuations propagators the term $Dq^2$ substitutes the kinetic energy $p^2/2m$.]

In the limit of small magnetic fields the Cooperons, like the quasi-particle Green functions, can be separated into the phase $\exp\{2ie\int_{\mathbf{r}'}^{\mathbf{r}}\mathbf{A}(\mathbf{r}_{\scriptscriptstyle1})d\mathbf{r}_{\scriptscriptstyle1}/c\}$ and the gauge invariant part at $\mathbf{H}=0$,  $\tilde{C}^{R,A}(\boldsymbol{\rho},\epsilon,\omega-\epsilon)=\left[\mp{i}(2\epsilon-\omega)\tau-D\boldsymbol{\nabla}_{\boldsymbol{\rho}}^2\right]^{-1}$, see Appendix~C in Ref.~\onlinecite{Khodas2003}. At a finite magnetic field, one may express the gauge invariant part of the Cooperon propagator using the Landau level quantization:
\begin{align}\label{eq:QKE-Cooperon}
\tilde{C}_N^{R,A}(\epsilon,\omega-\epsilon)=\left[\mp{i}(2\epsilon-\omega)\tau+\Omega_c\tau(N+1/2)\right]^{-1},
\end{align}
where N is the index of the Landau level.  Similarly, the propagator  of the superconducting fluctuations written in terms of the Landau levels becomes:
\begin{subequations}
\begin{align}\label{eq:QKE-L}
&\tilde{L}_N^{R,A}(\omega)=\frac{-1}{\nu}\hspace{-1mm}\left[\ln\left(\frac{T}{T_c}\right)+\psi_{\scriptscriptstyle{R,A}}\left(\omega,N\right)-\psi\left(\frac{1}{2}\right)+\varsigma\omega\right]^{-1}\hspace{-3mm};
\end{align}
\begin{align}\label{eq:QKE-Digamma}
\psi_{\scriptscriptstyle{R,A}}\left(\omega,N\right)=\psi\left(\frac{1}{2}\mp\frac{{i}\omega}{4\pi{T}}+\frac{\Omega_c(N+1/2)}{4\pi{T}}\right)\hspace{-1mm}
\end{align}
\end{subequations}
Here, $\psi(x)$ is the digamma function. The primary goal of this calculation is to analyze the measurements of the Nernst effect in superconducting films.~\cite{Aubin2006,Aubin2007} In such films the electron states are not quantized and therefore $\nu$ is the density of states of three dimensional electrons (as well as $D$). The parameter $\varsigma\propto1/(\nu\lambda\varepsilon_F)$  is important for understanding the difference in magnitude between the longitudinal and transverse Peltier coefficients. The longitudinal Peltier coefficient, $\alpha_{xx}$, contains an integral over the frequency that vanishes when $\varsigma=0$ while the integrand determining $\alpha_{xy}$ remains finite even in the absence of $\varsigma$.  As a result, in the expression for the Nernst coefficient given in  Eq.~\ref{eq:Intro-NernstCoefficient} the second term in the numerator is smaller than the first one by a factor of the order $T/(\nu\lambda\varepsilon_F)$.~\cite{Larkin1995}

\begin{figure}[pb]
\begin{flushright}\begin{minipage}{0.5\textwidth}  \centering
        \includegraphics[width=0.6\textwidth]{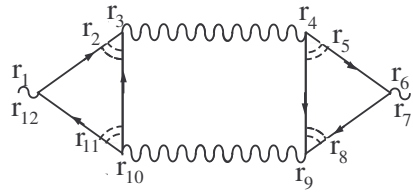}
                 \caption[0.4\textwidth]{\small  The Aslamazov-Larkin diagram.} \label{fig:AL}
\end{minipage}\end{flushright}
\end{figure}

Using the expressions for the quasi-particle Green functions, the Cooperons and the propagators of the superconducting fluctuations in the equilibrium state we may investigate the contributions of $\hat{G}_{\boldsymbol{\nabla}T}$ and $\hat{L}_{\boldsymbol{\nabla}T}$ to the current. Recall that we are interested in the transverse current. For illustration, let us show how to find the transverse current for one representative term out of the few contributions to the Aslamazov-Larkin diagram:
\begin{widetext}
\begin{align}\label{eq:QKE-ALFlux}\nonumber
&j_{e\hspace{1.5mm}x}^{con}(\mathbf{r}_{\scriptscriptstyle1})=\frac{e\boldsymbol{\nabla}_yT}{2T_0}\int\hspace{-1mm}\frac{d\epsilon{d}\epsilon'd\omega}{(2\pi)^3}\hspace{-1mm}\int\hspace{-1mm}{d\mathbf{r}_{\scriptscriptstyle2}...d\mathbf{r}_{\scriptscriptstyle12}}
\lim_{\mathbf{r}_{\scriptscriptstyle12}\rightarrow\mathbf{r}_{\scriptscriptstyle1}}\hspace{-1mm}\left(\frac{\boldsymbol{\nabla}_{\scriptscriptstyle1}^{x}}{2m}+\frac{ieHy_{\scriptscriptstyle1}}{4mc}-\frac{\boldsymbol{\nabla}_{\scriptscriptstyle12}^{x}}{2m}+\frac{ieHy_{\scriptscriptstyle12}}{4mc}\right)
\lim_{\mathbf{r}_{\scriptscriptstyle6}\rightarrow\mathbf{r}_{\scriptscriptstyle7}}\left(\frac{\boldsymbol{\nabla}_{\scriptscriptstyle7}^y}{2m}-\frac{ieHx_{\scriptscriptstyle7}}{4mc}-\frac{\boldsymbol{\nabla}_{\scriptscriptstyle6}^y}{2m}-\frac{ieHx_{\scriptscriptstyle6}}{4mc}\right)\\\nonumber&
g_{0}^R(\mathbf{r}_{\scriptscriptstyle1},\mathbf{r}_{\scriptscriptstyle2},\epsilon)g_{0}^A(\mathbf{r}_{\scriptscriptstyle11},\mathbf{r}_{\scriptscriptstyle2},\omega-\epsilon)g_{0}^R(\mathbf{r}_{\scriptscriptstyle11},\mathbf{r}_{\scriptscriptstyle12},\epsilon)
C_R(\mathbf{r}_{\scriptscriptstyle2},\mathbf{r}_{\scriptscriptstyle3},\epsilon,\omega-\epsilon)
C_R(\mathbf{r}_{\scriptscriptstyle10},\mathbf{r}_{\scriptscriptstyle11},\epsilon,\omega-\epsilon)
L_{eq}^{R}(\mathbf{r}_{\scriptscriptstyle3},\mathbf{r}_{\scriptscriptstyle4},\omega)L_{eq}^{A}(\mathbf{r}_{\scriptscriptstyle9},\mathbf{r}_{\scriptscriptstyle10},\omega)\\&g_{0}^R(\mathbf{r}_{\scriptscriptstyle5},\mathbf{r}_{\scriptscriptstyle6},\epsilon')
g_{0}^A(\mathbf{r}_{\scriptscriptstyle5},\mathbf{r}_{\scriptscriptstyle8},\epsilon')g_{0}^R(\mathbf{r}_{\scriptscriptstyle7},\mathbf{r}_{\scriptscriptstyle8},\epsilon')
C_R(\mathbf{r}_{\scriptscriptstyle4},\mathbf{r}_{\scriptscriptstyle5},\epsilon',\omega-\epsilon')C_R(\mathbf{r}_{\scriptscriptstyle8},\mathbf{r}_{\scriptscriptstyle9},\epsilon',\omega-\epsilon')F(\epsilon,\epsilon',\omega).
\end{align}
In Fig.~\ref{fig:AL} we indicate the spatial coordinates corresponding to the expression given above. Since in this part of the calculation we concentrate on the integration over the spatial coordinates, we collect all the frequency dependent factors into the function $F(\epsilon,\epsilon',\omega)=\epsilon\left[\tanh\left({\epsilon}/{2T}\right)-\tanh\left(({\epsilon-\omega})/{2T}\right)\right]\tanh\left(({\omega-\epsilon'})/{2T}\right)
{\partial{n_P(\omega)}}/{\partial\omega}$ and leave them aside for a while.

Next, we rewrite the Cooperons and the propagators of the superconducting fluctuations using the basis of the Landau levels states, $\varphi_{N,n}(\mathbf{r})=R_{N,n}(r)e^{in\phi}/\sqrt{2\pi}$
(where $R_{N,n}(r)$ are the  generalized Laguerre polynomials). In addition, we separate the quasi-particles Green functions into the phases and the gauge invariant Green functions. Then, following the flux technique introduced in Ref.~\onlinecite{Khodas2003}, we rearrange Eq.~\ref{eq:QKE-ALFlux} as:
\begin{align}\label{eq:QKE-ALFlux2}
&j_{e\hspace{1.5mm}x}^{con}(\mathbf{r}_{\scriptscriptstyle1})=\frac{e\boldsymbol{\nabla}_yT}{4\pi{T}_0\ell_H^2}\int\frac{d\epsilon{d}\epsilon'd\omega}{(2\pi)^3}\sum_{N,M}\int{d\mathbf{r}_{\scriptscriptstyle2}...d\mathbf{r}_{\scriptscriptstyle12}}
e^{-ie\mathbf{H}(\mathbf{r}_{\scriptscriptstyle11}-\mathbf{r}_{\scriptscriptstyle1})\times(\mathbf{r}_{\scriptscriptstyle1}-\mathbf{r}_{\scriptscriptstyle2})/2c}e^{-ie\mathbf{H}(\mathbf{r}_{\scriptscriptstyle5}-\mathbf{r}_{\scriptscriptstyle6})\times(\mathbf{r}_{\scriptscriptstyle6}-\mathbf{r}_{\scriptscriptstyle8})/2c}\\\nonumber
&\lim_{\mathbf{r}_{\scriptscriptstyle12}\rightarrow\mathbf{r}_{\scriptscriptstyle1}}\left(\frac{\boldsymbol{\nabla}_{\scriptscriptstyle1}^{x}}{2m}+\frac{ieH(y_{\scriptscriptstyle1}-y_{\scriptscriptstyle2})}{4mc}-\frac{\boldsymbol{\nabla}_{\scriptscriptstyle12}^x}{2m}-\frac{ieH(y_{\scriptscriptstyle11}-y_{\scriptscriptstyle12})}{4mc}\right)
\tilde{g}_{0}^R(\mathbf{r}_{\scriptscriptstyle1}-\mathbf{r}_{\scriptscriptstyle2},\epsilon)
\tilde{g}_{0}^A(\mathbf{r}_{\scriptscriptstyle11}-\mathbf{r}_{\scriptscriptstyle2},\omega-\epsilon)\tilde{g}_{0}^R(\mathbf{r}_{\scriptscriptstyle11}-\mathbf{r}_{\scriptscriptstyle12},\epsilon)
\\\nonumber
&\lim_{\mathbf{r}_{\scriptscriptstyle6}\rightarrow\mathbf{r}_{\scriptscriptstyle7}}
\left(\frac{\boldsymbol{\nabla}_{\scriptscriptstyle7}^y}{2m}-\frac{ieH(x_{\scriptscriptstyle7}-x_{\scriptscriptstyle8})}{4mc}-\frac{\boldsymbol{\nabla}_{\scriptscriptstyle6}^y}{2m}+\frac{ieH(x_{\scriptscriptstyle5}-x_{\scriptscriptstyle6})}{4mc}\right)
\tilde{g}_{0}^R(\mathbf{r}_{\scriptscriptstyle5}-\mathbf{r}_{\scriptscriptstyle6},\epsilon')\tilde{g}_{0}^A(\mathbf{r}_{\scriptscriptstyle5}-\mathbf{r}_{\scriptscriptstyle8},\epsilon')\tilde{g}_{0}^R(\mathbf{r}_{\scriptscriptstyle7}-\mathbf{r}_{\scriptscriptstyle8},\epsilon')\\\nonumber
&e^{-ie\mathbf{H}(\mathbf{r}_{\scriptscriptstyle8}-\mathbf{r}_{\scriptscriptstyle11})\times(\mathbf{r}_{\scriptscriptstyle11}-\mathbf{r}_{\scriptscriptstyle2})/c-
ie\mathbf{H}(\mathbf{r}_{\scriptscriptstyle2}-\mathbf{r}_{\scriptscriptstyle5})\times(\mathbf{r}_{\scriptscriptstyle5}-\mathbf{r}_{\scriptscriptstyle8})/c}
\varphi_{N,0}(\mathbf{r}_{\scriptscriptstyle2}-\mathbf{r}_{\scriptscriptstyle5})\varphi_{M,0}(\mathbf{r}_{\scriptscriptstyle8}-\mathbf{r}_{\scriptscriptstyle11})
C_N^{R}(\epsilon,\omega-\epsilon)C_M^2(\epsilon,\omega-\epsilon)\\\nonumber&
L_{N}^{R}(\omega)L_{M}^{A}(\omega)
C_N^R(\epsilon',\omega-\epsilon')C_M^R(\epsilon',\omega-\epsilon')F(\epsilon,\epsilon',\omega),
\end{align}
\end{widetext}
where $\ell_H=\sqrt{c/2eH}$ is the magnetic length in the Cooper channel. In the last step we used the orthogonality of the generalized Laguerre polynomials (an example for the treatment of the propagators in this basis can be found in Ref.~\onlinecite{Laikhtman1994}). The first two exponents in Eq.~\ref{eq:QKE-ALFlux2} contain the magnetic fluxes accumulated in the triangles $(\mathbf{r}_{\scriptscriptstyle1},\mathbf{r}_{\scriptscriptstyle2},\mathbf{r}_{\scriptscriptstyle11})$ and $(\mathbf{r}_{\scriptscriptstyle5},\mathbf{r}_{\scriptscriptstyle6},\mathbf{r}_{\scriptscriptstyle8})$, respectively. One way to get the transverse current, is to extract the magnetic field from these two fluxes or from the diamagnetic terms. As a result the transverse current appears with the coefficient $\omega_c\tau$. We neglect these terms; we will see that when the magnetic field responsible for turning the current to the transverse direction is extracted from the Cooperons or the propagators of the superconducting fluctuations one gets a much larger factor of the order $\Omega_C/T$. Therefore, the integration over the coordinates of the two triangles can be done with the quasi-particle Green functions taken at $\mathbf{H}=0$:
\begin{align}\label{eq:QKE-ALFlux3}
&j_{e\hspace{1.5mm}x}^{con}=-\frac{e\boldsymbol{\nabla}_yT}{8\pi^2T_0\ell_H^2}\nu^2\tau^4\int{d\epsilon{d}\epsilon'd\omega}\int{d\mathbf{r}}\sum_{N,M}\\\nonumber
&
\left[2D\left(\frac{\partial}{\partial{x}}+\frac{ieHy}{c}\right)\varphi_{N,0}(\mathbf{r})\right]
\left[2D\left(\frac{\partial}{\partial{y}}-\frac{ieHx}{c}\right)\varphi_{M,0}(\mathbf{r})\right]\\\nonumber
&C_N^R(\epsilon,\omega-\epsilon)C_M^R(\epsilon,\omega-\epsilon)
L_{N}^{R}(\omega)L_{M}^{A}(\omega)
C_N^R(\epsilon',\omega-\epsilon')\\\nonumber&C_M^R(\epsilon',\omega-\epsilon')F(\epsilon,\epsilon',\omega).
\end{align}
The integral over the coordinate corresponds to the matrix element of the velocity operators $\langle{N,0}|V_xV_y|M,0\rangle$, where $|M,0\rangle=\varphi_{M,0}$ is the quantum state of a particle with a mass equal to $1/2D$ in the $M$ Landau level and zero angular momentum in the $z$-direction. Using the known properties of the Laguerre polynomials, the matrix element can be written as $\langle{N,0}|V_xV_y|M,0\rangle=2iD^2[(N+1)\delta_{N,M-1}-(M+1)\delta_{M,N-1}]/\ell_H^2$. Finally, the contribution to the current becomes:
\begin{align}\label{eq:QKE-ALFlux4}\nonumber
&j_{e\hspace{1.5mm}x}^{con}(\mathbf{r}_{\scriptscriptstyle1})=-i\frac{e\boldsymbol{\nabla}_yT}{4\pi^2T_0\ell_H^4}\nu^2D^2\tau^4\int{d\epsilon{d}\epsilon'd\omega}\sum_{N=0}^{\infty}
(N+1)\\\nonumber&C_N^R(\epsilon,\omega-\epsilon)C_{N+1}^R(\epsilon,\omega-\epsilon)C_N^R(\epsilon',\omega-\epsilon')C_{N+1}^R(\epsilon',\omega-\epsilon')\\
&\left[L_{N}^{R}(\omega)L_{N+1}^{A}(\omega)-L_{N+1}^{R}(\omega)L_{N}^{A}(\omega)\right]
F(\epsilon,\epsilon',\omega).
\end{align}
In the limit $\mathbf{H}\rightarrow0$ when the quantization of the collective modes can be neglected, one may replace the Cooperons and the propagators of the superconducting fluctuations  in Eq.~\ref{eq:QKE-ALFlux} by the product of the phase terms (with charge $2e$) and the corresponding propagators in the absence of a magnetic field. Then, the contribution to the current at vanishingly small magnetic field can be found by employing the flux technique of Ref.~\onlinecite{Khodas2003}. One may check that the same result is obtained when the transformation from the discrete sum into an integral over a continuous variable is performed in Eq.~\ref{eq:QKE-ALFlux4}.

Let us conclude with a remark regarding the diagrammatic interpretation of the different contributions to $\mathbf{j}_{e}^{con}$. As already mentioned, the analytical structure and the expressions for the vertices of these diagrams were found from the quantum kinetic equation. In principle, the same diagrams can be calculated using the Kubo formula. However, if for simplicity one uses in the Kubo formula the heat current operator of non-interacting electrons described in Eq.~\ref{eq:CurrentNonIntFrequency}, the resulting expressions for these diagrams differ from those obtained in the quantum kinetic approach. Most important, as one can see from Eq.~\ref{eq:Pelt-JETransInv}, in the quantum kinetic approach the frequency accompanies the renormalized velocity, so that the expression for the electric current is generally of the form $eg(\epsilon)v_i(\epsilon)g(\epsilon)\epsilon{v}_j(\epsilon)\boldsymbol{\nabla}_jT/T$. In other words, the frequency appears together with the velocity that was already renormalized by the interaction. On the other hand, owing to the fact that the frequency in the simplified version of the Kubo formula is attached to the external vertex before the renormalization of the velocity, the expression for the current has a totaly different structure.

This is also the proper place to explain what is so unique in the superconducting fluctuations in the diffusive limit that leads to the giant Nernst effect.  As any transverse current coefficient, $\alpha_{xy}$ contains a difference of two almost equal terms. In addition, like all thermoelectric coefficients the integral over the frequency in $\alpha$ contains a factor of the quasi-particle frequency. Consequently, as discussed in Appendix~\ref{sec:p-hSymmtery} the contribution of the quasi-particles to the transverse Peltier coefficient  includes two small parameters. The first, is the usual $\omega_c\tau$ that appears in all transverse currents. The second is a reminiscence of the fact that the frequency factor (that in the Boltzmann equation is converted into the energy) is responsible for the vanishing of  the Peltier coefficient under the approximation of a constant density of states. When a non-constant density of states is considered, the integration over the energy yields another small parameter proportional to $T/\varepsilon_F$. Now we turn to the contribution of the superconducting fluctuations to the transverse component of $\mathbf{j}_{e}^{con}$, and   consider Eq.~\ref{eq:QKE-ALFlux4} as a representative example. For the moment we ignore the factor $\epsilon$ in $F(\epsilon,\epsilon',\omega)$ associated with the thermoelectric current. Then, the difference between the two almost identical terms results in an odd integrand with respect to the frequency of the superconducting fluctuations, $\omega$, which potentially may lead to the vanishing of $\alpha_{xy}$. So, how can the superconducting fluctuations induce a strong Nernst signal? The explanation lies in the fact that the Cooperons accompanying the superconducting fluctuation depend on the frequency of the incoming/outgoig quasi-particles and not only on the frequency $\omega$ carried by the fluctuations (Eq.~\ref{eq:QKE-Cooperon}). The dependence of the Cooperons on $\epsilon$ combined with the frequency factor $\epsilon$ in $F(\epsilon,\epsilon',\omega)$ save the situation. This is because the integration over $\epsilon$ results in an integrand that is an even function of $\omega$ and, hence, there is no longer danger that the transverse Peltier coefficient vanishes. We shall see that instead of the two small parameters obtained for the quasi-particles, the contribution of the superconducting fluctuations includes only one. Because of the extra-sensitivity of these fluctuations to the magnetic field this parameter is $\Omega_c/T_0$.

\section{Final expressions for the transverse component of $\mathbf{j}_{e}^{con}$}\label{sec:FinalExpression}

If one examines Eq.~\ref{eq:Pelt-JETransInv} which presents the general expression for the contributions to the electric current from $\hat{G}_{\boldsymbol{\nabla}T}$ and $\hat{L}_{\boldsymbol{\nabla}T}$, one may notice that not all the terms contain the derivative of a Fermi distribution function. As one may expect, the terms in which the Fermi distribution function is not differentiated contribute only to the transverse component of $\mathbf{j}_{e}^{con}$ and not to the longitudinal one. After integration over the Fermion degrees of freedom  (the frequency $\epsilon$ and the coordinates of the quasi-particles Green functions), the terms proportional to $\partial{n}_F(\epsilon)/\partial\epsilon$  give two non-vanishing contributions. The first one corresponds to the Aslamazov-Larkin diagram presented in Fig.~\ref{fig:diagrams}(c):
\begin{align}\label{eq:QKE-JETransInv2}\nonumber
&j_{e\hspace{1.5mm}i}^{con1}=\varepsilon_{ij}\frac{e\boldsymbol{\nabla}_j{T}}{16\pi^2T_0}\nu^2\hspace{-1mm}\int\hspace{-1mm}{d\omega}\sum_{N=0}^{\infty}(N+1)\frac{\partial{n_P(\omega)}}{\partial\omega}\\\nonumber
&\left[L_{N}^{R}(\omega)L_{N+1}^{A}(\omega)-L_{N+1}^{R}(\omega)L_{N}^{A}(\omega)\right]\\\nonumber
&\left[\psi_{\scriptscriptstyle{R}}\left(\omega,N\right)-
\psi_{\scriptscriptstyle{R}}\left(\omega,N+1\right)
+\psi_{\scriptscriptstyle{A}}\left(\omega,N\right)-
\psi_{\scriptscriptstyle{A}}\left(\omega,N+1\right)
\right]\\\nonumber&
\left[\Omega_c(N+1/2)\left(\psi_{\scriptscriptstyle{R}}\left(\omega,N\right)-
\psi_{\scriptscriptstyle{A}}\left(\omega,N\right)\right)\right.\\
&\left.-\Omega_c(N+3/2)
\left(\psi_{\scriptscriptstyle{R}}\left(\omega,N+1\right)-
\psi_{\scriptscriptstyle{A}}\left(\omega,N+1\right)\right)
\right],
\end{align}
where the upper index in $j^{con1}$ enumerates the contribution to the current, and $\varepsilon_{ij}$ is the anti-symmetric tensor. The second contribution generated by terms with the derivative $\partial{n}_F(\epsilon)/\partial\epsilon$ corresponds to the diagram with three Cooperons shown in Fig.~\ref{fig:diagrams}(a):
\begin{widetext}
\begin{align}\label{eq:QKE-JETransInv3}
&j_{e\hspace{1.5mm}i}^{con2}=-\varepsilon_{ij}\frac{e\boldsymbol{\nabla}_j{T}}{4\pi^2T_0}\nu\int{d\omega}\sum_{N=0}^{\infty}\Omega_c(N+1)
\left\{\frac{1}{4}L_N^R(\omega)\frac{\partial{n_P(\omega)}}{\partial\omega}
\left[\frac{i\omega+\Omega_c(N+1/2)}{4\pi{T_0}}
\left(\phantom{\frac{1}{!}}\hspace{-3mm}\psi_{\scriptscriptstyle{A}}'\left(\omega,N\right)
-\psi_{\scriptscriptstyle{R}}'\left(\omega,N\right)\phantom{\frac{1}{!}}\hspace{-3mm}\right)\right.\right.\\\nonumber&
\left.\left.+
\frac{i\omega+\Omega_c(N+3/2)}{\Omega_c}
\left(\phantom{\frac{1}{!}}\hspace{-3mm}\psi_{\scriptscriptstyle{A}}\left(\omega,N\right)
-\psi_{\scriptscriptstyle{R}}\left(\omega,N\right)
-\psi_{\scriptscriptstyle{A}}\left(\omega,N+1\right)
+\psi_{\scriptscriptstyle{R}}\left(\omega,N+1\right)
\phantom{\frac{1}{!}}\hspace{-3mm}\right)\right]\right.\\\nonumber&\left.
+\frac{i}{2}L_N^An_P(\omega)
\left[\frac{i\omega+\Omega_c(N+1/2)}{(4\pi{T_0})^2}
\psi_{\scriptscriptstyle{A}}''\left(\omega,N\right)
+\frac{i\omega+\Omega_c(N+3/2)}{4\pi{T_0}\Omega_c}
\left(\psi_{\scriptscriptstyle{A}}'\left(\omega,N\right)
-\psi_{\scriptscriptstyle{A}}'\left(\omega,N+1\right)
\right)\right]
+N\leftrightarrow{N+1}\phantom{\frac{1}{1}}\hspace{-2mm}
\right\}+c.c.
\end{align}
\end{widetext}
Here $\psi_{\scriptscriptstyle{R,A}}'$ and $\psi_{\scriptscriptstyle{R,A}}''$ correspond to the first and second derivatives of the digamma function defined in Eq.~\ref{eq:QKE-Digamma}. The notation $N\leftrightarrow{N+1}$ means that $N$ is replaced by $N+1$ and the other way around in all the terms inside the curly brackets.  Notice that there are no contributions proportional to the derivative of the distribution function which can be attributed to the diagram shown in Fig~\ref{fig:diagrams}(b).

Next, we discuss the group of terms that are proportional to $n_F(\epsilon)$. The diagrammatic interpretation of these terms, which are generated by Eq.~\ref{eq:Pelt-JETransInv}, includes all three diagrams presented in Fig.~\ref{fig:diagrams}. However, one may check that the contributions from the diagrams shown in Fig.~\ref{fig:diagrams}(b) and~\ref{fig:diagrams}(c) are cancelled by a part of the contribution from the diagram given in Fig.~\ref{fig:diagrams}(a). The remaining contribution is:
\begin{align}\label{eq:QKE-JETransInv4}
&j_{e\hspace{1.5mm}i}^{con3}=-i\varepsilon_{ij}\frac{e\boldsymbol{\nabla}_j{T}}{4\pi^2{T}_0}\nu\int{d\omega}\sum_{N=0}^{\infty}(N+1)n_P(\omega)
\\\nonumber
&\left\{\left[L_{N}^{R}(\omega)+L_{N+1}^{R}(\omega)\right]\left[\psi_{\scriptscriptstyle{R}}\left(\omega,N\right)-
\psi_{\scriptscriptstyle{R}}\left(\omega,N+1\right)\right]\right.\\\nonumber
&\left.+
\frac{\Omega_c}{4\pi{T}}L_{N}^{R}(\omega)\psi_{\scriptscriptstyle{R}}'\left(\omega,N\right)
+\frac{\Omega_c}{4\pi{T_0}}L_{N+1}^{R}(\omega)\psi_{\scriptscriptstyle{R}}'\left(\omega,N+1\right)
\right\}
\end{align}
In the derivation of the different contributions to $\mathbf{j}_{e}^{con}$  we used the following identities for products of the distribution functions:
\begin{align}\label{eq:Pelt-identites}
&n_F(\epsilon)n_F(\omega-\epsilon)=n_P(\omega)[n_F(\epsilon-\omega)-n_F(\epsilon)];\\\nonumber
&\frac{\partial{n}_F(\omega-\epsilon)}{\partial\omega}n_F(\epsilon)=\frac{\partial{n_P(\omega)}}{\partial\omega}\left[n_F(\epsilon-\omega)-n_F(\epsilon)\right]\\\nonumber
&\hspace{30mm}-\frac{\partial{n_F(\omega-\epsilon)}}{\partial\omega}n_P(\omega).
\end{align}

Further analysis of $\mathbf{j}_{e}^{con}$ at arbitrary temperatures and magnetic fields shows that in $j_{e\hspace{1.5mm}i}^{con1}$ and $j_{e\hspace{1.5mm}i}^{con2}$ the integration over the frequency  accumulates at $\omega\sim{T}\ll1/\tau$. As a consequence of the narrow range of the integration, the final expressions for these two contributions vanish in the limit $T\rightarrow0$. In contrast,  in $j_{e\hspace{1.5mm}i}^{con3}$ the integration over the frequency is not limited to small frequencies and, hence, the outcome of the integration depends logarithmically on the scattering rate $1/\tau$ which acts as an ultraviolet cutoff. In addition, as the temperature goes to zero there is even a more serious problem with this term, because its pre-factor is proportional to $\Omega_c/T_0$. Such a dependence on the temperature violates the third law of thermodynamics. [The connection between the third law of thermodynamics and the Nernst effect was discussed in the Introduction.] We shall see that the dangerous parts in $j_{e\hspace{1.5mm}i}^{con3}$ are cancelled out by the magnetization current that up to now we have not yet considered.

\section{The magnetization current and the third law of thermodynamics}\label{sec:Magnetization}

In this section we examine the magnetization current given in Eq.~\ref{eq:Current-Jmag}. In general, we need to insert the  $\boldsymbol{\nabla}T$-dependent part of the Green function, $\hat{G}_{loc-eq}+\hat{G}_{\boldsymbol{\nabla}T}$, into Eq.~\ref{eq:Current-Jmag}. Since after the averaging over the disorder $G_{\boldsymbol{\nabla}T}(\mathbf{r}\rightarrow\mathbf{r}',\epsilon)$ is translationally invariant, it is clear that this part of the Green function does not contribute to the magnetization current. On the other hand, the explicit dependence of the local equilibrium Green function on the center of mass coordinate leads to a non-zero contribution to the magnetization current:
\begin{align}\label{eq:QKE-Jmag2}
\mathbf{j}_e^{mag}=2ic\boldsymbol{\nabla}_{\mathbf{R}}&\times\mathbf{M}(\mathbf{R})\int\frac{d\epsilon}{2\pi}\lim_{\boldsymbol{\rho}\rightarrow0}\epsilon\frac{\mathbf{R}\boldsymbol{\nabla}T}{T_0}\\\nonumber&\frac{\partial{g}_{eq}^{<}(\boldsymbol{\rho},\epsilon;\mathbf{A},imp)}{\partial\epsilon}
\end{align}
Thus, $G_{\boldsymbol{\nabla}T}$ and $G_{loc-eq}$ are complementary to each other; while the first contributes only to $\mathbf{j}_{e}^{con}$, the other one fully determines $\mathbf{j}_{e}^{mag}$.
One should recall that we are looking for a current that does not vanish after spatial averaging, i.e., after integration with respect to the center of mass coordinate $\mathbf{R}$. Since in the process of averaging over $\mathbf{R}$ we may integrate by parts, the magnetization current can be written as
\begin{align}\label{eq:QKE-Jmag3}
j_{e\hspace{1.5mm}i}^{mag}&=2i\varepsilon_{ij}cM_z\lim_{\boldsymbol{\rho}\rightarrow0}\int\frac{d\epsilon}{2\pi}\frac{\boldsymbol{\nabla}_jT}{T_0}g_{eq}^{<}(\boldsymbol{\rho},\epsilon;\mathbf{A},imp).
\end{align}
Here we integrated by parts over the frequency as well. One may recognize that $\mathbf{j}_{e}^{mag}$ is directly related to the magnetization density at equilibrium:
\begin{align}\label{eq:QKE-JM}
j_{e\hspace{1.5mm}i}^{mag}&=-\varepsilon_{ij}c\langle{M_z}\rangle\frac{\boldsymbol{\nabla}_jT}{T_0}.
\end{align}
The result demonstrates the strength of the quantum kinetic approach. This method provides a way to derive both components of the current without engaging any thermodynamical arguments.

\begin{figure}[pt]
\begin{flushright}\begin{minipage}{0.5\textwidth}  \centering
        \includegraphics[width=0.9\textwidth]{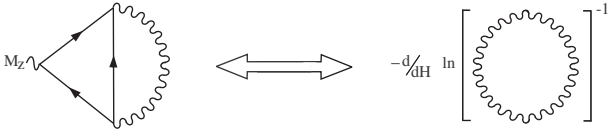} \hspace{0.05in}
                 \caption[0.4\textwidth]{\small An illustration of the relation between the magnetization current term that is obtained from the local equilibrium Green function and the thermodynamic diagram for the magnetization.} \label{fig:Magnetization}
\end{minipage}\end{flushright}
\end{figure}

Actually, at this point one may employ in Eq.~\ref{eq:QKE-JM} the known expression for the magnetization in the presence of superconducting fluctuation. Still, since we are interested in the interplay between the quasi-particle excitations and the fluctuations of the superconducting order parameter, let us derive the expression for the first order correction to the magnetization induced by the fluctuations starting with Eq.~\ref{eq:QKE-Jmag3}. Using the standard identities for the Keldysh Green function at equilibrium, one gets:
\begin{align}\label{eq:QKE-Jmag4}
&j_{e\hspace{1.5mm}i}^{mag}=ic\varepsilon_{ij}\frac{\boldsymbol{\nabla}_jT}{T_0}\lim_{\mathbf{r}'\rightarrow\mathbf{r}}\int\frac{d\epsilon}{2\pi}d\mathbf{r}_{\scriptscriptstyle1}d\mathbf{r}_{\scriptscriptstyle2}\left[M_{z}(\mathbf{r})+M_z(\mathbf{r}')\right]
\\\nonumber&\times{n}_{F}(\epsilon)
\left[g_0^{A}(\mathbf{r}-\mathbf{r}_{\scriptscriptstyle1},\epsilon;\mathbf{A},imp)\sigma^{A}(\mathbf{r}_{\scriptscriptstyle1}-\mathbf{r}_{\scriptscriptstyle2},\epsilon;\mathbf{A},imp)\right.\\\nonumber
&\left.\times{g}_0^{A}(\mathbf{r}_{\scriptscriptstyle2}-\mathbf{r}',\epsilon;\mathbf{A},imp)-
g_0^{R}(\mathbf{r}-\mathbf{r}_{\scriptscriptstyle1},\epsilon;\mathbf{A},imp)\right.\\\nonumber&\left.\times\sigma^{R}(\mathbf{r}_{\scriptscriptstyle1}-\mathbf{r}_{\scriptscriptstyle2},\epsilon;\mathbf{A},imp)g_0^{R}(\mathbf{r}_{\scriptscriptstyle2}-\mathbf{r}',\epsilon;\mathbf{A},imp)\right].
\end{align}
Here, for convenience, we returned to the initial coordinates.
Next, we use the fact that the equilibrium Green function in the absence of fluctuations satisfies the following identity $-\partial{g}_{0}^{R,A}/\partial{\mathbf{H}}={g}_{0}^{R,A}\mathbf{M}{g}_{0}^{R,A}$. Therefore, the expression for the magnetization current can be rewritten as
\begin{align}\label{eq:QKE-Jmag5}
j_{e\hspace{1.5mm}i}^{mag}&=-2ic\varepsilon_{ij}\frac{\boldsymbol{\nabla}_jT}{T_0}\lim_{\mathbf{r}'\rightarrow\mathbf{r}}\int\frac{d\epsilon}{2\pi}d\mathbf{r}_{\scriptscriptstyle1}
n_{F}(\epsilon)
\\\nonumber&\left[\frac{\partial{g}_0^{A}(\mathbf{r}-\mathbf{r}_{\scriptscriptstyle1},\epsilon;\mathbf{A},imp)}{\partial{H}_z}
\sigma^{A}(\mathbf{r}_{\scriptscriptstyle1}-\mathbf{r}',\epsilon;\mathbf{A},imp)\right.\\\nonumber&\left.
-
\frac{\partial{g}_0^{R}(\mathbf{r}-\mathbf{r}_{\scriptscriptstyle1},\epsilon;\mathbf{A},imp)}{\partial{H_z}}\sigma^{R}(\mathbf{r}_{\scriptscriptstyle1}-\mathbf{r}_{\scriptscriptstyle2},\epsilon;\mathbf{A},imp)\right].
\end{align}
Finally, using the explicit expression for the self-energy and rearranging all the terms we reformulate the expression for the magnetization current in terms of the propagator of the superconducting fluctuations:
\begin{align}\label{eq:QKE-Jmag6}
&j_{e\hspace{1.5mm}i}^{mag}=-ic\varepsilon_{ij}\frac{\boldsymbol{\nabla}_jT}{T_0}\lim_{\mathbf{r}'\rightarrow\mathbf{r}}\int\frac{d\omega}{2\pi}d\mathbf{r}_{\scriptscriptstyle1}
n_{P}(\omega)\\\nonumber&
\left[\frac{\partial{\Pi}^{R}(\mathbf{r}-\mathbf{r}_{\scriptscriptstyle1},\omega;\mathbf{A},imp)}{\partial{H}_z}\L^{R}(\mathbf{r}_{\scriptscriptstyle1}-\mathbf{r}',\omega;\mathbf{A},imp)\right.\\\nonumber&\left.
-
\frac{\partial{\Pi}^{A}(\mathbf{r}-\mathbf{r}_{\scriptscriptstyle1},\omega;\mathbf{A},imp)}{\partial{H_z}}L^{A}(\mathbf{r}_{\scriptscriptstyle1}-\mathbf{r}_{\scriptscriptstyle2},\omega;\mathbf{A},imp)\right]\\\nonumber
&=-ic\varepsilon_{ij}\frac{\boldsymbol{\nabla}_jT}{T_0}\frac{\partial}{\partial{H_z}}\lim_{\boldsymbol{\rho}\rightarrow0}\int\frac{d\omega}{2\pi}n_{P}(\omega)
\\\nonumber&\left[\ln{L}_{R}^{-1}(\boldsymbol{\rho},\omega;\mathbf{A},imp)
-\ln{L}_{A}^{-1}(\boldsymbol{\rho},\omega;\mathbf{A},imp)\right].
\end{align}
The transition from Eq.~\ref{eq:QKE-Jmag4} to the last line in Eq.~\ref{eq:QKE-Jmag6} is illustrated in Fig.~\ref{fig:Magnetization}. Averaging over the disorder and transforming from the expression for the propagator as a function of the coordinates to the basis of Landau levels, one obtains the known expression for the correction to the magnetization in the lowest order with respect to the fluctuations:~\cite{Varlamov,Galitski2001}
\begin{align}\label{eq:QKE-Jmag7}
j_{e\hspace{1.5mm}i}^{mag}=\varepsilon_{ij}\frac{\boldsymbol{\nabla}_jT}{T_0}&
\frac{\partial}{\partial{H}}\frac{eH}{\pi}\int\frac{d\omega}{2\pi{i}}\sum_{N=0}^{\infty}n_P(\omega)\\\nonumber&\left[\ln\left(L_N^R(\omega)\right)^{-1}
-\ln\left(L_N^A(\omega)\right)^{-1}\right].
\end{align}
The discussion of higher order corrections to the magnetization is given in Ref.~\onlinecite{Galitski2001}.

Similar to $\mathbf{j}_{e}^{con3}$ in Eq.~\ref{eq:QKE-JETransInv4}, the integration over the frequency in the magnetization current is restricted by the scattering rate, and at low temperature $\mathbf{j}_e^{mag}$ also diverges as $\Omega_c/T$. The opposite sign of the magnetization current relative to $\mathbf{j}_{e}^{con3}$ suggests that these dangerous parts may cancel each other making the Nernst signal compatible with the third law of thermodynamics. Another hint for this cancellation is the similar analytical structure of $\mathbf{j}_{e}^{mag}$ and $\mathbf{j}_{e}^{con3}$. [All the terms in Eq.~\ref{eq:QKE-JETransInv4} and in Eq.~\ref{eq:QKE-Jmag7} are a product of either retarded or advanced functions only.] In Appendix~\ref{App:Magnetization} we show that the diverging parts of the magnetization current indeed identically cancel out the diverging parts of $\mathbf{j}_{e}^{con3}$. We demonstrate that the total current is independent of $\tau$ in the whole temperature range $T\ll1/\tau$. As a result, the Nernst signal is regular at $T\rightarrow0$. Moreover, the contributions which are constant with respect to the temperature also vanish and the remaining terms are linear in $T$.

\section{The phase diagram for the Nernst effect - comparison between the theoretical results and the experiment}\label{sec:phaseDiagram}

\begin{figure}[tp]
\begin{flushright}\begin{minipage}{0.5\textwidth}  \centering
        \includegraphics[width=0.7\textwidth]{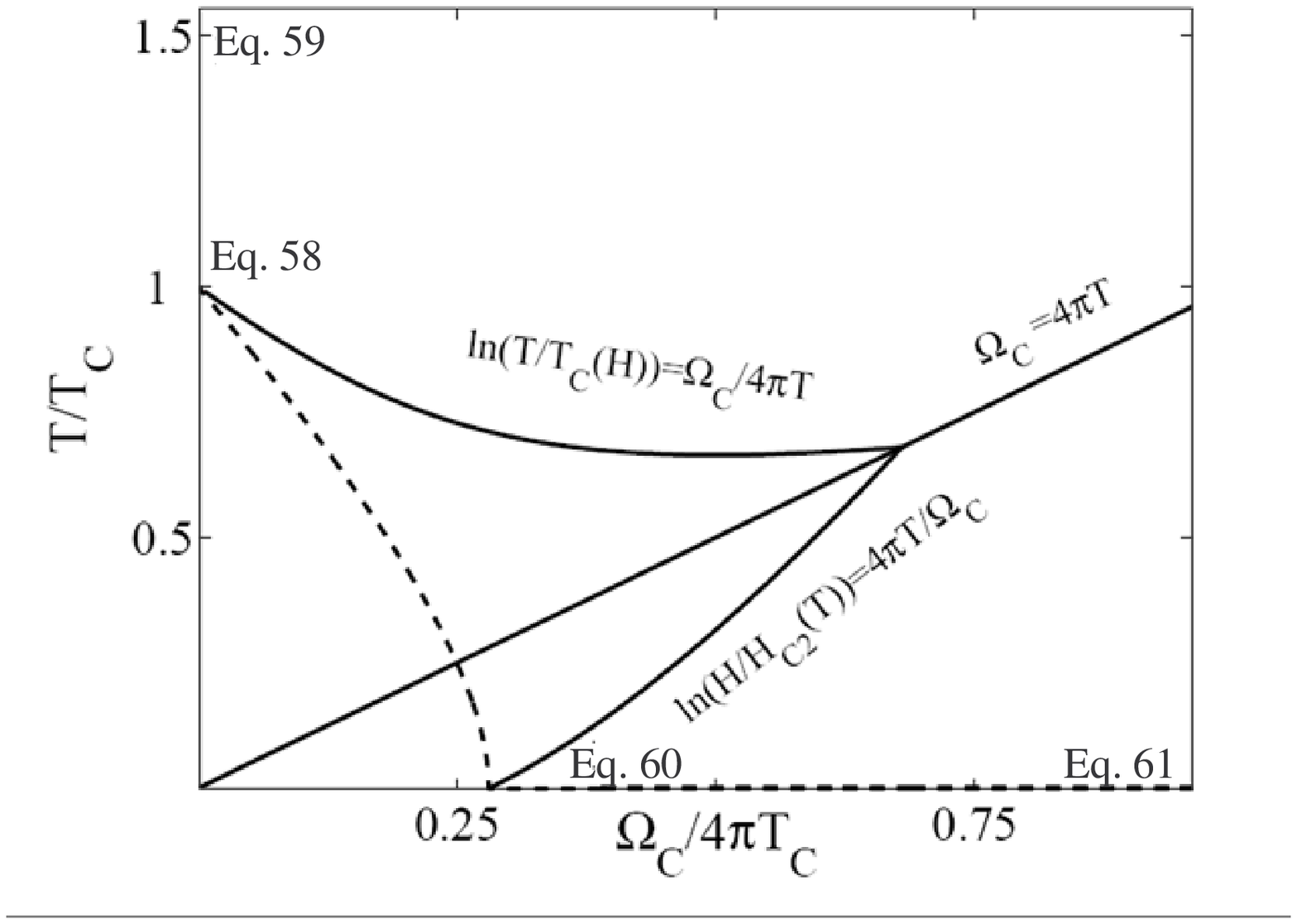} \hspace{0.05in}
                 \caption[0.4\textwidth]{\small The phase diagram for the Peltier coefficient $\alpha_{xy}$. We indicate the equations in the text which give the corresponding expressions for $\alpha_{xy}$ in the different limits. $\Omega_c=4eHD/c$ is the cyclotron frequency for the fluctuations of the superconducting order parameter in the diffusive regime.} \label{fig:PhaseDiagram}
\end{minipage}\end{flushright}
\end{figure}

In the following part we present the theoretical expressions for the transverse Peltier coefficient for a superconducting film in the normal state for various regions of the temperature and the magnetic field.  The phase diagram for the Peltier coefficient is plotted in Fig.~\ref{fig:PhaseDiagram}. In the area below the  line  $\ln(T/T_c(H))=\Omega_c/4\pi{T}$ the Landau level quantization of the superconducting fluctuations becomes essential. The line $\ln(H/H_{c_{\scriptscriptstyle2}}(T))=4\pi{T}/\Omega_c$ separates the regions of classical and quantum fluctuations. From now on $T_0$ is replaced by $T$ which represents the spatially-averaged temperature.

For a small magnetic field, $\Omega_c\ll{T}$, close to the transition temperature ($T\approx{T_c}$) the leading contribution to $\alpha_{xy}$ is given by the Aslamazov-Larkin term (see Fig.~\ref{fig:diagrams} (c)) and the magnetization current:
\begin{align}\label{eq:closetoTc}
&\alpha_{xy}\approx\frac{e\Omega_c}{192T\ln{T/T_c(H)}}.
\end{align}
In the previous section we discussed in details the importance of the magnetization current in cancelling the quantum contributions to the Nernst signal. In the vicinity of $T_c$ one can interpret the expression in Eq.~\ref{eq:closetoTc} in terms of the classical picture in which the Cooper pairs with a finite lifetime are responsible for the thermoelectric current. The magnetization current is just equal to $-2/3$ of the leading order contribution from the Aslamazov-Larkin term. Note that unlike the electric conductivity, $\sigma _{xx}$, for which the anomalous Maki-Thompson~\cite{Maki1968} and the Aslamazov-Larkin terms yield comparable corrections, the contribution from the anomalous Maki-Thompson term to the Nernst signal is $\sim ({T}/\varepsilon _{F})^{2}\ll 1$ smaller than the one given by Eq.~\ref{eq:closetoTc}. Therefore, it is natural that in the vicinity of $T_{c}$ our result coincides with the expression~\cite{Ussishkin2002,Varlamov} obtained phenomenologically from the time dependent
Ginzburg-Landau equation (TDGL).

When temperature is increased further away from the critical temperature, the sum of the contributions to the transverse  Peltier coefficient from all the diagrams and the magnetization current yields:
\begin{align}\label{eq:farfromTc}
&\alpha_{xy}\approx\frac{e\Omega_c}{24\pi^2T\ln{T/T_c}}.
\end{align}
A comparison between the transverse Peltier coefficient in the vicinity of $T_c$ (Eq.~\ref{eq:closetoTc}) and far from the transition (Eq.~\ref{eq:farfromTc}) reveals that the two expressions differ only by a numerical coefficient. The similarity between the expressions for $\alpha_{xy}$ in the two different limits is not seen in paraconductivity. This is a consequence of the cancelation of the quantum contributions to $\mathbf{j}_{e}^{con}$ by the magnetization current, which is specific for the Nernst effect. Away from the critical region $T\gg{T}_c$, the quantum nature of the fluctuations reveals itself in contributions to $\mathbf{j}_{e}^{con3}$ and $\mathbf{j}_{e}^{mag}$ that contain an integration over a wide interval of frequencies between $T$ and $1/\tau$. As a result, these terms become of the order $\ln(\ln1/T\tau)-\ln(\ln{T}/Tc)$. However, as we show in Appendix~\ref{App:Magnetization} these $\tau$-dependent terms in $\mathbf{j}_{e}^{con3}$ and $\mathbf{j}_{e}^{mag}$ cancel each other.~\cite{comment} The Peltier coefficient far from $T_c$ demonstrates how the third law of thermodynamics constrains the magnitude of the Nernst signal not only at $T\rightarrow0$ but also at high temperatures, $T\gg{T}_c$.

\begin{figure}[tp]
\begin{flushright}\begin{minipage}{0.5\textwidth}  \centering \subfigure[]{
        \label{fig:35nm} 
        \includegraphics[width=0.75\textwidth]{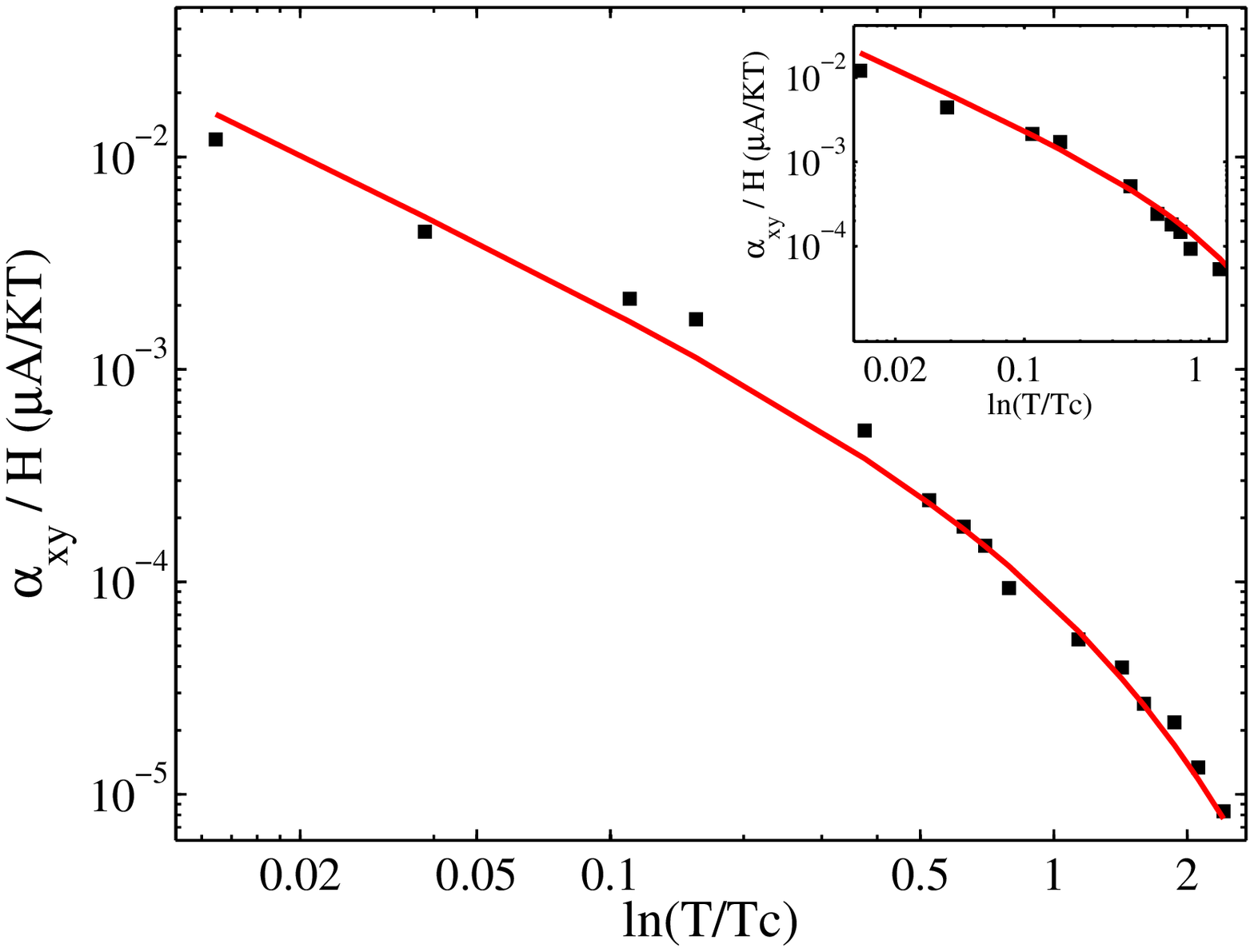}} \hspace{0.05in}%
        \subfigure[]{
        \label{fig:12.5nm} 
        \includegraphics[width=0.75\textwidth]{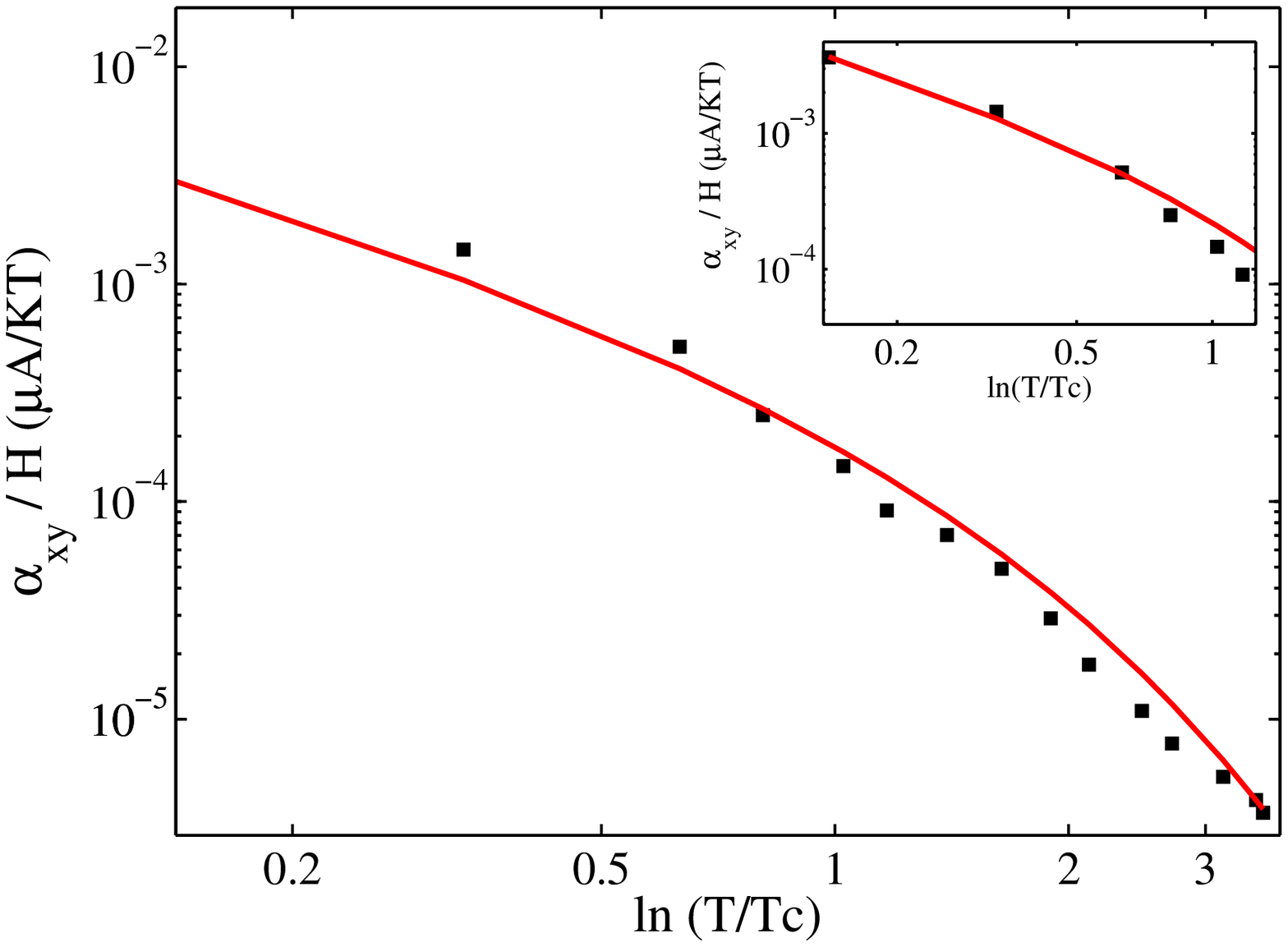}}
          \caption[0.4\textwidth]{\small The transverse Peltier coefficient $\alpha_{xy}$ divided by the magnetic field $H$ as a function of $\ln{T/T_c}$ in the limit $H\rightarrow0$ for films of thicknesses (a) $35{\rm nm}$ and (b) $12.5{\rm nm}$. The experimental data of Ref.~\onlinecite{Aubin2007} is presented by the black squares and the solid line corresponds to the theoretical curve given by Eq.~\ref{eq:farfromTc}.  The insert presents the fitting of the data in the vicinity of $T_c$ with Eq.~\ref{eq:closetoTc}.} \label{fig:fittingT=Tc}
\end{minipage}\end{flushright}
\end{figure}

The comparison of our result with the experimental observation of Ref.~\onlinecite{Aubin2007} for two $Nb_{0.15}Si_{0.85}$ films of thicknesses $35{\rm nm}$ and $12.5{\rm nm}$ (with critical temperatures $T_c=380{\rm mK}$ and $T_c=165{\rm mK}$, correspondingly) is given in Fig.~\ref{fig:fittingT=Tc}. The Peltier coefficient  depends on the mean field temperature of the superconducting transition, $T_c^{MF}$, and on the diffusion coefficient through  $\Omega_c$. Throughout the paper we fit the data using the same diffusion coefficient $D=0.187{\rm cm^2/sec}$ which is within the measurement accuracy of the value that is extracted from the experiment in Ref.~\onlinecite{Aubin2007}.

The cancelation of the terms proportional to $\Omega_c/T$ in the limit $T\ll{\Omega_c}$ was described in the previous section. (Without this cancellation  we would get a finite Nernst effect in the limit $T\rightarrow0$ and the third law of thermodynamics would be violated.) After the cancellation, the remaining contributions to $\alpha_{xy}$ in the limit $T\rightarrow0$ are linear in the temperature:
\begin{align}\label{eq:alphaB=Bc}
&\alpha_{xy}\approx-\frac{eT\ln3}{3\Omega_c(\ln{H/H_{c_{\scriptscriptstyle2}}(T)})^2}\hspace{5mm}\hbox{for $H\approx{H_{c_{\scriptscriptstyle2}}}$},
\end{align}
and
\begin{align}\label{eq:alphaB<Bc}
\alpha_{xy}\approx\frac{2eT}{3\Omega_c\ln{H/H_{c_{\scriptscriptstyle2}}}}\hspace{5mm}\hbox{for $H\gg{H_{c_{\scriptscriptstyle2}}}$}.
\end{align}
Similar to the limit $\Omega_c<T$ the integrals determining the final expression for $\alpha$ accumulate at low frequency. This situation is not typical for fluctuations induced by a quantum phase transition. Notice that $\alpha_{xy}$  changes its sign in this region. Since the transverse signal is non-dissipative the sign of the effect is not fixed. As mentioned before, in the vicinity of $T_c$ for $\Omega_c\ll{T}$, the main contribution to the Peltier coefficient is from the Aslamazov-Larkin term and the magnetization current. The magnetization current is opposite in sign to the Aslamazov-Larkin terms and equals $2/3$ of it. When crossing to the region $\ln(T/T_c(H))<\Omega_c/T$ (see the phase diagram in Fig.~\ref{fig:PhaseDiagram}) the contribution from the magnetization current grows. To the first order in $\Omega_c/T$ the magnetization current cancels the Aslamazov-Larkin term and the Peltier coefficient turns out to be proportional to $O[(\Omega_c/T)^2]$. Lowering further the temperature and increasing the magnetic field one reaches the region $\Omega_c>{T}$ and $\ln(H/H_{c_{\scriptscriptstyle2}}(T))<T/\Omega_c$. In this region the magnetization current becomes dominant. Since the magnetization current gives a contribution that is opposite in sign to the Aslamazov-Larkin term, we obtain that the Peltier coefficient is negative.

\begin{figure}[pt]
\begin{flushright}\begin{minipage}{0.5\textwidth}  \centering
        \includegraphics[width=0.75\textwidth]{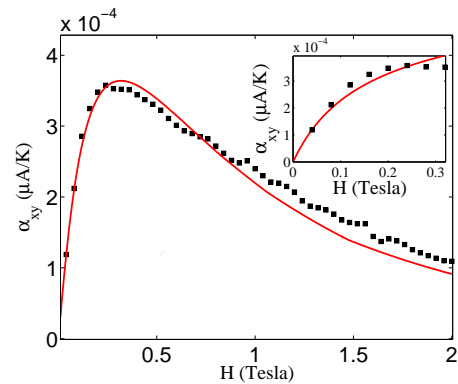} \hspace{0.05in}
                 \caption[0.4\textwidth]{\small The transverse Peltier coefficient $\alpha_{xy}$
 as a function of the magnetic field measured at $T=410{\rm mK}$. The black squares correspond to the experimental data of Ref.~8 while the solid line describes the theoretical result. The arrow on the phase diagram illustrates the direction of the measurement. In the insert the low magnetic field data is fitted with the theoretical curve given by Eq.~\ref{eq:closetoTc}.} \label{fig:fittingB}
\end{minipage}\end{flushright}
\end{figure}

In Fig.~\ref{fig:fittingB} we plot the Peltier coefficient for the $35{\rm nm}$ film as a function of the magnetic field at a temperature higher than $T_c$. We take $T_c^{MF}=385{\rm mK}$ to be slightly higher than the measured critical temperature anticipating a small suppression of the temperature of the transition by fluctuations. [The data in Fig.~\ref{fig:fittingB}, unlike the data in Fig.~\ref{fig:35nm}, is presented in linear rather than a logarithmic scale. Therefore, this fit is much more sensitive to the input parameters compared to the one in Fig.~\ref{fig:35nm}. As a result, the small deviation of $T_c^{MF}$ from the measured $T_c$ can be noticed. For consistency, we use the same value of $T_c^{MF}$ also in Fig.~\ref{fig:35nm}.] Fig.~\ref{fig:fittingB} demonstrates the agreement between the theoretical expressions and the experimental observation for a broad range of magnetic fields. In addition, we show that the experimental data is well described by Eq.~\ref{eq:closetoTc} in the limit of vanishing magnetic field; see inset of Fig.~\ref{fig:fittingB}. Since Eq.~\ref{eq:closetoTc} is valid in the limit $\Omega_c\ll{T}$,  it can describe only the first few point in the measurement. In order to fit the entire range of the magnetic field we had to include higher order terms in $\Omega_c/T$. For that we needed to sum the contributions from all diagrams and the magnetization current. We performed the calculation assuming that  $\ln(T/T_c(H))\ll1$; therefore the theoretical curve starts to deviate from the measured data when $\ln(T/T_c(H))$ is no longer small ($H\approx1Tesla$).

\section{Summary}

We demonstrated that the contribution from the fluctuations of the superconducting order parameter to the Nernst effect in disordered films is dominant and can be observed far away from the transition. We showed that the important role of the magnetization current is in cancelling the quantum contributions, thus making the Nernst signal compatible with the third law of thermodynamics. The third law of thermodynamics constrains the magnitude of the Nernst signal not only at low temperatures, but also far from $T_c$. As a consequence of this constraint the phase diagram is less rich and diverse than one expects in the vicinity of a quantum phase transition.

The Nernst effect provides an excellent opportunity to test the use of the quantum kinetic approach in the description of thermoelectric transport phenomena. We showed that in this scheme we get automatically all contributions to the Nernst coefficient as response to the temperature gradient, in particular the one from the magnetization current. This is an advantage of the quantum kinetic approach but it is not the only one.  This method also allows us to verify the Onsager relations; a comprehensive discussion of this issue is presented in Appendix~\ref{sec:OR}. The fact that the we were able to find independently the two off-diagonal components of the conductivity tensor, $\alpha_{ij}$ and $\tilde{\alpha}_{ij}$, and verify that they are connected through the Onsager relation assures that the quantum kinetic approach developed in this paper and in Ref.~\onlinecite{QKE} gives the correct expressions for the electric and heat currents.

Finally, we should remark that our results for the Peltier coefficient differ in few aspects from those obtained recently in Ref.~\onlinecite{Skvortsov2008} using the Kubo formula. As we already discussed in the end of Sec.~\ref{sec:Peltier}, the simplified Kubo formula cannot give the correct electric current as a response to a temperature gradient. Therefore, the claim of the authors of Ref.~\onlinecite{Skvortsov2008} that the difference between the Nernst signal calculated using the   simplified Kubo formula and the quantum kinetic approach is only in the numerical coefficients is unacceptable. The expression given in Ref.~\onlinecite{Skvortsov2008} for $\alpha_{xy}$ in the vicinity of $T_c$  cannot fit the experimental data, and it also contradicts the phenomenological result of the TDGL.~\cite{Ussishkin2002} The only fit of the experimental data presented in Ref.~\onlinecite{Skvortsov2008} is a logarithmic plot of the Nernst signal as a function of  temperature using the formula for temperatures not too close to $T_c$. Such a logarithmic plot is not very sensitive to the numerical coefficients. The striking agreement between our results and the experimental data, in particular our ability to obtain the non-trivial dependence of the Nernst signal on the magnetic field and the fact that we reproduced the phenomenological result~\cite{Ussishkin2002} reinforces us in the correctness of our method.

\begin{acknowledgments}
We thank  O.~Entin-Wohlman and the condensed matter theory group in TAMU for their interest in this work and for the extended discussions. The research was supported by the US-Israel BSF.
\end{acknowledgments}

\appendix

\section{Onsager's Relations}\label{sec:OR}

In this Appendix we compare the electric current arising as a response to a temperature gradient and the heat current generated by an electric field. We verify that for the Gaussian fluctuations of the superconducting order parameter, the two expression are connected through the Onsager relations,~\cite{Onsager1931} $\tilde{\alpha}_{ij}(\mathbf{B})=T_0{\alpha}_{ji}(-\mathbf{B})$. [In a similar way, in Ref.~\onlinecite{QKE} we  demonstrated the Onsager relations for the longitudinal current in the presence of the Coulomb interaction.]

The derivation of the electric current induced by the temperature gradient was presented in Sec.~\ref{sec:QKE}, where we discussed the two contributions to the current. The first one, $\mathbf{j}_{e}^{con}$, was found using the continuity equation and its expression before the expansion in the superconducting fluctuations is given in Eq.~\ref{eq:Pelt-JETransInv}. The second contribution, analyzed in Sec.~\ref{sec:Magnetization}, is from the magnetization current. This term contributes only to the transverse current, and it can be written as $j_{e\hspace{1.5mm}i}^{mag}=c\varepsilon_{i,j}M_z(-\boldsymbol{\nabla}_jT/T_0)$.

Next, we sketch the derivation of the heat current as a response to a uniform electric field. For this purpose we use the heat continuity equation, $\dot{Q}(\mathbf{r},t)+\boldsymbol{\nabla}\mathbf{j}_{h}^{con}(\mathbf{r},t)=\mathbf{j}_e^{con}\mathbf{E}$. The product $\mathbf{j}_e^{con}\mathbf{E}$ describes the work performed by the electric field on the current. Since the electric field cannot perform any work on the magnetization current, only $\mathbf{j}_{e}^{con}$ enters to the RHS of the continuity equation.~\cite{Jackson}  The heat density in the presence of an electro-magnetic field is a function of the magnetization and the electro-chemical potential:
\begin{align}\label{eq:OR-HeatDensity}
dQ(\mathbf{r},t)=dh(\mathbf{r},t)-(\mu-e\varphi(\mathbf{r})){dn}(\mathbf{r},t)+\mathbf{M}d\mathbf{H}(\mathbf{r},t),
\end{align}
where $h(\mathbf{r},t)$ is the Hamiltonian density.  To find the time derivative of the magnetic field, we turn to the  Maxwell equation $\dot{\mathbf{H}}=-c\boldsymbol{\nabla}\times\mathbf{E}$. Thus, the heat current is described by the following equation:
\begin{align}\label{eq:OR-HeatCon}
\boldsymbol{\nabla}\mathbf{j}_{h}
=-dh(\mathbf{r},t)+(\mu-e\varphi){dn}(\mathbf{r},t)-\mathbf{j}_e^{con}\boldsymbol{\nabla}\varphi+c\mathbf{M}\boldsymbol{\nabla}\times\mathbf{E}\\\nonumber
\end{align}
In Ref.~\onlinecite{QKE} we showed that for $\mathbf{H}=0$ the expression for the heat current found from the continuity equation is
\begin{align}\label{eq:OR-JhCon}\nonumber
&\mathbf{j}_{h}^{con}(\mathbf{H}=0)=\lim_{\begin{array}{c}\scriptstyle{ \mathbf{r}'\rightarrow\mathbf{r}} \\ \scriptstyle{t'\rightarrow{t}} \\
\end{array}}\hspace{-1mm}\left(\partial_{t}+ie\varphi(\mathbf{r})-\partial_{t'}+ie\varphi(\mathbf{r}')\right)\\
&\int{d\mathbf{r}_{\scriptscriptstyle1}dt_{\scriptscriptstyle1}}\left[\mathbf{\hat{\underline{\underline{v}}}}(\mathbf{r},t;\mathbf{r}_{\scriptscriptstyle1},t_{\scriptscriptstyle1})\hat{\underline{\underline{G}}}(\mathbf{r}_{\scriptscriptstyle1},t_{\scriptscriptstyle1};\mathbf{r}',t')\right]^{<}
+h.c.
\end{align}
This result was obtained for the Coulomb interaction but it also holds for the superconducting fluctuations because, unlike the charge, there is no principle difference between the way the fluctuations in the density and Cooper channels  carry heat. Here, we use the fact that in the presence of an interaction field, such as $\Delta$, which does not have its own dynamics,  the heat current can be formulated in terms of the quasi-particle Green function alone. This is compatible with the observation that according to the kinetic equation given in Eq.~\ref{eq:QKE-DEV}, the temperature gradient is coupled to $\hat{\underline{\underline{L}}}$ only through the quasi-particle Green functions inside $\hat{\underline{\underline{\Pi}}}$. [When interactions with dynamic fields like phonons are studied the heat current acquires additional terms.]

Although we restrict our derivation to the regime of linear response with respect to the electric field, the source term $\mathbf{j}_e^{con}\mathbf{E}$ is still important. This is because the source term makes the heat current to be gauge invariant as Eq.~\ref{eq:OR-JhCon} reveals. In principle, there may be an additional contribution to the heat current from the charge current carried by the superconducting fluctuations (corresponding to the RHS of Eq.~\ref{eq:Current-ContinuityEq}). This issue is not addressed here because such contributions are beyond the linear response.

When we consider the effect of applying a magnetic field the expression for the heat current given in Eq.~\ref{eq:OR-JhCon} has to be modified. The first change is simply to include the vector potential in the velocity as it is shown in Eq.~\ref{eq:QKE-velocityQP}. We denote the contribution from the heat current given in Eq.~\ref{eq:OR-JhCon} with the modified velocity as $\mathbf{j}_{h}^{con1}$. Besides, there is an additional contribution to the heat current, $\mathbf{j}_{h}^{con2}$, from the last term in Eq.~\ref{eq:OR-HeatDensity} that contains the magnetization:
\begin{align}\label{eq:OR-HeatMagCont}
\boldsymbol{\nabla}\mathbf{j}_{h}^{con2}=c\mathbf{M}(\boldsymbol{\nabla}\times\mathbf{E})=c\boldsymbol{\nabla}(\mathbf{E}\times\mathbf{M}).
\end{align}
Here we used the fact that under the condition of a constant magnetic field $\boldsymbol{\nabla}\times\mathbf{M}=0$. For the setup of the Nernst measurement (see Fig.~\ref{fig:NernstSetup}), in which the magnetic field is aligned along the $z$ directions, the contribution of the magnetization to the heat current is
\begin{align}\label{eq:OR-JhMag}
\mathbf{j}_{h\hspace{1.5mm}i}^{con2}=
c\varepsilon_{ij}E_jM_z.
\end{align}
At this stage, one may wonder whether there is a contribution to the transverse current that cannot be found from the continuity equation, i.e., a term of the form $\boldsymbol{\nabla}\times\mathbf{W}$. In the case of the electric current generated by the temperature gradient, we saw the term of this kind is the magnetization current. This term does not vanish and contributes to the transport electric current because the non-uniform temperature induces a coordinate dependent magnetization (see Sec.~\ref{sec:Magnetization}). However, in the presence of a constant electric field, the system remains uniform. Therefore, the quantity that we denoted by $\mathbf{W}$ should be independent of the spatial coordinate and, hence, $\boldsymbol{\nabla}\times\mathbf{W}=0$.

Let us compare between the electric current as a response to a temperature gradient and the heat current generated by a electric field when a magnetic field is applied. One may immediately notice that  $\mathbf{j}_{h}^{con2}$ and $\mathbf{j}_{e}^{mag}$ (given in Eqs.~\ref{eq:OR-JhMag} and~\ref{eq:QKE-JM}, respectively) satisfy the Onsager relations:
\begin{align}
\frac{j_{h\hspace{1.5mm}i}^{con2}(\mathbf{H})}{E_j}=T_0\frac{j_{e\hspace{1.5mm}j}^{mag}(-\mathbf{H})}{-\boldsymbol{\nabla}_iT}.
\end{align}
It is interesting that these two terms coincide although they seem to have a different origin. The contribution of the magnetization, $\boldsymbol{\nabla}\times\mathbf{M}$, to the electric current cannot be found from the continuity equation, and it is non-zero due to the dependence of $\hat{G}_{loc-eq}$ on the center of mass coordinate. When the response to a uniform electric field is considered, the Green functions are independent of the center of mass coordinate. Still, there is an equivalent contribution to the heat current arising from the continuity equation.

Now, we show that $\mathbf{j}_{e}^{con}$ and $\mathbf{j}_{h}^{con1}$ also satisfy the Onsager relations.
In order to find $\mathbf{j}_{h}^{con1}$ we need to know the expression for the electric field dependent propagators. The electric field dependent Green function can be written in the following form:~\cite{QKE}
\begin{align}\label{eq:OR-G_E}
&\hat{G}_{\mathbf{E}}(\boldsymbol{\rho},\epsilon;\mathbf{A},imp)=\hat{g}_{eq}\left(\epsilon\right)
\hat{\Sigma}_{\mathbf{E}}\left(\epsilon\right)\hat{g}_{eq}\left(\epsilon\right)\\\nonumber
&-\frac{ie\mathbf{E}}{2}\left[\frac{\partial\hat{g}_{eq}\left(\epsilon\right)}{\partial\epsilon}\mathbf{\hat{v}}_{eq}(\epsilon)\hat{g}_{eq}\left(\epsilon\right)
-\hat{g}_{eq}\left(\epsilon\right)\mathbf{\hat{v}}_{eq}(\epsilon)\frac{\partial\hat{g}_{eq}\left(\epsilon\right)}{\partial\epsilon}
\right].
\end{align}
The above equation is similar to  Eq.~\ref{eq:QKE-G_TransInv2} for $\hat{G}_{\boldsymbol{\nabla}T}$. Since   $\Sigma_{\mathbf{E}}$ contains also the electric field dependent propagator of the superconducting fluctuations, we have to find the equation for $\hat{L}_{\mathbf{E}}$. Owing to the fact that the superconducting fluctuations carry charge, their coupling to the electric field is more complicated than their dependence on the temperature gradient described in Eqs.~\ref{eq:QKE-V_loceq} and~\ref{eq:QKE-V_GradT}:
\begin{align}\label{eq:OR-L_E}
&\hat{L}_{\mathbf{E}}(\boldsymbol{\rho},\omega;\mathbf{A},imp)=-\hat{L}_{eq}\left(\omega\right)
\hat{\Pi}_{\mathbf{E}}\left(\omega\right)\hat{L}_{eq}\left(\omega\right)\\\nonumber
&+ie\mathbf{E}\left[\frac{\partial\hat{L}_{eq}\left(\omega\right)}{\partial\omega}\boldsymbol{\hat{\mathcal{V}}}_{eq}(\omega)\hat{L}_{eq}\left(\omega\right)
-\hat{L}_{eq}\left(\omega\right)\boldsymbol{\hat{\mathcal{V}}}_{eq}(\omega)\frac{\partial\hat{L}_{eq}\left(\omega\right)}{\partial\omega}
\right].
\end{align}
Inserting the expression for $G_{\mathbf{E}}$ into Eq.~\ref{eq:OR-JhCon} and extracting the lesser component, we get:
\begin{widetext}
\begin{align}\label{eq:OR-Jh}\nonumber
&{j}_{h\hspace{1.5mm}i}^{con1}=
\frac{eE_{j}}{2}\int\frac{d\epsilon}{2\pi}\epsilon\frac{\partial{n_F(\epsilon)}}{\partial\epsilon}\left[v_{i}^R(\epsilon)g_{eq}^{R}(\epsilon)v_{j}^A(\epsilon)g_{eq}^A(\epsilon)
+v_{i}^R(\epsilon)g_{eq}^{R}(\epsilon)v_{j}^R(\epsilon)g_{eq}^A(\epsilon)-
v_{i}^R(\epsilon)g_{eq}^{R}(\epsilon)v_{j}^R(\epsilon)g_{eq}^R(\epsilon)\right.\\\nonumber&\left.-
g_{eq}^{R}(\epsilon)v_{j}^R(\epsilon)g_{eq}^R(\epsilon)v_{i}^A(\epsilon)
\right]+
eE_j\int\frac{d\epsilon}{2\pi}\epsilon{n}_F(\epsilon)\left[v_{i}^R(\epsilon)\frac{\partial{g}_{eq}^R(\epsilon)}{\partial\epsilon}v_{j}^R(\epsilon)g_{eq}^R(\epsilon)-
v_{i}^R(\epsilon){g}_{eq}^R(\epsilon)v_{j}^R(\epsilon)\frac{\partial{g}_{eq}^R(\epsilon)}{\partial\epsilon}\right]\\
&+i\int\frac{d\epsilon}{2\pi}\epsilon{v}_{i}^R(\epsilon)g_{eq}^R(\epsilon)\left[\Sigma_{\mathbf{E}}^{<}(\epsilon)(1-n_F(\epsilon))+\Sigma_{\mathbf{E}}^{>}(\epsilon)n_F(\epsilon)\right](g_{eq}^{R}(\epsilon)-g_{eq}^A(\epsilon))+c.c.
\end{align}
\end{widetext}
The fulfilment of Onsager relation demands microscopic reversibility, which in our case implies that $\hat{G}(\mathbf{r},\mathbf{r}',\epsilon;\mathbf{H})=\hat{G}(\mathbf{r}',\mathbf{r},\epsilon;-\mathbf{H})$ and $\hat{L}(\mathbf{r},\mathbf{r}',\epsilon;\mathbf{H})=\hat{L}(\mathbf{r}',\mathbf{r},\epsilon;-\mathbf{H})$. Since the currents contain a trace over the coordinates, it is obvious that the first two terms in Eq.~\ref{eq:OR-Jh} and the first two terms in Eq.~\ref{eq:Pelt-JETransInv} are connected through the Onsager relations.

Let us examine the remaining terms in $\mathbf{j}_{e}^{con}$ and $\mathbf{j}_{h}^{con1}$. The electric current contains not only the contribution of $\Sigma_{\boldsymbol{\nabla}T}$ but also terms with $\hat{\Pi}_{\boldsymbol{\nabla}T}$, while the remaining part of the heat current contains only $\Sigma_{\mathbf{E}}$.  Actually, when we treat $\hat{\Sigma}_{\mathbf{E}}$ we must consider the possibility that $\mathbf{E}$ enters also through $\hat{L}$. Then, as we show later, since the equation for $\hat{L}_{E}$ includes more terms than the equation for $\hat{L}_{\boldsymbol{\nabla}{T}}$ (see Eqs.~\ref{eq:OR-L_E} and~\ref{eq:QKE-V_GradT}), the corresponding contributions to the electric and heat currents  coincide.


Let us start with the contributions to the currents  in which $\hat{\Sigma}$ depends on the electric filed/temperature gradient through $\hat{G}$ rather than $\hat{L}$. Since we are interested in the effect of Gaussian fluctuations, we can use the expressions for the $G_{\mathbf{E}}$ and $G_{\boldsymbol{\nabla}T}$ in the absence of interactions:
\begin{align}\label{eq:OR-GT}\nonumber
&G_{\mathbf{F}}^{<}(\epsilon)\hspace{-0.5mm}=\hspace{-0.5mm}\frac{i}{2}F_j(\epsilon){n}_F(\epsilon)\hspace{-1mm}\left[\frac{\partial{g_0^{R}(\epsilon)}}{\partial\epsilon}v_0^jg_0^{R}(\epsilon)
-g_0^{R}(\epsilon)v_0^j\frac{\partial{g}_0^{R}(\epsilon)}{\partial\epsilon}\right]\\
&+\frac{i}{2}F_j(\epsilon)\frac{\partial{n}_F(\epsilon)}{\partial\epsilon}g_0^{R}(\epsilon)v_0^j\left[g_0^{A}(\epsilon)-g_0^{R}(\epsilon)\right]-c.c,
\end{align}
where $F_j(\epsilon)$ is equal to $eE_j$ and $\epsilon\boldsymbol{\nabla}_jT/T_0$, respectively. The only difference between the above equation for the lesser component of  $\hat{G}_{\mathbf{F}}$ and the one for the greater component is that in the latter the distribution function, $n_{F}(\epsilon)$ should be replaced by $n_{F}(\epsilon)-1$. (In Eq.~\ref{eq:OR-GT} and below we start to place the spatial direction indices also as superscripts.)

\begin{figure}[h]
\begin{flushright}\begin{minipage}{0.5\textwidth}  \centering
        \includegraphics[width=0.5\textwidth]{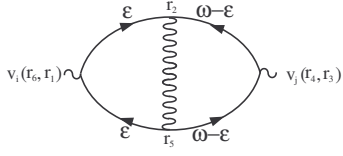}
                 \caption[0.4\textwidth]{\small A diagrammatic representation of the contribution to the current written in Eq.~\ref{eq:OR-JeG}. For simplicity the scattering by impurities is not indicated.}
                 \label{fig:Je1}
\end{minipage}\end{flushright}
\end{figure}

Using the identities given in Eq.~\ref{eq:Pelt-identites}, one may notice that for the discussed contributions to the currents (denoted as $\mathbf{j}_{e,h}(\hat{G}_{\mathbf{F}},\mathbf{H})$) only the terms proportional to the derivative of the distribution function in Eq.~\ref{eq:OR-GT} do not vanish. As a result we get:
\begin{widetext}
\begin{align}\label{eq:OR-JeG}
j_{e/h}^{i}&(\hat{G}_{\mathbf{F}},\mathbf{H})
=-\frac{i}{2}\int\frac{d\epsilon{d}\omega}{(2\pi)^{2}}d\mathbf{r}_{\scriptscriptstyle2}...d\mathbf{r}_{\scriptscriptstyle6}f_{e/h}(\epsilon)F_j(\omega-\epsilon){v}_{0}^i(\mathbf{r}_{\scriptscriptstyle6},\mathbf{r}_{\scriptscriptstyle1})
\left[g_{0}^R(\mathbf{r}_{\scriptscriptstyle1},\mathbf{r}_{\scriptscriptstyle2},\epsilon)-g_{0}^A(\mathbf{r}_{\scriptscriptstyle1},\mathbf{r}_{\scriptscriptstyle2},\epsilon)\right]\\\nonumber
&\left[L_{eq}^{R}(\mathbf{r}_{\scriptscriptstyle2},\mathbf{r}_{\scriptscriptstyle5},\omega)-L_{eq}^{A}(\mathbf{r}_{\scriptscriptstyle2},\mathbf{r}_{\scriptscriptstyle5},\omega)\right]
{v}_{0}^j(\mathbf{r}_{\scriptscriptstyle4},\mathbf{r}_{\scriptscriptstyle3})
\left[g_{0}^R(\mathbf{r}_{\scriptscriptstyle5},\mathbf{r}_{\scriptscriptstyle4},\omega-\epsilon)-g_{0}^A(\mathbf{r}_{\scriptscriptstyle5},\mathbf{r}_{\scriptscriptstyle4},\omega-\epsilon)\right]
\left[g_{0}^R(\mathbf{r}_{\scriptscriptstyle3},\mathbf{r}_{\scriptscriptstyle2},\omega-\epsilon)-g_{0}^A(\mathbf{r}_{\scriptscriptstyle3},\mathbf{r}_{\scriptscriptstyle2},\omega-\epsilon)\right]\\\nonumber
&\left[g_{0}^R(\mathbf{r}_{\scriptscriptstyle6},\mathbf{r}_{\scriptscriptstyle1},\epsilon)-g_{0}^A(\mathbf{r}_{\scriptscriptstyle6},\mathbf{r}_{\scriptscriptstyle1},\epsilon)\right]
\frac{\partial{n_P(\omega)}}{\partial\omega}[n_F(\epsilon-\omega)-n_F(\epsilon)].
\end{align}
\end{widetext}
Here, $f_{e}(\epsilon)=-e$ and $f_{h}(\epsilon)=\epsilon$. The diagrammatic representation of the above expression is presented in Fig.~\ref{fig:Je1}. Recall that the bare velocity  $\mathbf{v}_0(\mathbf{r},\mathbf{r}')\propto\delta(\mathbf{r-r}')$.  Finally, we shall change the frequency as follows $\epsilon\rightarrow\omega-\epsilon$:
\begin{widetext}
\begin{align}\label{eq:OR-JeG2}
j_{e/h}^{i}&(\hat{G}_{\mathbf{F}},\mathbf{H})
=-\frac{i}{2}\int\frac{d\epsilon{d}\omega}{(2\pi)^{2}}d\mathbf{r}_{\scriptscriptstyle2}...d\mathbf{r}_{\scriptscriptstyle6}f_{e/h}(\omega-\epsilon)F_j(\epsilon){v}_{0}^i(\mathbf{r}_{\scriptscriptstyle6},\mathbf{r}_{\scriptscriptstyle1})
\left[g_{0}^R(\mathbf{r}_{\scriptscriptstyle1},\mathbf{r}_{\scriptscriptstyle2},\omega-\epsilon)-g_{0}^A(\mathbf{r}_{\scriptscriptstyle1},\mathbf{r}_{\scriptscriptstyle2},\omega-\epsilon)\right]\\\nonumber
&\left[L_{eq}^{R}(\mathbf{r}_{\scriptscriptstyle2},\mathbf{r}_{\scriptscriptstyle5},\omega)-L_{eq}^{A}(\mathbf{r}_{\scriptscriptstyle2},\mathbf{r}_{\scriptscriptstyle5},\omega)\right]
{v}_{0}^j(\mathbf{r}_{\scriptscriptstyle4},\mathbf{r}_{\scriptscriptstyle3})
\left[g_{0}^R(\mathbf{r}_{\scriptscriptstyle5},\mathbf{r}_{\scriptscriptstyle4},\epsilon)-g_{0}^A(\mathbf{r}_{\scriptscriptstyle5},\mathbf{r}_{\scriptscriptstyle4},\epsilon)\right]
\left[g_{0}^R(\mathbf{r}_{\scriptscriptstyle3},\mathbf{r}_{\scriptscriptstyle2},\epsilon)-g_{0}^A(\mathbf{r}_{\scriptscriptstyle3},\mathbf{r}_{\scriptscriptstyle2},\epsilon)\right]\\\nonumber
&\left[g_{0}^R(\mathbf{r}_{\scriptscriptstyle6},\mathbf{r}_{\scriptscriptstyle1},\omega-\epsilon)-g_{0}^A(\mathbf{r}_{\scriptscriptstyle6},\mathbf{r}_{\scriptscriptstyle1},\omega-\epsilon)\right]
\frac{\partial{n_P(\omega)}}{\partial\omega}[n_F(\epsilon-\omega)-n_F(\epsilon)].
\end{align}
\end{widetext}
Under the condition of a microscopic reversibility mentioned previously, we see that the electric current created by a temperature gradient as given in Eq.~\ref{eq:OR-JeG} (with $f_{e}(\epsilon)=-e$ and $F_j(\omega-\epsilon)=(\omega-\epsilon)\boldsymbol{\nabla}_jT/T_0$) and the heat current genereted by an electric field as described in Eq.~\ref{eq:OR-JeG2} (with $f_{h}(\omega-\epsilon)=\omega-\epsilon$ and $F_j(\epsilon)=E_j$) satisfy the Onsager relations. Thus, the microscopic reversibility and the Onsager relations emerging from it correspond to reading the diagram in Fig.~\ref{fig:Je1} from right to left instead of left to right (i.e., reading it in Hebrew instead of English).

Next, we shall examine the contribution to the currents in which the propagator of the superconducting fluctuations (or the polarization operator) depends on the electric field/ temperature gradient. We start with the corresponding contribution to the heat current as a response to an electric field, $\mathbf{j}_h(L_{\mathbf{E}},\mathbf{H})$. Using the identities for the distribution functions in Eq.~\ref{eq:Pelt-identites}, this term can can be written as:
\begin{align}\label{eq:OR-JhL}\nonumber
&j_h^{i}(L_{\mathbf{E}},\mathbf{H})=\hspace{-1.5mm}\int\hspace{-1mm}\frac{d\epsilon{d}\omega}{(2\pi)^2}d\mathbf{r}_{\scriptscriptstyle2}...d\mathbf{r}_{\scriptscriptstyle4}\epsilon{v}_{0}^i(\mathbf{r}_{\scriptscriptstyle4},\mathbf{r}_{\scriptscriptstyle1})
\left[n_F(\epsilon)-n_F(\epsilon-\omega)\right]\\\nonumber
&\left[g_{eq}^A(\mathbf{r}_{\scriptscriptstyle1},\mathbf{r}_{\scriptscriptstyle2},\epsilon)-g_{eq}^R(\mathbf{r}_{\scriptscriptstyle1},\mathbf{r}_{\scriptscriptstyle2},\epsilon)\right]
\left[g_{eq}^A(\mathbf{r}_{\scriptscriptstyle3},\mathbf{r}_{\scriptscriptstyle4},\epsilon)-g_{eq}^R(\mathbf{r}_{\scriptscriptstyle3},\mathbf{r}_{\scriptscriptstyle4},\epsilon)\right]\\\nonumber
&\left[g_{eq}^A(\mathbf{r}_{\scriptscriptstyle3},\mathbf{r}_{\scriptscriptstyle2},\omega\epsilon)-g_{eq}^R(\mathbf{r}_{\scriptscriptstyle3},\mathbf{r}_{\scriptscriptstyle2},\omega-\epsilon)\right]
\\
&\left[(1+n_P(\omega))L_{\mathbf{E}}^{<}(\mathbf{r}_{\scriptscriptstyle2},\mathbf{r}_{\scriptscriptstyle3},\omega)
-n_P(\omega)L_{\mathbf{E}}^{>}(\mathbf{r}_{\scriptscriptstyle2},\mathbf{r}_{\scriptscriptstyle3},\omega)\right].
\end{align}
Using the definition of $\hat{\Pi}_{\boldsymbol{\nabla}T}$, one may notice that the expression for the heat current can be rewritten in the form:
\begin{align}\label{eq:OR-JhL2}
j_h^{i}&(L_{\mathbf{E}},\mathbf{H})=\frac{T_0}{\boldsymbol{\nabla}_iT}\int\frac{d\omega}{2\pi}d\mathbf{r}'\left(\frac{\partial{n_P(\omega)}}{\partial\omega}\right)^{-1}\\\nonumber
&\times\left[\Pi_{\boldsymbol{\nabla}_iT}^{<}(\mathbf{r}',\mathbf{r},\omega)(1+n_P(\omega))-\Pi_{\boldsymbol{\nabla}_iT}^{>}(\mathbf{r}',\mathbf{r},\omega)n_P(\omega)\right]\\\nonumber
&\times\left[(1+n_P(\omega))L_{E_j}^{<}(\mathbf{r},\mathbf{r}',\omega)
-n_P(\omega)L_{E_j}^{>}(\mathbf{r},\mathbf{r}',\omega)\right].
\end{align}

Let us now turn to the equivalent contributions to the electric current created  by a temperature gradient, $\mathbf{j}_e(L_{\boldsymbol{\nabla}T},\mathbf{H})$. Performing manipulations similar to those in the expression for the heat current we get:
\begin{widetext}
\begin{align}\label{eq:OR-JeL}
j_e^{j}&(L_{\boldsymbol{\nabla}T},\mathbf{H})=-ie\int\frac{d\omega}{2\pi}d\mathbf{r}_{\scriptscriptstyle2}..d\mathbf{r}_{\scriptscriptstyle4}\left[
(L_{eq}^{R}(\mathbf{r}_{\scriptscriptstyle3},\mathbf{r}_{\scriptscriptstyle4},\omega)-L_{eq}^{A}(\mathbf{r}_{\scriptscriptstyle3},\mathbf{r}_{\scriptscriptstyle4},\omega))\mathcal{V}_j^R(\mathbf{r}_{\scriptscriptstyle4},\mathbf{r}_{\scriptscriptstyle1},\omega)L_{eq}(\mathbf{r}_{\scriptscriptstyle1},\mathbf{r}_{\scriptscriptstyle2},\omega)
-L_{eq}^A(\mathbf{r}_{\scriptscriptstyle3},\mathbf{r}_{\scriptscriptstyle4},\omega)\mathcal{V}_j^A(\mathbf{r}_{\scriptscriptstyle4},\mathbf{r}_{\scriptscriptstyle1},\omega)\right.\\\nonumber
&\left.\times(L_{eq}^{R}(\mathbf{r}_{\scriptscriptstyle1},\mathbf{r}_{\scriptscriptstyle2},\omega)-L_{eq}^{A}(\mathbf{r}_{\scriptscriptstyle1},\mathbf{r}_{\scriptscriptstyle2},\omega))
\right]\left[(1+n_P(\omega))\Pi_{\boldsymbol{\nabla}_iT}^{<}(\mathbf{r}_{\scriptscriptstyle2},\mathbf{r}_{\scriptscriptstyle3},\omega)-
n_P(\omega)\Pi_{\boldsymbol{\nabla}_iT}^{>}(\mathbf{r}_{\scriptscriptstyle2},\mathbf{r}_{\scriptscriptstyle3},\omega)\right]\\\nonumber
&+e\int\frac{d\epsilon{d}\omega}{(2\pi)^2}d\mathbf{r}_{\scriptscriptstyle2}..d\mathbf{r}_{\scriptscriptstyle4}v_0^{j}(\mathbf{r}_{\scriptscriptstyle4},\mathbf{r}_{\scriptscriptstyle1})\left[g_0^{A}(\mathbf{r}_{\scriptscriptstyle1},\mathbf{r}_{\scriptscriptstyle2},\epsilon)-g_0^{R}(\mathbf{r}_{\scriptscriptstyle1},\mathbf{r}_{\scriptscriptstyle2},\epsilon)\right]
\left[g_0^{A}(\mathbf{r}_{\scriptscriptstyle3},\mathbf{r}_{\scriptscriptstyle2},\omega-\epsilon)-g_0^{R}(\mathbf{r}_{\scriptscriptstyle3},\mathbf{r}_{\scriptscriptstyle2},\omega-\epsilon)\right]\\\nonumber
&\times\left[(1+n_P(\omega))L_{\boldsymbol{\nabla}_iT}^{<}(\mathbf{r}_{\scriptscriptstyle2},\mathbf{r}_{\scriptscriptstyle3},\omega)-n_P(\omega)L_{\boldsymbol{\nabla}_iT}^{>}(\mathbf{r}_{\scriptscriptstyle2},\mathbf{r}_{\scriptscriptstyle3},\omega)\right]
[n_F(\epsilon)-n_F(\epsilon-\omega)]
\left[g_0^{A}(\mathbf{r}_{\scriptscriptstyle3},\mathbf{r}_{\scriptscriptstyle4},\epsilon)-g_0^{R}(\mathbf{r}_{\scriptscriptstyle3},\mathbf{r}_{\scriptscriptstyle4},\epsilon)\right].
\end{align}
\end{widetext}
Keeping in mind the definition of the polarization operator, we may rewrite the second integral in terms of $\hat{\Pi}_{\mathbf{E}}(\omega)$. Then we may collect the two contributions into a more compact expression:
\begin{widetext}
\begin{align}\label{eq:OR-JeL2}
&j_e^{j}(L_{\boldsymbol{\nabla}T},\mathbf{H})=-\frac{1}{E_j}\int\frac{d\omega}{2\pi}d\mathbf{r}_{\scriptscriptstyle2}...d\mathbf{r}_{\scriptscriptstyle4}\left[(1+n_P(\omega))\Pi_{\boldsymbol{\nabla}_iT}^{<}(\mathbf{r}_{\scriptscriptstyle3},\mathbf{r}_{\scriptscriptstyle4},\omega)
-n_P(\omega)\Pi_{\boldsymbol{\nabla}_iT}^{>}(\mathbf{r}_{\scriptscriptstyle3},\mathbf{r}_{\scriptscriptstyle4},\omega)\right]\left(\frac{\partial{n_P(\omega)}}{\partial\omega}\right)^{-1}\\\nonumber
&\times\left\{
-ieE_j\frac{\partial{n_P(\omega)}}{\partial\omega}\left[L_{eq}^{A}(\mathbf{r}_{\scriptscriptstyle4},\mathbf{r}_{\scriptscriptstyle1},\omega)
\mathcal{V}_i^A(\mathbf{r}_{\scriptscriptstyle1},\mathbf{r}_{\scriptscriptstyle2},\omega)
\left(L_{eq}^{R}(\mathbf{r}_{\scriptscriptstyle2},\mathbf{r}_{\scriptscriptstyle3},\omega)-L_{eq}^{A}(\mathbf{r}_{\scriptscriptstyle2},\mathbf{r}_{\scriptscriptstyle3},\omega)\right)
-\left(L_{eq}^{R}(\mathbf{r}_{\scriptscriptstyle4},\mathbf{r}_{\scriptscriptstyle1},\omega)-L_{eq}^{A}(\mathbf{r}_{\scriptscriptstyle4},\mathbf{r}_{\scriptscriptstyle1},\omega)\right)\right.\right.\\\nonumber
&\left.\left.\mathcal{V}_i^R(\mathbf{r}_{\scriptscriptstyle1},\mathbf{r}_{\scriptscriptstyle2},\omega)L_{eq}^{R}(\mathbf{r}_{\scriptscriptstyle2},\mathbf{r}_{\scriptscriptstyle3},\omega)
\right]+L_{eq}^{A}(\mathbf{r}_{\scriptscriptstyle4},\mathbf{r}_{\scriptscriptstyle1},\omega)
\left[(1+n_P(\omega))\Pi_{E_j}^{<}(\mathbf{r}_{\scriptscriptstyle1},\mathbf{r}_{\scriptscriptstyle2},\omega)-n_P(\omega)\Pi_{E_j}^{>}(\mathbf{r}_{\scriptscriptstyle1},\mathbf{r}_{\scriptscriptstyle2},\omega)\right]L_{eq}^{R}(\mathbf{r}_{\scriptscriptstyle2},\mathbf{r}_{\scriptscriptstyle3},\omega)\phantom{\frac{1}{1}}\hspace{-3mm}\right\}.\\\nonumber
\end{align}
\end{widetext}
The expression inside the curly brackets in the above equation can be written as $(1+n_P(\omega))L_{\mathbf{E}}^{<}(\omega)-n_P(\omega)L_{\mathbf{E}}^{>}(\omega)$. To obtain this identity one should find the lesser and greater components of $\hat{L}$ from Eq.~\ref{eq:OR-L_E}. A simple calculation reveals that in this combination of $L_{\mathbf{E}}^{<}$ and $L_{\mathbf{E}}^{>}$ only the terms proportional to the derivative of the Bose distribution function and those including $\Pi_{\mathbf{E}}^{<,>}$ (which also contain $\partial{n_P(\omega)}/\partial\omega$) give a non-zero contribution. Once again, if we invert the direction of the propagation of all the ingredients in Eq.~\ref{eq:OR-JeL2} and also the direction of the magnetic field we get that the expression for this last contribution to the electric current becomes:
\begin{align}\label{eq:OR-JeL3}
j_e^{j}&(L_{\boldsymbol{\nabla}T},-\mathbf{H})=-\frac{1}{E_j}\int\frac{d\omega}{2\pi}d\mathbf{r}'\left(\frac{\partial{n_P(\omega)}}{\partial\omega}\right)^{-1}\\\nonumber
&\times\left[\Pi_{\boldsymbol{\nabla}_iT}^{<}(\mathbf{r}',\mathbf{r},\omega)(1+n_P(\omega))-\Pi_{\boldsymbol{\nabla}_iT}^{>}(\mathbf{r}',\mathbf{r},\omega)n_P(\omega)\right]\\\nonumber
&\times\left[(1+n_P(\omega))L_{E_j}^{<}(\mathbf{r},\mathbf{r}',\omega)
-n_P(\omega)L_{E_j}^{>}(\mathbf{r},\mathbf{r}',\omega)\right].
\end{align}
Comparing  $j_e^{j}(L_{\boldsymbol{\nabla}_iT},-\mathbf{B})$ given above and $j_h^{i}(E_j,\mathbf{B})$ presented in Eq.~\ref{eq:OR-JhL2}, one immediately sees that they are indeed  connected by the Onsager relations.

In conclusion,  we demonstrated the Onsager relation for the Gaussian fluctuations of the superconducting order parameter (i.e., to the leading order in $(\varepsilon_F\tau)^{-1}$). The structure of the expressions for the electric and heat currents, Eqs.~\ref{eq:Pelt-JETransInv} and~\ref{eq:OR-Jh}, indicates that the same is true for any order. [An example for a general proof of the Onsager relations is given in Sec.~6 of Ref.~\onlinecite{QKE}.]

\section{Cancelation of the thermoelectric current in the limit $T\rightarrow0$}\label{App:Magnetization}

In this Appendix we demonstrate how the contribution to the transverse component of $\mathbf{j}_{e}^{con}$ that does not vanish when $T\rightarrow0$ is cancelled by the magnetization current given in Eq.~\ref{eq:QKE-JM}. From the three terms constituting $\mathbf{j}_{e}^{con}$, which are described in Sec.~\ref{sec:Peltier}, the one that remains at low temperatures, $\mathbf{j}_{e}^{con3}$, is presented in Eq.~\ref{eq:QKE-JETransInv4}. Let us restore the general expression from which $\mathbf{j}_{e}^{con3}$ originates:
\begin{align}\label{eq:App-Mag-Jcon}
&j_{e\hspace{1.5mm}i}^{con3}=\frac{e\boldsymbol{\nabla}_jT}{T_0}\int\frac{d\epsilon}{2\pi}d\mathbf{r}_{\scriptscriptstyle2}...d\mathbf{r}_{\scriptscriptstyle6}n_{F}(\epsilon)
\left[v_0^i(\mathbf{r}_{\scriptscriptstyle6},\mathbf{r}_{\scriptscriptstyle1})\right.\\\nonumber&\left.\times{g}_0^{R}(\mathbf{r}_{\scriptscriptstyle1},\mathbf{r}_{\scriptscriptstyle2},\epsilon)
\sigma_{eq}^{R}(\mathbf{r}_{\scriptscriptstyle2},\mathbf{r}_{\scriptscriptstyle3},\epsilon)g_0^{R}(\mathbf{r}_{\scriptscriptstyle3},\mathbf{r}_{\scriptscriptstyle4},\epsilon)
v_0^j(\mathbf{r}_{\scriptscriptstyle4},\mathbf{r}_{\scriptscriptstyle5})g_0^{R}(\mathbf{r}_{\scriptscriptstyle5},\mathbf{r}_{\scriptscriptstyle6},\epsilon)\right.\\\nonumber
&\left.-v_0^i(\mathbf{r}_{\scriptscriptstyle6},\mathbf{r}_{\scriptscriptstyle1})g_0^{R}(\mathbf{r}_{\scriptscriptstyle1},\mathbf{r}_{\scriptscriptstyle2},\epsilon)
v_0^j(\mathbf{r}_{\scriptscriptstyle2},\mathbf{r}_{\scriptscriptstyle3})g_0^{R}(\mathbf{r}_{\scriptscriptstyle3},\mathbf{r}_{\scriptscriptstyle4},\epsilon)\sigma_{eq}^{R}(\mathbf{r}_{\scriptscriptstyle4},\mathbf{r}_{\scriptscriptstyle5},\epsilon)\right.\\\nonumber&\left.\times{g}_0^{R}(\mathbf{r}_{\scriptscriptstyle5},\mathbf{r}_{\scriptscriptstyle6},\epsilon)
\right]+c.c.
\end{align}
Here we use the notation $\mathbf{v}_0(\mathbf{r},\mathbf{r}')$ for the bare velocity (see Eq.~\ref{eq:QKE-velocityQP} with $\hat{\sigma}_{eq}$ taken to be zero). Following Refs.~\onlinecite{Streda1977} and~\onlinecite{Streda1982} and recalling that we are interested in the current averaged over the space, we may replace the convolution $g_0^{R}(\epsilon)\mathbf{v}_0g_0^{R}(\epsilon)$ in the above equation with $-i(\mathbf{r}-\mathbf{r}')g_0^{R}(\mathbf{r},\mathbf{r}',\epsilon)$. Then, the contribution to the current from $\mathbf{j}_{e}^{con3}$ is:
\begin{align}\label{eq:App-Mag-Jcon2}
&j_{e\hspace{1.5mm}i}^{con3}=-i\frac{e\boldsymbol{\nabla}_jT}{T_0}\int\frac{d\epsilon}{2\pi}d\mathbf{r}_{\scriptscriptstyle2}...d\mathbf{r}_{\scriptscriptstyle4}n_{F}(\epsilon)\\\nonumber&
\left[v_0^j(\mathbf{r}_{\scriptscriptstyle4},\mathbf{r}_{\scriptscriptstyle1})(\mathbf{r}_{\scriptscriptstyle1}-\mathbf{r}_{\scriptscriptstyle2})_i{g}_0^{R}(\mathbf{r}_{\scriptscriptstyle1},\mathbf{r}_{\scriptscriptstyle2},\epsilon)
\sigma_{eq}^{R}(\mathbf{r}_{\scriptscriptstyle2},\mathbf{r}_{\scriptscriptstyle3},\epsilon)g_0^{R}(\mathbf{r}_{\scriptscriptstyle3},\mathbf{r}_{\scriptscriptstyle4},\epsilon)
\right.\\\nonumber
&\left.-v_0^i(\mathbf{r}_{\scriptscriptstyle4},\mathbf{r}_{\scriptscriptstyle1})(\mathbf{r}_{\scriptscriptstyle1}-\mathbf{r}_{\scriptscriptstyle2})_jg_0^{R}(\mathbf{r}_{\scriptscriptstyle1},\mathbf{r}_{\scriptscriptstyle2},\epsilon)\sigma_{eq}^{R}(\mathbf{r}_{\scriptscriptstyle2},\mathbf{r}_{\scriptscriptstyle3},\epsilon){g}_0^{R}(\mathbf{r}_{\scriptscriptstyle3},\mathbf{r}_{\scriptscriptstyle4},\epsilon)
\phantom{v_0^j}\hspace{-3mm}\right]\\\nonumber&+c.c.
\end{align}
The above expression can be rewritten using the magnetization $\mathbf{M}$ defined below Eq.~\ref{eq:Current-Jmag}:
\begin{align}\label{eq:App-Mag-Jcon3}\nonumber
&j_{e\hspace{1.5mm}i}^{con3}=-2i\varepsilon_{ij}\frac{\boldsymbol{\nabla}_jT}{T_0}cM_z\lim_{\boldsymbol{\rho}\rightarrow0}\int\frac{d\epsilon}{2\pi}{g}_{eq}^{<}(\boldsymbol{\rho},\epsilon;\mathbf{A},imp)\\\nonumber&
+i\epsilon_{ij}\frac{e\boldsymbol{\nabla}_jT}{2T_0}\int\frac{d\epsilon}{2\pi}d\mathbf{r}_{\scriptscriptstyle2}...d\mathbf{r}_{\scriptscriptstyle4}n_F(\epsilon)
\varepsilon_{z\alpha\beta}(\mathbf{r}_{\scriptscriptstyle2}+\mathbf{r}_{\scriptscriptstyle3})_{\alpha}{v}_0^{\beta}(\mathbf{r}_{\scriptscriptstyle4},\mathbf{r}_{\scriptscriptstyle1})\\\nonumber
&\left[g_0^{R}(\mathbf{r}_{\scriptscriptstyle1},\mathbf{r}_{\scriptscriptstyle2},\epsilon)\sigma_{eq}^{R}(\mathbf{r}_{\scriptscriptstyle2},\mathbf{r}_{\scriptscriptstyle3},\epsilon){g}_0^{R}(\mathbf{r}_{\scriptscriptstyle3},\mathbf{r}_{\scriptscriptstyle4},\epsilon)\right.\\&\left.
-g_0^{A}(\mathbf{r}_{\scriptscriptstyle1},\mathbf{r}_{\scriptscriptstyle2},\epsilon)\sigma_{eq}^{A}(\mathbf{r}_{\scriptscriptstyle2},\mathbf{r}_{\scriptscriptstyle3},\epsilon){g}_0^{A}(\mathbf{r}_{\scriptscriptstyle3},\mathbf{r}_{\scriptscriptstyle4},\epsilon)\right].
\end{align}
Obviously, when we add the magnetization current presented in Eq.~\ref{eq:QKE-Jmag3} to $\mathbf{j}_{e}^{con3}$, only the second term in Eq.~\ref{eq:App-Mag-Jcon3} remains. We shall denote this remaining term as $\mathbf{j}_{e}^{rem}$.

\begin{figure}[pt]
\begin{flushright}\begin{minipage}{0.5\textwidth}  \centering
        \includegraphics[width=0.6\textwidth]{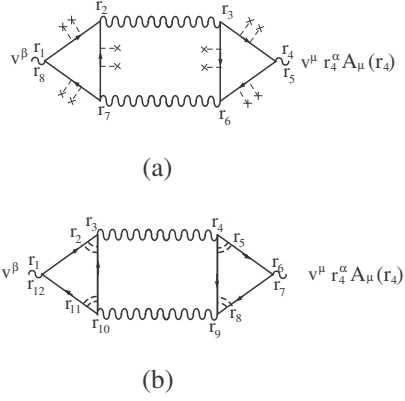}
                 \caption[0.4\textwidth]{\small (a) The contribution to $\mathbf{j}_{e}^{rem}$ (a) before and (b) after averaging over the disorder.} \label{fig:Jrem}
\end{minipage}\end{flushright}
\end{figure}

Next, we replace $\sigma_{eq}$ with its explicit expression given in Eq.~\ref{eq:Pelt-SelfEnergy}:
\begin{align}\label{eq:App-Mag-Jcon4}\nonumber
&j_{e\hspace{1.5mm}i}^{rem}=\epsilon_{ij}\frac{e\boldsymbol{\nabla}_jT}{2T_0}\int\frac{d\epsilon{d}\omega}{(2\pi)^{2}}d\mathbf{r}_{\scriptscriptstyle2}...d\mathbf{r}_{\scriptscriptstyle4}n_P(\omega)
\varepsilon_{z\alpha\beta}(\mathbf{r}_{\scriptscriptstyle2}+\mathbf{r}_{\scriptscriptstyle3})_{\alpha}\\\nonumber&\times{v}_0^{\beta}(\mathbf{r}_{\scriptscriptstyle4},\mathbf{r}_{\scriptscriptstyle1})
g_0^{R}(\mathbf{r}_{\scriptscriptstyle1},\mathbf{r}_{\scriptscriptstyle2},\epsilon)g_0^{A}(\mathbf{r}_{\scriptscriptstyle3},\mathbf{r}_{\scriptscriptstyle2},\omega\epsilon)
g_0^{R}(\mathbf{r}_{\scriptscriptstyle3},\mathbf{r}_{\scriptscriptstyle4},\epsilon)\\\nonumber&\times
\left[L^{R}(\mathbf{r}_{\scriptscriptstyle2},\mathbf{r}_{\scriptscriptstyle3},\omega)n_F(\epsilon-\omega)
-L^{A}(\mathbf{r}_{\scriptscriptstyle2},\mathbf{r}_{\scriptscriptstyle3},\omega)n_F(\epsilon)
\right]\\&+c.c.
\end{align}
Here, we dropped terms with three retarded (advanced) quasi-particles Green functions and we used the identities for the products of distribution functions presented in Eq.~\ref{eq:Pelt-identites}. We show now that $\mathbf{j}_{e}^{rem}$ contributes to the current only at finite temperature, $\mathbf{j}_{e}^{rem}\xrightarrow[T\rightarrow0]{}0$. We demonstrate that $\mathbf{j}_{e}^{rem}$ is closely related to the contribution of the Aslamazov-Larkin diagram to the magnetic susceptibility.~\cite{Aslamazov1974} To calculate $\mathbf{j}_{e}^{rem}$,  we allow the magnetic field to depend on the coordinate; in the end of the derivation we shall take the limit of a uniform magnetic field.

Let us first concentrate on the product $(\mathbf{r}+\mathbf{r}')L^{R,A}(\mathbf{r},\mathbf{r}',\omega)$. Since the magnetic field varies in space, the dependence of the propagator $L$ on the center of mass coordinate is through the magnetic field. The result of acting with the operator $R_{\alpha}$ on $L(\mathbf{R};\boldsymbol{\rho},\omega)$ is equal to the derivative with respect to the vector potential $A_{\beta}(\mathbf{r})$ multiplied by $r_{\alpha}A_{\beta}(\mathbf{r})$:
\begin{align}\label{eq:App-Mag-derivative}\nonumber
&(\mathbf{r}+\mathbf{r}')_{\alpha}L^{R}(\mathbf{r},\mathbf{r}',\omega)=-4i\frac{e}{c}\int\frac{d\epsilon'}{2\pi}d\mathbf{r}_{\scriptscriptstyle1}d\mathbf{r}_{\scriptscriptstyle2}d\mathbf{r}_{\scriptscriptstyle3}
\tanh\left(\frac{\omega-\epsilon'}{2T}\right)\\\nonumber&\times
L^{R}(\mathbf{r},\mathbf{r}_{\scriptscriptstyle1},\omega)v_{0}^{\beta}(\mathbf{r}_{\scriptscriptstyle2},\mathbf{r}_{\scriptscriptstyle3})r_{\scriptscriptstyle2}^{\alpha}A_{\beta}(\mathbf{r}_{\scriptscriptstyle2})
g^{R}(\mathbf{r}_{\scriptscriptstyle1},\mathbf{r}_{\scriptscriptstyle2},\epsilon')g^{R}(\mathbf{r}_{\scriptscriptstyle3},\mathbf{r}_{\scriptscriptstyle4},\epsilon')\\&\times
g^{A}(\mathbf{r}_{\scriptscriptstyle1},\mathbf{r}_{\scriptscriptstyle4},\omega-\epsilon')
L^{R}(\mathbf{r}_{\scriptscriptstyle4},\mathbf{r}',\omega).
\end{align}
Now, we may return to the limit of a constant magnetic field, $\mathbf{H}(\mathbf{r})=H\hat{z}$, and set the vector potential to be $\mathbf{A}(\mathbf{r})=\mathbf{H}(\mathbf{r})\times\mathbf{r}/2$. Inserting the expression given in Eq.~\ref{eq:App-Mag-derivative} into $j_{e\hspace{1.5mm}i}^{rem}$ presented in Eq.~\ref{eq:App-Mag-Jcon4}, we obtain:
\begin{widetext}
\begin{align}\label{eq:App-Mag-Jrem0}\nonumber
&j_{e\hspace{1.5mm}i}^{rem}=-i\epsilon_{ij}\frac{e^2\boldsymbol{\nabla}_jT}{T_0c}\int\frac{d\epsilon{d}\epsilon'd\omega}{(2\pi)^3}d\mathbf{r}_{\scriptscriptstyle2}...d\mathbf{r}_{\scriptscriptstyle8}n_P(\omega)
\tanh\left(\frac{\omega-\epsilon'}{2T}\right)\varepsilon_{z\alpha\beta}v_0^{\beta}(\mathbf{r}_{\scriptscriptstyle8},\mathbf{r}_{\scriptscriptstyle1})
v_{0}^{\mu}(\mathbf{r}_{\scriptscriptstyle4},\mathbf{r}_{\scriptscriptstyle5})r_{\scriptscriptstyle4}^{\alpha}\varepsilon_{z\nu\mu}H(\mathbf{r}_{\scriptscriptstyle4})\mathbf{r}_{\scriptscriptstyle4}^{\nu}
g_0^{R}(\mathbf{r}_{\scriptscriptstyle1},\mathbf{r}_{\scriptscriptstyle2},\epsilon)\\\nonumber
&\times
g_0^{A}(\mathbf{r}_{\scriptscriptstyle3},\mathbf{r}_{\scriptscriptstyle2},\omega-\epsilon)
g_0^{R}(\mathbf{r}_{\scriptscriptstyle3},\mathbf{r}_{\scriptscriptstyle4},\epsilon)
\left\{L^{R}(\mathbf{r}_{\scriptscriptstyle2},\mathbf{r}_{\scriptscriptstyle3},\omega)
g^{R}(\mathbf{r}_{\scriptscriptstyle3},\mathbf{r}_{\scriptscriptstyle4},\epsilon')g^{R}(\mathbf{r}_{\scriptscriptstyle5},\mathbf{r}_{\scriptscriptstyle6},\epsilon')
g^{A}(\mathbf{r}_{\scriptscriptstyle3},\mathbf{r}_{\scriptscriptstyle6},\omega-\epsilon')
L^{R}(\mathbf{r}_{\scriptscriptstyle6},\mathbf{r}_{\scriptscriptstyle7},\omega)n_F(\epsilon-\omega)\right.\\&\left.
+L^{A}(\mathbf{r}_{\scriptscriptstyle3},\mathbf{r}_{\scriptscriptstyle2},\omega)
g^{A}(\mathbf{r}_{\scriptscriptstyle3},\mathbf{r}_{\scriptscriptstyle4},\epsilon')
g^{A}(\mathbf{r}_{\scriptscriptstyle5},\mathbf{r}_{\scriptscriptstyle6},\epsilon')
g^{R}(\mathbf{r}_{\scriptscriptstyle3},\mathbf{r}_{\scriptscriptstyle6},\omega-\epsilon')
L^{A}(\mathbf{r}_{\scriptscriptstyle6},\mathbf{r}_{\scriptscriptstyle7},\omega)n_F(\epsilon)
\right\}+c.c.
\end{align}
\end{widetext}
The diagrammatic interpretation of the above expression is presented in Fig.~\ref{fig:Jrem}(a). Unlike the superconducting fluctuations, the electrons are considered to be three dimensional. Then, for an isotropic system we may rewrite the product of the antisymmetric tensors as: $\varepsilon_{z\alpha\beta}\varepsilon_{z\nu\mu}=(\delta_{\nu,\alpha}\delta_{\mu,\beta}-\delta_{\nu,\beta}\delta_{\mu,\alpha})/3$. Now we average over the disorder  (see Fig.~\ref{fig:Jrem}(b)) and represent all the propagators (Green functions, propagators of the superconducting fluctuations and Cooperons) as a product of the phases and the gauge invariant terms. All the phases are collected into the function $e^{i\Phi}$. After performing the Fourier transform, the equation for $j_{e\hspace{1.5mm}i}^{rem}$ becomes:
\begin{widetext}
\begin{align}\label{eq:App-Mag-Jrem1}\nonumber
&j_{e\hspace{1.5mm}i}^{rem}=-i\epsilon_{ij}\frac{e^2\boldsymbol{\nabla}_jT}{3T_0c}(\delta_{\nu,\alpha}\delta_{\mu,\beta}-\delta_{\nu,\beta}\delta_{\mu,\alpha})\int\frac{d\epsilon{d}\epsilon'd\omega}{(2\pi)^3}\frac{d\mathbf{k}d\mathbf{k}'d\mathbf{q}d\mathbf{Q}}{(2\pi)^{4d}}n_P(\omega)
\tanh\left(\frac{\omega-\epsilon'}{2T}\right)
e^{i\Phi}\frac{\partial^2H(\mathbf{Q})}{\partial{Q}^{\alpha}\partial{Q}^{\nu}}
\tilde{g}_0^{R}(\mathbf{k},\epsilon)\\\nonumber&\times\tilde{v}_0^{\beta}(\mathbf{k},\mathbf{k}+\mathbf{Q})
\tilde{g}_0^{R}(\mathbf{k}+\mathbf{Q},\epsilon)
\tilde{g}_0^{A}(\mathbf{q}-\mathbf{k},\omega-\epsilon)
\left\{\tilde{C}^{R}(\mathbf{q}+\mathbf{Q},\epsilon,\omega-\epsilon)\tilde{L}^{R}(\mathbf{q}+\mathbf{Q},\omega)\tilde{C}^{R}(\mathbf{q}+\mathbf{Q},\epsilon',\omega-\epsilon')
\right.\\\nonumber&\left.\times\tilde{g}^{R}(\mathbf{k}'+\mathbf{Q},\epsilon')
\tilde{v}_{0}^{\mu}(\mathbf{k}'+\mathbf{Q},\mathbf{k}')\tilde{g}^{R}(\mathbf{k}',\epsilon')
\tilde{g}^{A}(\mathbf{q}-\mathbf{k}',\omega-\epsilon')
\tilde{C}^{R}(\mathbf{q},\epsilon',\omega-\epsilon')L^{R}(\mathbf{q},\omega)\tilde{C}^{R}(\mathbf{q},\epsilon,\omega-\epsilon)n_F(\epsilon-\omega)
\right.\\\nonumber&\left.+\tilde{C}^{R}(\mathbf{q}+\mathbf{Q},\epsilon,\omega-\epsilon)\tilde{L}^{A}(\mathbf{q}+\mathbf{Q},\omega)\tilde{C}^{A}(\mathbf{q}+\mathbf{Q},\epsilon',\omega-\epsilon')
\tilde{g}^{A}(\mathbf{k}'+\mathbf{Q},\epsilon')
\tilde{v}_{0}^{\mu}(\mathbf{k}'+\mathbf{Q},\mathbf{k}')\tilde{g}^{A}(\mathbf{k}',\epsilon')
\tilde{g}^{R}(\mathbf{q}-\mathbf{k}',\omega-\epsilon')\right.\\&\left.\times
\tilde{C}^{A}(\mathbf{q},\epsilon',\omega-\epsilon')\tilde{L}^{A}(\mathbf{q},\omega)\tilde{C}^{R}(\mathbf{q},\epsilon,\omega-\epsilon)n_F(\epsilon)
\right\}+c.c.
\end{align}
\end{widetext}
Here, $\mathbf{H}(\mathbf{Q})=(2\pi)^{d}H\delta(\mathbf{Q})\hat{z}$ and $\mathbf{v}_{0}(\mathbf{k}',\mathbf{k})=\left(\frac{\mathbf{k}}{2m}-i\frac{e\mathbf{H}}{c}\times\frac{\overrightarrow{\partial}}{\partial\mathbf{k}}+
\frac{\mathbf{k}'}{2m}+i\frac{e\mathbf{H}}{c}\times\frac{\overleftarrow{\partial}}{\partial\mathbf{k}'}\right)$ is the Fourier transform of the velocity; the arrows above the derivatives indicate on which of the Green functions the derivative acts.

Let us transfer the derivatives with respect to the momentum $\mathbf{Q}$ from the magnetic field to  the rest of the expression using integration by parts. Since $\Phi$ which collects all the phases contains only  derivatives with respect to the momenta, it will not be differentiated in the course of this operation. In the diffusive limit, the main contribution to the current is obtained when the derivatives with respect to $\mathbf{Q}$ act on the propagators of the collective modes, either $L$ or $C$ (here we rely on the arguments given below Eq.~\ref{eq:QKE-ALFlux2}). Moreover, it follows from the tensor structure of Eq.~\ref{eq:App-Mag-Jrem1} that only terms in which $\partial^2/\partial{Q}_{\alpha}\partial{Q}_{\nu}=2\delta_{\alpha,\nu}\partial/\partial{Q}^2$ survive. [One should keep in mind that
the gauge invariant propagators of the superconducting fluctuations and Cooperons depend on the square of the momentum.] Then, $j_{e\hspace{1.5mm}i}^{rem}$ can be written as:
\begin{widetext}
\begin{align}\label{eq:App-Mag-Jrem2}\nonumber
&j_{e\hspace{1.5mm}i}^{rem}=-2i\epsilon_{ij}\frac{e^2\boldsymbol{\nabla}_jT}{3T_0c}H\int\frac{d\epsilon{d}\epsilon'd\omega}{(2\pi)^3}\frac{d\mathbf{k}d\mathbf{k}'d\mathbf{q}}{(2\pi)^{3d}}n_P(\omega)
\tanh\left(\frac{\omega-\epsilon'}{2T}\right)
e^{i\Phi}
\tilde{g}_0^{R}(\mathbf{k},\epsilon)\tilde{v}_0^{\beta}(\mathbf{k},\mathbf{k})\\\nonumber&\times
\tilde{g}_0^{R}(\mathbf{k},\epsilon)
\tilde{g}_0^{A}(\mathbf{q}-\mathbf{k},\omega-\epsilon)
\left\{\frac{\partial}{\partial{q^2}}\left[\tilde{C}^{R}(\mathbf{q},\epsilon,\omega-\epsilon)\tilde{L}^{R}(\mathbf{q},\omega)\tilde{C}^{R}(\mathbf{q},\epsilon',\omega-\epsilon')\right]
\tilde{g}^{R}(\mathbf{k}',\epsilon')
\tilde{v}_{0}^{\mu}(\mathbf{k}',\mathbf{k}')\tilde{g}^{R}(\mathbf{k}',\epsilon')\right.\\\nonumber&\left.\times
\tilde{g}^{A}(\mathbf{q}-\mathbf{k}',\omega-\epsilon')
\tilde{C}^{R}(\mathbf{q},\epsilon',\omega-\epsilon')L^{R}(\mathbf{q},\omega)\tilde{C}^{R}(\mathbf{q},\epsilon,\omega-\epsilon)n_F(\epsilon-\omega)
+\frac{\partial}{\partial{q^2}}\left[\tilde{C}^{R}(\mathbf{q},\epsilon,\omega-\epsilon)\tilde{L}^{A}(\mathbf{q},\omega)\tilde{C}^{A}(\mathbf{q},\epsilon',\omega-\epsilon')\right]
\right.\\&\left.\times\tilde{g}^{A}(\mathbf{k}',\epsilon')
\tilde{v}_{0}^{\mu}(\mathbf{k}',\mathbf{k}')\tilde{g}^{A}(\mathbf{k}',\epsilon')
\tilde{g}^{R}(\mathbf{q}-\mathbf{k}',\omega-\epsilon')
\tilde{C}^{A}(\mathbf{q},\epsilon',\omega-\epsilon')\tilde{L}^{A}(\mathbf{q},\omega)\tilde{C}^{R}(\mathbf{q},\epsilon,\omega-\epsilon)n_F(\epsilon)
\right\}+c.c.
\end{align}
\end{widetext}

The next step is to integrate over the electronic degrees of freedom and to transform to the basis of the Landau levels. This can be performed following the explanation presented in Sec.~\ref{sec:Peltier}. The only difference is in the matrix elements for the Landau levels. While in the calculation presented in the main text the matrix element is $\langle{N,0}|V_{x}V_{y}|M,0\rangle$, here we have $\langle{N,0}|V_{x}^2+V_{y}^2|M,0\rangle=4D^2\left[(N+1)\delta_{N,M-1}+(M+1)\delta_{M,N-1}\right]/\ell_H^2$. Finally, after replacing the derivative with respect to $q^2$ with a derivative with respect to the index of the Landau levels, the expression for $j_{i}^{rem}$ acquires the form:
\begin{widetext}
\begin{align}\label{eq:App-Mag-Jrem3}\nonumber
&j_{e\hspace{1.5mm}i}^{rem}=-4i\pi\frac{\nu^2{D}^2}{3\ell_{H}^4}\tau^4\varepsilon_{ij}\frac{e\boldsymbol{\nabla}_jT}{T_0}\int\frac{d\epsilon{d}\omega{d}\epsilon'}{(2\pi)^3}
\sum_{N}(N+1)\tanh\left(\frac{\omega-\epsilon'}{2T}\right)
n_p(\omega)
\frac{\partial}{\partial{N}}\left[
n_F(\epsilon-\omega)C_{N}^{R}(\epsilon,\omega-\epsilon)L_N^{R}(\omega)
\right.\\\nonumber&\left.\times{C}_N^{R}(\epsilon',\omega-\epsilon')
C_{N+1}^{R}(\epsilon',\omega-\epsilon')
L_{N+1}^{R}(\omega)
C_{N+1}^{R}(\epsilon,\omega-\epsilon)+
n_F(\epsilon)C_{N}^{R}(\epsilon,\omega-\epsilon)L_N^{A}(\omega)
C_N^{A}(\epsilon',\omega-\epsilon')
C_{N+1}^{R}(\epsilon',\omega-\epsilon')\right.\\&\left.\times
L_{N+1}^{A}(\omega)
C_{N+1}^{A}(\epsilon,\omega-\epsilon)\right]
+c.c.
\end{align}
\end{widetext}
To recognize that the above expression goes to zero in the limit $T\rightarrow0$, we have to integrate over the fermionic frequencies $\epsilon$ and $\epsilon'$:
\begin{align}\label{eq:App-Mag-Jrem3}
&j_{e\hspace{1.5mm}i}^{rem}=\frac{i}{24\pi^2}\nu^2\varepsilon_{ij}\frac{e\boldsymbol{\nabla}_jT}{T_0}\int{d\omega}
\sum_{N}(N+1)
n_p(\omega)\\\nonumber
&\times\frac{\partial}{\partial{N}}
L_N^{R}(\omega)L_{N+1}^{R}(\omega)\left[\psi_{N}^{R}\left(\frac{1}{2}-\frac{i\omega}{4\pi{T_0}}+\frac{\Omega_C(N+1/2)}{4\pi{T}_0}\right)\right.\\\nonumber&\left.
-\psi_{N}^{R}\left(\frac{1}{2}-\frac{i\omega}{4\pi{T_0}}+\frac{\Omega_C(N+3/2)}{4\pi{T}_0}\right)\right]^2
+c.c.
\end{align}
now, we can exploit the fact that in the expression differentiated with respect to N the argument $N$ stands together with the frequency, $-i\omega+\Omega_cN$. Replacing the derivative $\partial/\partial{N}$ by a derivative with respect to the frequency and integrating by parts, one immediately gets that the above integral includes the factor $\partial{n}_{P}(\omega)/\partial\omega$ and, hence, vanishes at zero temperature.

To summaries, we  showed that in accordance with the third law of thermodynamics, the transverse thermoelectric transport coefficients $\alpha_{xy}$ and $\tilde{\alpha}_{xy}$ go to zero in the limit $T\rightarrow0$. [As to the longitudinal coefficients, both $\mathbf{j}_{e}^{con3}$ and $\mathbf{j}_{e}^{mag}$ do not appear while the remaining contributions vanish independently at $T\rightarrow0$.] It follows from this result that at finite $T$, one can obtain the temperature dependence of the thermoelectric current by substituting $n_P(\omega)$ with $n_P(\omega)+\Theta(-\omega)$ in all the expressions determining $\mathbf{j}_{e}^{con}$ and $\mathbf{j}_{e}^{mag}$.

\section{The quasi-particles contribution to the thermoelectric transport coefficients}\label{sec:p-hSymmtery}

In this Appendix we discuss the role of the particle-hole asymmetry and the constant density of states approximation in determining the quasi-particles contribution to the thermoelectric transport coefficients. In metallic conductors the quasi-particle excitations yield a negligible contribution to the Nernst effect and to its counterpart, the Ettingshausen effect. Let us consider a system with an electron and hole conducting bands that have a particle-hole symmetry, i.e., the bands are identical and the two species of particles differ only in their charge. We shall describe the deviation of the distribution functions of the two species ($\delta{f}_e$ and $\delta{f}_h$) from their equilibrium value in the linear response to a temperature gradient. For that we write the classical Boltzamann equation in the relaxation time approximation:
\begin{equation}
\frac{\delta{f}_{e,h}(\epsilon_{\mathbf{k}})}{\tau}=\frac{\partial{f_0}(\epsilon_{\mathbf{k}})}{\partial{T}}\mathbf{v}_{\mathbf{k}}\boldsymbol{\nabla}T\mp\frac{e\mathbf{v}_{\mathbf{k}}\times\mathbf{H}}{c}\frac{\partial{f_0(\epsilon_{\mathbf{k}})}}{\partial\mathbf{k}}.
\label{eq:PHS-BoltzmannEq}
\end{equation}%
Here the equilibrium (Fermi-Dirac) distribution function is denoted by $f_0(\epsilon_k)$, and $\mathbf{v}_k$ is the velocity of the particles.

The electric current is the sum of the electric currents due to the electrons and the holes:
\begin{equation}
\mathbf{j}_e^{total}=-2e\int\frac{d\mathbf{k}}{(2\pi)^d}\mathbf{v}_{\mathbf{k}}\delta{f}_{e}(\epsilon_{\mathbf{k}})+2e\int\frac{d\mathbf{k}}{(2\pi)^d}\mathbf{v}_{\mathbf{k}}\delta{f}_{h}(\epsilon_{\mathbf{k}}).
\label{eq:PHS-Currentp+h}
\end{equation}%
Notice that the factor of $2$ results form the sum over the two spin directions. For simplicity we only examine the limit of vanishingly small magnetic field. In order to determine whether a current vanishes in the particle-hole symmetric system we just need to count the powers of the electric charge; an odd power means cancellation of the two contributions to the current.

We start from the longitudinal electric current induced by the temperature gradient. In the limit $\mathbf{H}\rightarrow0$, the longitudinal current is independent of the magnetic field:
\begin{equation}
\mathbf{j}_e^{x}=2e\int\frac{d\mathbf{k}}{(2\pi)^d}\frac{\partial{f}_{0}(\epsilon_{\mathbf{k}})}{\partial\epsilon_{\mathbf{k}}}\left[\epsilon_{\mathbf{k}}D_e-\epsilon_{\mathbf{k}}D_h\right]\frac{\boldsymbol{\nabla}_xT}{T_0}=0,
\label{eq:PHS-Longitudinal}
\end{equation}%
where $D_e=D_h\equiv{D}=v_k^2\tau/d$ with $d$ the dimension of the system. Since the expression includes only one power of the charge, there is no longitudinal electric current unless particle-hole asymmetry is introduced.

The transverse current is obtained when the Lorentz force in the Boltzmann equation acts on the distribution function. Therefore, the expression for the transverse current contains an additional power of the charge:
\begin{equation}
\mathbf{j}_e^{y}=2e\int\frac{d\mathbf{k}}{(2\pi)^d}\frac{\partial{f}_{0}(\epsilon_{\mathbf{k}})}{\partial\epsilon_{\mathbf{k}}}\epsilon_{\mathbf{k}}D\left[\omega_c\tau-(-\omega_c\tau)\right]\frac{\boldsymbol{\nabla}_xT}{T_0}\neq0.
\label{eq:PHS-Transverse}
\end{equation}
The additional charge enters through the cyclotron frequency $\omega_c={eH/m^{*}c}$. The even power of the charge means that the particle-hole symmetry does not constrain the Nernst effect.

Now, we look at the contribution for the transverse electric current in a metal with only one conducting band. We use the approximate of a constant density of states which is standard for Fermi liquid systems. Under this approximation the expression for the transverse current is:
\begin{equation}
\mathbf{j}_e^{y}=2e\nu_0D(\omega_c\tau)\frac{\boldsymbol{\nabla}{T}}{T_0}\int{d\epsilon_{\mathbf{k}}}\frac{\partial{f}_{0}(\epsilon_{\mathbf{k}})}{\partial\epsilon_{\mathbf{k}}}\epsilon_{\mathbf{k}}.
\label{eq:PHS-ConstDOS}
\end{equation}
Since near the Fermi energy the integrand is an odd function of the energy, the resulting current is zero. Therefore, under the approximation of a constant density of states at the Fermi energy this contribution vanishes.~\cite{Sondheimer1948} One may conclude that in metallic systems with high Fermi energy the contribution of the quasi-particles to the Nernst signal includes a small factor related to the deviation from the constant density of states which is of the order $T/\varepsilon_F$. In semi-metals like $Bi$ where the constant density of states approximation cannot be used, a large Nernst signal was measured.~\cite{Behnia2007}

Let us compare the magnitudes of the transverse Peltier coefficient generated by the quasi-particles and by the superconducting fluctuations in a film of thickness $a$. The first is of the order $\sim(\omega_c\tau)e\nu{D}a{T}/\varepsilon_F$ for $\omega_c\tau\ll1$ while the second one is of the order $\sim{e}\Omega_c/T$ for $\Omega_c/T\ll1$ and $\sim{eT}/\Omega_c$ for the opposite limit. Thus in the limit of vanishing small magnetic field the ratio between the contribution of the quasi-particles and the fluctuations is $\alpha_{xy}^{qp}/\alpha_{xy}^{fl}\sim(k_Fa)T^2\tau/\varepsilon_F$. At higher magnetic fields (but still in the limit $\omega_c\tau\ll1$) this ratio becomes $\alpha_{xy}^{qp}/\alpha_{xy}^{fl}\sim(k_Fa)\varepsilon_F\tau(\omega_c\tau)^2\ll1$. Under the condition of the experiment,~\cite{Aubin2006,Aubin2007} the ratio $\alpha_{xy}^{qp}/\alpha_{xy}^{fl}\ll1$ up to $T\lesssim100T_c$ and $H\lesssim100H_{c_2}$. The reason why the Nernst signal generated by the superconducting fluctuations dominates the one produced by the quasi-particles was explained in the end of Sec.~\ref{sec:Peltier}.


\begin{thebibliography}{}

\bibitem{Zaanen2007} J.~Zaanen, \textit{Nature}~\textbf{448}, 1000 (2007).

\bibitem{Sachdev2007} S.~A.~Hartnoll, P.~K.~Kovtun, M.~Muller, and S.~Sachdev,
\textit{Phys.~Rev.~B}~\textbf{76}, 144502 (2007).

\bibitem{Ong2000} Z.~A.~Xu, N.~P.~Ong, Y.~Wang, T.~Kakeshita, and S.~Uchida, \textit{Nature}~\textbf{406}, 486 (2000).

\bibitem{Ong2005} Y.~Wang, L.~Li, and N.~P.~Ong, \textit{Phys.~Rev.~B}~\textbf{ 73}, 024510 (2006).

\bibitem{Aubin2006} A.~Pourret, H.~Aubin, J.~Lesueur, C.~A.~Marrache-Kikuchi, L.~Bergée, L.~Dumoulin, and
K.~Behnia, \textit{Nat.~Phys.}~\textbf{2}, 683 (2006).

\bibitem{Aubin2007} A.~Pourret, H.~Aubin, J.~Lesueur, C.~A.~Marrache-Kikuchi, L.~Berg´e, L.~Dumoulin, and K.~Behnia, \textit{Phys.~Rev.~B}~\textbf{76}, 214504 (2007).


\bibitem{Anderson2007} P.~W.~Anderson,
\textit{Nat.~Phys.}~\textbf{3}, 160 (2007).

\bibitem{Podolsky2007} D.~Podolsky, S.~Raghu, and A.~Vishwanath,
\textit{Phys.~Rev.~Lett.}~\textbf{99}, 117004 (2007).

\bibitem{Huse2004} S.~Mukerjee, and D.~A.~Huse,
\textit{Phys.~Rev.~B}~\textbf{70}, 014506 (2004).


\bibitem{Sondheimer1948} E.~Sondheimer,
\textit{Proc. Roy. Soc. (London)}~\textbf{A193}, 484 (1948).


\bibitem{Aslamazov1968} L.~G.~Aslamazov, and A.~I.~Larkin, \textit{Fiz. Tverd. Tela}~\textbf{10}, 1104 (1968) [\textit{Sov. Phys. Solid State}~\textbf{10}, 875 (1968)].

\bibitem{Maki1968} K.~Maki,
\textit{Prog.~Theor.~Phys.}~\textbf{40}, 193 (1968).

\bibitem{Varlamov} A.~I.~Larkin, and  A.~A.~Varlamov,
\textit{Theory of fluctuations in superconductors},
(Carendon, Oxford, 2005).

\bibitem{Kubo1957} R.~Kubo,
\textit{J.~Phys.~Soc.~Jpn.}~\textbf{12}, 570 (1957).

\bibitem{Luttinger1964} J.~M.~Luttinger,\textit{Phys.~Rev.}~\textbf{135}, A1505  (1964).

\bibitem{commentH} Notice that Eq.~\ref{eq:CurrentNonInt} can be rewriten using the commutation relation for the field operators as:
$\mathbf{j}_h(\tau)=\frac{1}{2}\sum_{\mathbf{p}}\frac{\partial\varepsilon_{\mathbf{p}}}{\partial\mathbf{p}}\{c_{\mathbf{p}}^{\dag}[c_{\mathbf{p}},H]-[c_{\mathbf{p}}^{\dag},H]c_{\mathbf{p}}\}$, where $H$ is the Hamiltonian.

\bibitem{QKE} K.~Michaeli, and A.~M.~Finkel'stein, \textit{Phys.~Rev.~B}~\textbf{80}, 115111  (2009).


\bibitem{Obraztsov1965} Y.~N.~Obraztsov,
\textit{Fiz.~Tverd.~Tela.}~\textbf{7}, 573 (1965) [\textit{Sov.~Phys.~Solid State}~\textbf{7}, 455-461 (1965)].


\bibitem{Streda1977} L.~Smrcka, and P.~Streda,
\textit{J.~Phys.~C:~Solid~State~Phys.}~\textbf{10}, 2153  (1977).

\bibitem{Halperin1997} N.~R.~Cooper, B.~I.~Halperin, and I.~M.~Ruzin,
\textit{Phys.~Rev.~B}~\textbf{55}, 2344  (1997).


\bibitem{Keldysh1964} L.~V.~Keldysh, \textit{Zh.~Eksp.~Teor.~Fiz.}~\textbf{47}, 1515 (1964) [\emph{Sov.~Phys.~JETP}~\textbf{20}, 1018 (1965)].


\bibitem{Rammer1986} J.~Rammer, and H.~Smith, \textit{Rev.~Mod.~Phys.}~\textbf{58}, 323 (1986).


\bibitem{Haug} H.~Haug, and A.-P.~Jauho,  \textit{Quantum Kinetics in Transport and Optics of Semiconductors},
(Springer, Berlin, 1996).


\bibitem{Hu1976}  C.-R.~Hu,
\textit{Phys.~Rev.~B}~\textbf{14}, 4834  (1976).


\bibitem{KM2008}  K.~Michaeli, and A.~M.~Finkel'stein, \emph{Europhys.~Lett.}~\textbf{86}, 27007 (2009).


\bibitem{Reizer2008}  A.~Sergeev, M.~Y.~Reizer, and V.~Mitin, \textit{Phys.~Rev.~B}~\textbf{77}, 064501 (2008).



\bibitem{Khodas2003} M.~A.~Khodas, and A.~M.~Finkel'stein,
\textit{Phys.~Rev.~B.}~\textbf{68}, 155114 (2003).


\bibitem{Larkin1995} A.~G.~Aronov, S.~Hikami, and A.~I.~Larkin,
 \textit{Phys.~Rev.~B}~\textbf{51}, 3880  (1995).

\bibitem{Laikhtman1994} B.~Laikhtman, and E.~L.~Altshuler,
\textit{Ann.~Phys.}~\textbf{232}, 332 (1994).

\bibitem{Galitski2001} V.~M.~Galitski, and A.~I.~Larkin,
\textit{Phys.~Rev.~B}~\textbf{63}, 174506 (2001).



\bibitem{Ussishkin2002} I.~Ussishkin, S.~L.~Sondhi, and D.~A.~Huse,
\textit{Phys.~Rev.~Lett.}~\textbf{89}, 287001 (2002).

\bibitem{comment} Besides the genral proof presented in Appendix~\ref{App:Magnetization}, we checked straightforwardly the cancelation of the $\tau$-dependent terms originating from Eqs.~\ref{eq:QKE-JETransInv4} and~\ref{eq:QKE-Jmag7}.

\bibitem{Skvortsov2008} M.~N.~Serbyn, M.~A.~Skvortsov, A.~A.~Varlamov, and V.~Galitski,
\textit{Phys.~Rev.~Lett.}~\textbf{102}, 067001 (2009).


\bibitem{Onsager1931} L.~Onsager, \textit{Phys.~Rev.}~\textbf{37}, 405 (1931); L.~Onsager, \textit{Phys.~Rev.}~\textbf{38}, 2265 (1931).


\bibitem{Jackson} J.~D.~Jackson,
\textit{Classical electrodynamics},
(Wiley, New York, 1962).



\bibitem{Streda1982} P.~Streda, \textit{J.~Phys.~C:~Solid~State~Phys.}~\textbf{15}, L717  (1982).

\bibitem{Aslamazov1974} L.~G.~Aslamazov, and A.~I.~Larkin, \textit{Zh.~Eksp.~Teor.~Fiz.}~\textbf{67}, 647 (1974) [\emph{Sov.~Phys.~JETP}~\textbf{40}, 321 (1975)].




\bibitem{Behnia2007} K.~Behnia, M.-A.~Measson, and Y.~Kopelevich,
\textit{Phys.~Rev.~Lett.}~\textbf{98}, 166602 (2007).




\end{thebibliography}
\end{document}